\documentclass[12pt,preprint]{aastex}
\usepackage{txfonts}

\newcommand{\dmfb}{$\Delta m_{15}(B)$}
\newcommand{\dmf}{$\Delta m_{15}$}
\newcommand{\bvi}{\hbox{$BV\!I$ }}
\newcommand{\vi}{\hbox{$V\!-\!I$}}               
\newcommand{\ri}{\hbox{$R\!-\!I$}}
\newcommand{\bvri}{\hbox{$BV\!RI$ }}
\newcommand{\ubvri}{\hbox{$U\!BV\!RI$ }}          

\newcommand{\kms}{{\rm km \; s^{-1}}}
\newcommand{\kmsmpc}{{\rm km \; s^{-1} \; Mpc^{-1}}}

\slugcomment{to appear in \emph{The Astrophysical Journal}}
\shorttitle{MLCS2k2 Distances to SN Ia}
\shortauthors{Jha, Riess, \& Kirshner}

\begin{document}
\title{Improved Distances to Type Ia Supernovae with \\ Multicolor Light
  Curve Shapes: MLCS2k2}

\author{Saurabh~Jha\altaffilmark{1,2}, Adam~G.~Riess\altaffilmark{3}, 
and Robert~P.~Kirshner\altaffilmark{4}}
\altaffiltext{1}{Miller Institute and Department of Astronomy, 601
Campbell Hall, University of California, Berkeley, CA 94720}
\altaffiltext{2}{present address: Kavli Institute for Particle
  Astrophysics and Cosmology, Stanford Linear Accelerator Center, 2575
  Sand Hill Road MS 29, Menlo Park, CA 94025}
\altaffiltext{3}{Space Telescope Science Institute, 3700 San
Martin Drive, Baltimore, MD 21218}
\altaffiltext{4}{Harvard-Smithsonian Center for Astrophysics, 60
Garden Street, Cambridge, MA 02138}
\email{saurabh@slac.stanford.edu}

\begin{abstract}
We present an updated version of the Multicolor Light Curve Shape
method to measure distances to type Ia supernovae (SN Ia),
incorporating new procedures for $K$-correction and extinction
corrections. We also develop a simple model to disentangle intrinsic
color variations and reddening by dust, and expand the method to
incorporate $U$-band light curves and to more easily accommodate prior
constraints on any of the model parameters. We apply this method to
133 nearby SN Ia, including 95 objects in the Hubble flow ($cz \geq
2500 \; \kms$), which give an intrinsic dispersion of less than 7\% in
distance. The Hubble flow sample, which is of critical importance to
all cosmological uses of SN Ia, is the largest ever presented with
homogeneous distances. We find the Hubble flow supernovae with $H_0
d_{\rm SN} \geq 7400 \; \kms$ yield an expansion rate that is 6.5
$\pm$ 1.8\% lower than the rate determined from supernovae within that
distance, and this can have a large effect on measurements of the dark
energy equation of state with SN Ia. Peculiar velocities of SN Ia host
galaxies in the rest frame of the Local Group are consistent with the
dipole measured in the Cosmic Microwave Background. Direct fits of SN
Ia that are significantly reddened by dust in their host galaxies
suggest their mean extinction law may be described by $R_V \simeq
2.7$, but optical colors alone provide weak constraints on $R_V$.
\end{abstract}

\keywords{cosmology: observations --- distance scale --- galaxies:
  distances and redshifts --- supernovae: general}

\section{Introduction}

The cosmological applications of type Ia supernovae (SN Ia) result
from precise distances to these calibrated candles. It has been well
established that the intrinsic luminosity of SN Ia is correlated with
the shape of their optical light curves (Phillips~1993; Hamuy et
al.~1995; Riess, Press, \& Kirshner 1995a; Hamuy et al.~1996a; Riess,
Press, \& Kirshner 1996a). Determining a precise distance requires
well-observed SN Ia light curves in multiple passbands, to constrain
the intrinsic luminosity and extinction by dust along the line of
sight to each SN Ia.

A number of methods have been developed to measure calibrated
distances from SN Ia multicolor light curves, with each enjoying a
similar level of success. The first of these was introduced by
Phillips~(1993), who noted that the parameter \dmfb, the amount by
which a SN Ia declined in the $B$-band during the first fifteen days
after maximum light, was well correlated with SN Ia intrinsic
luminosity. The \dmf\ method was transformed by Hamuy et al.~(1996a),
in which \dmf\ became a parameter in a multi-dimensional fit to six
template \bvi light curves spanning a wide range in \dmfb\ as it was
originally defined. Phillips et al.~(1999) present the current version
of this method, which incorporates measurement of the extinction via
late-time \bv\ color measurements (roughly independent of \dmf) and
\bv\ and \vi\ measurements at maximum light (for the measured \dmf,
determined in an iterative fashion). Updates to this technique are
given by Germany et al.~(2004) and Prieto, Rest, \& Suntzeff (2006),
the latter developing a technique to describe a continuous family of
light curves parameterized by \dmf.

Tripp \& Branch~(1999) present a two-parameter method, empirically
correlating SN Ia luminosity to \dmf\ and maximum light \bv\ color,
but without regard to the source of the color variation (i.e.,
extinction or intrinsic variation). Parodi et al.~(2000) and Reindl et
al.~(2005) present similar studies, with empirical correlations
between SN Ia maximum light magnitudes in \bvi, \dmf\ and color.

Other methods include the stretch correction of Perlmutter et
al.~(1997), in which the SN Ia intrinsic luminosity is correlated with
a simple stretching of the time axis of a fiducial light curve. This
method has been presented in detail for the $B$ band (Goldhaber et
al.~2001), and can be extended straightforwardly to $U$ and $V$ (with
$R$ and $I$ posing more difficulty; Nugent, Kim, \& Perlmutter 2002;
Knop et al. 2003; Nobili et al.~2005; Jha et al.~2006).  Guy et
al.~(2005) use an innovative approach to constrain the spectral energy
distribution of SN Ia, parameterized continuously as a function of
color and stretch, and allows for the generation of light curve
templates in arbitrary passbands.  These methods all determine
distances by correlating a distance-dependent parameter (such as peak
magnitude in a particular passband, or an average magnitude difference
between the observations and templates) and one or more
distance-independent parameters (such as color, stretch, or \dmf).

Wang et al.~(2003) describe a novel technique which derives distances
based on ``light curves'' as a function of color rather than time,
which may have interesting implications for the physics behind the
observed correlations. Finally, Tonry et al.~(2003) follow a
non-parametric approach, by directly fitting an observed multicolor
light curve to a library of nearby SN Ia, and deriving an average
distance weighted by the goodness of fit to each object in the
library.

Here we describe an updated version of the Multicolor Light Curve
Shape method, denoted here as MLCS2k2. Riess, Press, \& Kirshner
(1995a) presented the first incarnation of this method (called LCS as
it was based only on $V$ band data), in which a continuum of template
light curves was created based on a ``training set'' of SN Ia with
known relative distances and the parameter $\Delta$, a particular SN's
under- or over-luminosity relative to some fiducial value. The
luminosity correction, $\Delta$, was used as a parameter in a
least-squares fit, resulting in a best-fit distance for each SN, along
with a quantitative estimate of the uncertainty. The MLCS method,
presented by Riess, Press, \& Kirshner (1996a) gave the details of the
model, as well as incorporating \bvri light curves and providing an
estimate of extinction by dust. Riess et al.~(1998a) updated the
training set, using reliable distances measured by the Hubble Law in
favor of other methods, added a quadratic ($\Delta^2$) term to create
the template light curves, and included the effects of covariance in
the model. The application of this version of MLCS to nearby SN Ia and
the Hubble Constant is given by Jha et al.~(1999a).

The MLCS approach has some advantages: by using a relatively large
training set to define a continuum of templates, the method becomes
less sensitive to peculiarities of individual objects. During the
training process, the variance and covariance in the residuals
determine the uncertainty in the model, and this is then used to
determine statistically reliable estimates of goodness-of-fit and
parameter uncertainties when the model is applied to other objects.
The model also attempts a physically motivated separation of
extinction by dust from intrinsic color variations (rather than
empirical correlations with color), and uses all of the light curves
(at all epochs) to constrain the extinction. Finally the model is
easily extended, for example, with the inclusion of a quadratic term
in $\Delta$, or the incorporation of the $U$-band presented here. The
major disadvantage is that the method requires accurate estimates of
luminosity and extinction for the training set sample in order to
construct reliable templates.

We have compiled modern multicolor photoelectric and CCD
Johnson/Cousins \ubvri data from a variety of sources. The precision
of SN Ia distances demands high quality photometry, and we have relied
most heavily on large, homogeneous data sets, such as the observations
presented by Hamuy et al.~(1996b) from the Cal\'an/Tololo survey (29
SN Ia), as well as those of Riess et al.~(1999) and Jha et al.~(2006),
consisting of 22 and 44 SN Ia, respectively, from the CfA monitoring
campaign. The complete sample we analyze consists of 133 SN Ia
described in Table \ref{ch4-tab-sninfo}. For a number of objects we
have combined data sets from different sources (see Jha et al.~2006
for some comparisons); in some cases we had to make subjective
assessments of the relative quality of inconsistent data.

The basic framework for MLCS2k2 was laid out by Jha
(2002)\footnote{Because the basic algorithms were designed in 2002, we
  continue to refer to this SN distance fitter as MLCS2k2, even though
  its implementation, applicability, and robustness have evolved
  substantially since then. Details of the model presented here, as
  well as recent updates, can be found at
  \url{http://astro.berkeley.edu/~saurabh/mlcs2k2/}} and it has
already been applied to SN Ia cosmology, including the silver and gold
samples of Riess et al.~(2004).  This work showed that the Universe
underwent a transition from an epoch of deceleration due to dark
matter to current acceleration driven by dark energy, whose inferred
properties are consistent with the cosmological constant.  In
addition, Riess et al.~(2005) use MLCS2k2 distances to show various
estimates of $H_0$ from SN Ia can be reconciled through the use of
modern Cepheid and supernova data, deriving $H_0$ = 73 $\pm$ 4
(statistical) $\pm$ 5 (systematic) $\kmsmpc$. In this paper, we focus
only on relative distances to nearby SN, independent of the zeropoint
derived from external measurements (e.g., Cepheid distances to SN Ia
hosts; Freedman et al.~2001; Sandage et al.~2006).


\begin{deluxetable}{lrrrrrcrrrccc}
\tabletypesize{\scriptsize}
\tablecolumns{13}
\singlespace
\tablewidth{0pt}
\tablecaption{Supernova and Host Galaxy Data \label{ch4-tab-sninfo}}
\tablehead{ \colhead{SN Ia} & \colhead{$\ell$\tablenotemark{a}} & 
\colhead{$b$\tablenotemark{a}} & 
\colhead{$cz_\sun$\tablenotemark{b}} & \colhead{$cz_{\rm LG}$\tablenotemark{b}} & 
\colhead{$cz_{\rm CMB}$\tablenotemark{b}} & \colhead{Morph\tablenotemark{c}} & 
\multicolumn{2}{c}{SN Offsets\tablenotemark{d}} & \colhead{$t_1$\tablenotemark{e}} & 
\colhead{Filters} & \colhead{$E(\bv)$\tablenotemark{f}} & \colhead{References} \\
 & \colhead{deg} & \colhead{deg} & \colhead{$\kms$} & 
\colhead{$\kms$} & \colhead{$\kms$} & & \colhead{\arcsec N} &
\colhead{\arcsec E} & \colhead{days} & & \colhead{mag} & }
\startdata
1972E  & 314.84 & $+$30.08 &   404 &   190 &   678 &  Sd/Irr &  $-$100.0 &   $-$38.0 &   $+$8.7 & UBV   & 0.056 & 1          \\
1980N  & 240.16 & $-$56.63 &  1760 &  1633 &  1653 &      S0 &   $-$20.0 &  $+$220.0 &   $-$1.0 & UBVRI & 0.021 & 2          \\
1981B  & 292.97 & $+$64.74 &  1808 &  1662 &  2151 &     Sbc &   $+$42.0 &   $+$39.0 &   $-$0.2 & UBVR  & 0.018 & 3,4        \\
1981D  & 240.21 & $-$56.69 &  1760 &  1633 &  1653 &      S0 &  $-$100.0 &   $-$20.0 &   $-$5.3 & UBV   & 0.021 & 2          \\
1986G  & 309.54 & $+$19.40 &   547 &   301 &   803 &      S0 &   $-$60.0 &  $+$120.0 &   $-$5.7 & UBVRI & 0.115 & 5          \\
1989B  & 241.92 & $+$64.42 &   703 &   567 &  1051 &      Sb &   $+$50.0 &   $-$15.0 &   $-$2.5 & UBVRI & 0.032 & 6          \\
1990N  & 294.36 & $+$75.98 &   978 &   885 &  1298 &     Sbc &    $-$1.0 &   $+$65.0 &  $-$10.4 & UBVRI & 0.026 & 7          \\
1990O  &  37.65 & $+$28.35 &  9193 &  9340 &  9175 &      Sa &    $-$3.9 &   $+$21.8 &   $+$0.6 & BVRI  & 0.093 & 8          \\
1990T  & 341.50 & $-$31.53 & 12112 & 12025 & 12013 &      Sa &    $-$1.9 &   $+$24.8 &  $+$14.9 & BVRI  & 0.053 & 8          \\
1990Y  & 232.64 & $-$53.85 & 11721 & 11597 & 11622 &       E &    $-$5.0 &    $+$1.0 &  $+$16.2 & BVRI  & 0.008 & 8          \\
1990af & 330.81 & $-$42.20 & 15169 & 15059 & 15056 &      S0 &    $+$7.4 &    $-$8.0 &   $-$3.1 & BV    & 0.035 & 8          \\
1991M  &  30.36 & $+$45.90 &  2180 &  2265 &  2277 &      Sc &   $+$60.0 &   $+$36.0 &   $+$1.8 & VRI   & 0.038 & 9          \\
1991S  & 214.06 & $+$57.43 & 16369 & 16263 & 16688 &      Sb &   $+$17.3 &    $+$4.4 &  $+$10.8 & BVRI  & 0.026 & 8,10       \\
1991T  & 292.60 & $+$65.19 &  1736 &  1591 &  2078 &     Sbc &   $+$44.0 &   $+$26.0 &  $-$11.1 & UBVRI & 0.022 & 7,10       \\
1991U  & 311.82 & $+$36.20 &  9503 &  9292 &  9801 &     Sbc &    $+$5.8 &    $-$2.2 &  $+$10.4 & BVRI  & 0.062 & 8          \\
1991ag & 342.56 & $-$31.60 &  4264 &  4182 &  4161 &      Sb &   $+$22.1 &    $-$4.4 &   $+$5.6 & BVRI  & 0.062 & 8          \\
1991bg & 278.23 & $+$74.46 &  1060 &   955 &  1391 &       E &   $-$57.0 &    $+$2.0 &   $+$0.9 & BVRI  & 0.041 & 11,12,13   \\
1992A  & 235.90 & $-$54.06 &  1877 &  1747 &  1781 &      S0 &   $+$62.0 &    $-$3.0 &   $-$7.0 & UBVRI & 0.018 & 10,14      \\
1992G  & 184.62 & $+$59.84 &  1565 &  1541 &  1827 &      Sc &   $-$10.0 &   $+$27.0 &   $+$3.0 & VRI   & 0.020 & 9          \\
1992J  & 263.55 & $+$23.50 & 13371 & 13077 & 13708 &    E/S0 &   $+$12.0 &   $-$11.9 &  $+$14.6 & BVI   & 0.057 & 8          \\
1992K  & 306.28 & $+$16.30 &  3087 &  2828 &  3340 &      Sb &   $-$15.4 &    $-$1.9 &  $+$13.6 & BVI   & 0.101 & 8          \\
1992P  & 295.62 & $+$73.10 &  7555 &  7449 &  7881 &      Sa &    $+$9.8 &    $-$4.3 &   $-$0.5 & BVI   & 0.021 & 8          \\
1992ae & 332.70 & $-$41.90 & 22544 & 22440 & 22426 &       E &    $+$4.0 &    $+$2.1 &   $+$2.5 & BV    & 0.036 & 8          \\
1992ag & 312.48 & $+$38.30 &  7465 &  7260 &  7765 &      Sa & {\phs}0.0 &    $-$3.0 &   $-$0.4 & BVI   & 0.097 & 8          \\
1992al & 347.33 & $-$38.40 &  4377 &  4324 &  4228 &      Sb &   $-$12.0 &   $+$19.0 &   $-$4.9 & BVRI  & 0.034 & 8          \\
1992aq &   1.78 & $-$65.30 & 30519 & 30536 & 30254 &      Sa &    $-$7.1 &    $+$2.4 &   $+$2.2 & BVI   & 0.012 & 8          \\
1992au & 319.11 & $-$65.80 & 18407 & 18338 & 18213 &       E &    $+$8.9 &   $+$21.2 &  $+$12.6 & BVI   & 0.017 & 8          \\
1992bc & 245.70 & $-$59.60 &  6056 &  5933 &  5936 &     Sab &    $-$2.0 &   $+$16.1 &  $-$10.1 & BVRI  & 0.022 & 8          \\
1992bg & 274.61 & $-$18.30 & 10553 & 10261 & 10697 &      Sa &    $+$5.8 &    $-$3.4 &   $+$4.3 & BVI   & 0.185 & 8          \\
1992bh & 267.85 & $-$37.30 & 13490 & 13254 & 13518 &     Sbc &    $-$3.6 &    $+$1.9 &   $-$0.6 & BVI   & 0.022 & 8          \\
1992bk & 265.01 & $-$48.90 & 17418 & 17229 & 17372 &       E &   $+$21.1 &   $+$11.9 &   $+$8.6 & BVI   & 0.015 & 8          \\
1992bl & 344.12 & $-$63.90 & 13101 & 13076 & 12871 &   S0/Sa &   $-$21.8 &   $+$15.3 &   $+$3.1 & BVI   & 0.011 & 8          \\
1992bo & 261.88 & $-$80.30 &  5666 &  5636 &  5435 &    E/S0 &   $-$54.7 &   $-$47.3 &   $-$7.5 & BVRI  & 0.027 & 8          \\
1992bp & 208.83 & $-$51.00 & 23773 & 23704 & 23646 &    E/S0 &    $-$1.4 &    $-$5.4 &   $-$1.6 & BVI   & 0.069 & 8          \\
1992br & 288.01 & $-$59.40 & 26442 & 26306 & 26319 &       E &    $-$6.3 &    $+$3.6 &   $+$1.8 & BV    & 0.026 & 8          \\
1992bs & 240.02 & $-$55.33 & 19097 & 18965 & 18998 &      Sb &    $+$3.6 &    $-$9.0 &   $+$2.5 & BV    & 0.011 & 8          \\
1993B  & 273.32 & $+$20.46 & 20866 & 20563 & 21191 &      Sb &    $+$5.4 &    $+$0.9 &   $+$3.6 & BVI   & 0.079 & 8          \\
1993H  & 318.22 & $+$30.33 &  7165 &  6962 &  7430 &      Sb &   $+$12.3 &    $+$1.0 &   $-$1.1 & BVRI  & 0.060 & 8,10       \\
1993L  &   5.95 & $-$64.37 &  1858 &  1885 &  1587 &      Sc &    $+$1.4 &   $-$25.0 &  $+$17.3 & BVRI  & 0.014 & 10         \\
1993O  & 312.41 & $+$28.92 & 15289 & 15065 & 15567 &    E/S0 &    $+$8.4 &   $-$14.1 &   $-$6.3 & BVI   & 0.053 & 8          \\
1993ac & 149.70 & $+$17.21 & 14800 & 14959 & 14784 &       E &   $+$31.0 &    $-$5.3 &   $+$7.0 & BVRI  & 0.163 & 15         \\
1993ae & 144.62 & $-$63.22 &  5712 &  5820 &  5410 &       E &   $+$22.7 &   $+$16.1 &  $+$13.8 & BVRI  & 0.038 & 15         \\
1993ag & 268.43 & $+$15.92 & 14690 & 14382 & 15003 &    E/S0 &    $-$6.1 &    $-$5.0 &   $-$1.7 & BVI   & 0.112 & 8          \\
1993ah &  25.88 & $-$76.77 &  8842 &  8892 &  8543 &      S0 &    $+$8.1 &    $-$0.9 &  $+$11.8 & BVI   & 0.020 & 8          \\
1994D  & 290.15 & $+$70.14 &   592 &   469 &   928 &      S0 &    $+$7.8 &    $-$9.0 &  $-$13.5 & UBVRI & 0.022 & 10,16,17,18\\
1994M  & 291.68 & $+$63.03 &  6943 &  6788 &  7289 &       E &   $-$28.2 &    $+$3.4 &   $+$3.1 & BVRI  & 0.023 & 15         \\
1994Q  &  64.38 & $+$39.67 &  8672 &  8871 &  8670 &      S0 &    $-$3.7 &    $-$0.1 &  $+$10.8 & BVRI  & 0.017 & 15         \\
1994S  & 187.37 & $+$85.14 &  4525 &  4501 &  4806 &     Sab &    $-$6.9 &   $-$14.0 &   $-$4.5 & BVRI  & 0.021 & 15         \\
1994T  & 318.01 & $+$59.83 & 10396 & 10265 & 10709 &      Sa &   $-$12.0 &    $+$3.8 &   $+$0.2 & BVRI  & 0.029 & 15         \\
1994ae & 225.34 & $+$59.66 &  1301 &  1175 &  1637 &      Sc &    $+$6.1 &   $-$29.7 &  $-$11.6 & UBVRI & 0.030 & 10,19      \\
1995D  & 230.02 & $+$39.65 &  1966 &  1774 &  2300 &      S0 &   $-$87.8 &   $+$11.8 &   $-$6.5 & BVRI  & 0.058 & 10,15      \\
1995E  & 141.99 & $+$30.26 &  3470 &  3638 &  3496 &      Sb &   $-$20.8 &    $+$7.6 &   $-$1.9 & BVRI  & 0.027 & 15         \\
1995ac &  58.69 & $-$55.04 & 14990 & 15157 & 14635 &      Sa &    $-$1.4 &    $-$0.9 &   $-$5.1 & BVRI  & 0.042 & 10,15      \\
1995ak & 169.65 & $-$48.98 &  6811 &  6875 &  6589 &     Sbc &    $+$0.8 &    $-$7.1 &   $+$3.6 & BVRI  & 0.038 & 15         \\
1995al & 192.17 & $+$50.83 &  1514 &  1465 &  1777 &     Sbc &    $-$3.1 &   $-$14.7 &   $-$3.9 & UBVRI & 0.014 & 15         \\
1995bd & 187.11 & $-$21.66 &  4377 &  4364 &  4326 &      Sa &    $-$1.0 &   $+$22.9 &   $-$8.4 & BVRI  & 0.498 & 15         \\
1996C  &  99.59 & $+$65.00 &  8094 &  8206 &  8244 &      Sa &   $+$13.2 &    $-$1.8 &   $+$2.4 & BVRI  & 0.013 & 15         \\
1996X  & 310.23 & $+$35.64 &  2032 &  1815 &  2333 &       E &   $-$31.7 &   $-$51.4 &   $-$3.0 & UBVRI & 0.069 & 15,20      \\
1996Z  & 253.60 & $+$22.55 &  2254 &  1971 &  2584 &      Sb &   $-$69.9 &    $+$2.0 &   $+$6.1 & BVR   & 0.064 & 15         \\
1996ab &  43.15 & $+$56.93 & 37109 & 37201 & 37239 &      Sa &    $+$0.6 &    $+$2.0 &   $+$1.0 & BV    & 0.032 & 15         \\
1996ai & 101.58 & $+$79.24 &   873 &   910 &  1101 &     Scd &    $+$2.9 &   $+$23.3 &   $-$1.1 & UBVRI & 0.014 & 15         \\
1996bk & 111.25 & $+$54.88 &  1985 &  2139 &  2085 &      S0 &    $-$9.7 &   $-$18.1 &   $+$3.0 & BVRI  & 0.018 & 15         \\
1996bl & 116.99 & $-$51.30 & 10793 & 10990 & 10447 &      Sc &    $+$5.6 &    $-$3.2 &   $-$2.4 & BVRI  & 0.092 & 15         \\
1996bo & 144.46 & $-$48.95 &  5182 &  5328 &  4893 &      Sc &    $-$2.1 &    $+$6.7 &   $-$6.2 & BVRI  & 0.077 & 10,15      \\
1996bv & 157.33 & $+$17.97 &  4996 &  5119 &  5013 &      Sa &    $+$2.0 &    $-$2.0 &   $+$5.1 & BVRI  & 0.105 & 15         \\
1997E  & 140.20 & $+$25.81 &  4001 &  4184 &  3997 &      S0 &   $+$57.0 &   $-$32.0 &   $-$2.6 & UBVRI & 0.124 & 21         \\
1997Y  & 124.77 & $+$62.37 &  4806 &  4911 &  4964 &      Sb &    $+$2.0 &    $-$8.0 &   $+$2.2 & UBVRI & 0.017 & 21         \\
1997bp & 301.15 & $+$51.21 &  2492 &  2301 &  2831 &  Sd/Irr &   $-$20.0 &   $-$15.0 &   $-$2.1 & UBVRI & 0.044 & 10,21      \\
1997bq & 136.29 & $+$39.48 &  2780 &  2943 &  2839 &     Sbc &   $-$60.0 &   $+$50.0 &  $-$10.3 & UBVRI & 0.024 & 21         \\
1997br & 311.84 & $+$40.32 &  2085 &  1884 &  2391 &  Sd/Irr &   $+$52.0 &   $-$21.0 &   $-$8.7 & UBVRI & 0.113 & 10,21,22   \\
1997cn &   9.14 & $+$69.51 &  4855 &  4846 &  5092 &       E &   $-$12.0 &    $+$7.0 &   $+$5.4 & UBVRI & 0.027 & 21,23      \\
1997cw & 113.09 & $-$49.48 &  5133 &  5342 &  4782 &     Sab &    $+$4.0 &    $+$8.0 &  $+$12.0 & UBVRI & 0.073 & 21         \\
1997dg & 103.61 & $-$33.98 &  9238 &  9507 &  8890 & \nodata & {\phs}0.0 &    $+$2.0 &   $+$0.8 & UBVRI & 0.078 & 21         \\
1997do & 171.00 & $+$25.26 &  3034 &  3084 &  3140 &     Sbc &    $-$4.0 &    $-$3.0 &   $-$6.1 & UBVRI & 0.063 & 21         \\
1997dt &  87.56 & $-$39.12 &  2194 &  2451 &  1828 &     Sbc &    $+$1.0 &    $-$9.0 &   $-$7.9 & UBVRI & 0.057 & 21         \\
1998D  &  63.78 & $+$72.91 &  3765 &  3825 &  3962 &      Sa &    $-$7.0 &   $-$26.0 &  $+$32.4 & UBVRI & 0.015 & 21         \\
1998V  &  43.94 & $+$13.34 &  5268 &  5464 &  5148 &      Sb &   $+$21.0 &   $-$21.0 &   $+$2.7 & UBVRI & 0.196 & 21         \\
1998ab & 124.86 & $+$75.19 &  8134 &  8181 &  8354 &      Sc &   $+$12.0 &    $+$2.0 &   $-$7.4 & UBVRI & 0.017 & 21         \\
1998aq & 138.83 & $+$60.26 &  1184 &  1274 &  1354 &      Sb &    $+$7.0 &   $-$18.0 &  $-$10.1 & UBVRI & 0.014 & 19         \\
1998bp &  43.64 & $+$20.48 &  3127 &  3312 &  3048 &       E &   $+$13.0 &    $-$1.0 &   $-$1.5 & UBVRI & 0.076 & 21         \\
1998bu & 234.41 & $+$57.01 &   855 &   702 &  1204 &     Sab &   $+$55.0 &    $+$4.0 &   $-$7.7 & UBVRI & 0.025 & 24,25      \\
1998co &  41.52 & $-$44.94 &  5418 &  5573 &  5094 &      S0 &    $+$5.0 &    $+$2.0 &   $+$3.0 & UBVRI & 0.043 & 21         \\
1998de & 122.03 & $-$35.24 &  4990 &  5228 &  4671 &      S0 &    $+$3.0 &   $+$72.0 &   $-$8.6 & UBVRI & 0.057 & 21,26      \\
1998dh &  82.82 & $-$50.64 &  2678 &  2892 &  2307 &     Sbc &   $+$10.0 &   $-$54.0 &   $-$7.8 & UBVRI & 0.068 & 21         \\
1998dk & 102.85 & $-$62.16 &  3963 &  4128 &  3609 &      Sc &    $+$3.0 &    $+$5.0 &  $+$18.9 & UBVRI & 0.044 & 21         \\
1998dm & 145.97 & $-$67.40 &  1968 &  2061 &  1668 &      Sc &   $-$37.0 &   $-$14.0 &   $+$9.9 & UBVRI & 0.044 & 21         \\
1998dx &  77.67 & $+$26.67 & 16197 & 16459 & 16102 &      Sb &   $-$12.0 &   $+$21.0 &   $+$1.2 & UBVRI & 0.041 & 21         \\
1998ec & 166.29 & $+$20.71 &  5966 &  6043 &  6032 &      Sb &   $-$20.0 &    $-$9.0 &  $+$13.3 & UBVRI & 0.085 & 21         \\
1998ef & 125.88 & $-$30.56 &  5319 &  5558 &  5020 &      Sa &    $-$2.0 &    $+$6.0 &   $-$7.3 & UBVRI & 0.073 & 21         \\
1998eg &  76.46 & $-$42.06 &  7423 &  7662 &  7056 &      Sc &   $-$25.0 &   $-$26.0 &   $-$0.1 & UBVRI & 0.123 & 21         \\
1998es & 143.18 & $-$55.18 &  3168 &  3301 &  2868 &   S0/Sa &   $+$11.0 & {\phs}0.0 &  $-$11.0 & UBVRI & 0.032 & 21,27      \\
1999X  & 186.58 & $+$39.59 &  7503 &  7474 &  7720 & \nodata &    $+$6.0 &    $+$4.0 &  $+$14.0 & UBVRI & 0.032 & 21         \\
1999aa & 202.72 & $+$30.31 &  4330 &  4227 &  4572 &      Sc &   $+$28.0 &    $+$1.0 &  $-$10.1 & UBVRI & 0.040 & 10,21,27,28\\
1999ac &  19.88 & $+$39.94 &  2848 &  2904 &  2943 &     Scd &   $-$30.0 &   $+$24.0 &  $-$14.4 & UBVRI & 0.046 & 21,27      \\
1999aw & 260.24 & $+$47.45 & 11392 & 11168 & 11763 & \nodata & {\phs}0.0 & {\phs}0.0 &   $-$8.2 & BVRI  & 0.032 & 29         \\
1999by & 166.91 & $+$44.11 &   657 &   704 &   827 &      Sb &   $+$91.0 &   $-$96.0 &  $-$10.9 & UBVRI & 0.016 & 27,30      \\
1999cc &  59.66 & $+$48.74 &  9392 &  9549 &  9452 &      Sc &    $+$2.0 &   $+$17.0 &   $-$2.7 & UBVRI & 0.023 & 21,31      \\
1999cl & 282.26 & $+$76.50 &  2281 &  2187 &  2605 &      Sb &   $+$23.0 &   $-$46.0 &   $-$6.5 & UBVRI & 0.038 & 21,31      \\
1999cp & 334.85 & $+$52.71 &  2845 &  2737 &  3115 &     Scd &   $+$23.0 &   $-$52.0 &   $-$7.4 & BVRI  & 0.024 & 28         \\
1999cw & 101.77 & $-$67.91 &  3725 &  3863 &  3380 &     Sab &    $-$2.0 &   $+$21.0 &  $+$24.0 & UBVRI & 0.036 & 21         \\
1999da &  89.73 & $+$32.64 &  3806 &  4059 &  3748 &       E &    $+$1.0 &   $-$71.0 &   $-$8.7 & BVRI  & 0.056 & 27,32      \\
1999dk & 137.35 & $-$47.46 &  4485 &  4654 &  4181 &      Sc &   $+$26.0 &    $+$4.0 &   $-$0.9 & UBVRI & 0.054 & 10,32      \\
1999dq & 152.83 & $-$35.87 &  4295 &  4436 &  4060 &      Sc &    $-$6.0 &    $-$4.0 &  $-$10.6 & UBVRI & 0.110 & 21,27      \\
1999ee &   6.50 & $-$55.93 &  3422 &  3451 &  3163 &     Sbc &   $-$10.0 &   $+$10.0 &   $-$9.5 & UBVRI & 0.020 & 33         \\
1999ef & 125.72 & $-$50.09 & 11733 & 11920 & 11402 &     Scd &   $-$10.0 &   $+$20.0 &   $+$6.8 & UBVRI & 0.087 & 21         \\
1999ej & 130.44 & $-$28.95 &  4114 &  4344 &  3831 &   S0/Sa &   $-$20.0 &   $+$18.0 &   $+$5.3 & UBVRI & 0.071 & 21         \\
1999ek & 189.40 &  $-$8.23 &  5253 &  5221 &  5278 &     Sbc &   $-$12.0 &   $-$12.0 &   $-$2.8 & UBVRI & 0.561 & 21,34      \\
1999gd & 198.83 & $+$33.98 &  5535 &  5451 &  5775 & \nodata &   $+$17.0 &    $+$7.0 &   $+$3.5 & UBVRI & 0.041 & 21         \\
1999gh & 255.04 & $+$23.73 &  2302 &  2019 &  2637 &       E &   $+$16.0 &   $+$52.0 &   $+$7.4 & UBVRI & 0.058 & 21         \\
1999gp & 143.25 & $-$19.50 &  8018 &  8215 &  7806 &      Sb &   $+$10.0 &   $-$11.0 &  $-$13.0 & UBVRI & 0.056 & 21,27,32   \\
2000B  & 166.35 & $+$22.79 &  5901 &  5976 &  5977 &       E &   $+$19.0 &   $-$14.0 &  $+$15.7 & UBVRI & 0.068 & 21         \\
2000E  & 100.89 & $+$14.84 &  1415 &  1711 &  1257 &     Sbc &   $-$26.7 &    $-$6.3 &  $-$15.5 & UBVRI & 0.364 & 35         \\
2000bh & 293.74 & $+$40.33 &  6905 &  6666 &  7248 & \nodata &   $-$11.0 &    $-$8.0 &   $+$5.4 & BVRI  & 0.048 & 36         \\
2000bk & 295.29 & $+$55.23 &  7628 &  7444 &  7976 &      S0 &   $-$10.0 &   $+$61.0 &  $+$12.3 & BVRI  & 0.025 & 32         \\
2000ca & 313.20 & $+$27.83 &  7080 &  6857 &  7352 &     Sbc &    $+$4.7 &    $+$0.6 &   $-$2.3 & UBVRI & 0.067 & 36         \\
2000ce & 149.10 & $+$32.00 &  4888 &  5025 &  4946 &      Sb &   $+$17.0 &   $+$15.0 &   $+$7.0 & UBVRI & 0.057 & 21,32      \\
2000cf &  99.88 & $+$42.16 & 10920 & 11137 & 10930 & \nodata &    $+$4.0 &    $+$3.0 &   $+$3.2 & UBVRI & 0.032 & 21,31      \\
2000cn &  53.44 & $+$23.31 &  7043 &  7257 &  6958 &     Scd &    $-$7.0 &    $-$7.0 &   $-$7.8 & UBVRI & 0.057 & 21         \\
2000cx & 136.50 & $-$52.48 &  2379 &  2536 &  2068 &      S0 &  $-$109.0 &   $-$23.0 &   $-$8.4 & UBVRI & 0.082 & 10,21,37,38\\
2000dk & 126.83 & $-$30.34 &  5228 &  5465 &  4931 &       E &    $+$9.0 &    $-$5.0 &   $-$4.5 & UBVRI & 0.070 & 21         \\
2000fa & 194.17 & $+$15.48 &  6378 &  6313 &  6533 &  Sd/Irr &    $+$4.0 &    $+$7.0 &  $-$10.1 & UBVRI & 0.069 & 21         \\
2001V  & 218.92 & $+$77.73 &  4539 &  4478 &  4846 &      Sb &   $+$28.0 &   $+$52.0 &  $-$12.9 & UBVRI & 0.020 & 39         \\
2001ay &  35.98 & $+$68.83 &  9067 &  9108 &  9266 &      Sb &    $+$9.0 &   $-$10.0 &   $-$3.0 & UBVRI & 0.019 & 40         \\
2001ba & 285.39 & $+$28.03 &  8819 &  8537 &  9152 &      Sb &   $-$22.0 &   $+$19.0 &   $-$3.5 & BVI   & 0.064 & 36         \\
2001bt & 337.32 & $-$25.87 &  4388 &  4275 &  4332 &      Sb &   $+$17.1 &   $-$14.1 &   $-$8.0 & BVRI  & 0.065 & 34         \\
2001cn & 329.65 & $-$24.05 &  4647 &  4498 &  4628 &      Sc &   $-$17.9 &    $-$2.6 &   $+$4.7 & UBVRI & 0.059 & 34         \\
2001cz & 302.11 & $+$23.29 &  4612 &  4350 &  4900 &     Sbc &   $-$31.4 &    $-$0.6 &   $-$6.3 & BVRI  & 0.092 & 34         \\
2001el & 251.52 & $-$51.40 &  1168 &  1002 &  1102 &     Scd &   $+$19.0 &   $-$22.0 &  $-$10.7 & UBVRI & 0.014 & 41         \\
2002bf & 156.46 & $+$50.08 &  7254 &  7327 &  7418 &      Sb &    $+$4.0 &    $+$0.6 &   $-$9.9 & BVRI  & 0.011 & 42         \\
2002bo & 213.04 & $+$54.85 &  1293 &  1184 &  1609 &      Sa &   $-$14.2 &   $+$11.6 &  $-$13.4 & UBVRI & 0.025 & 34,43,44   \\
2002cx & 318.71 & $+$69.14 &  7184 &  7085 &  7494 & \nodata &   $-$18.0 &   $+$11.0 &   $-$5.3 & BVRI  & 0.032 & 45         \\
2002er &  28.67 & $+$25.83 &  2568 &  2681 &  2563 &      Sc &    $+$4.7 &   $-$12.3 &   $-$7.5 & UBVRI & 0.157 & 46         \\
2003du & 101.18 & $+$53.21 &  1912 &  2081 &  1992 &  Sd/Irr &   $-$13.5 &    $-$8.8 &   $-$4.7 & BVRI  & 0.010 & 42         \\
\enddata

\tablenotetext{a}{Host-galaxy galactic longitude and latitude, from
the NASA/IPAC Extragalactic Database (NED); 
\url{http://nedwww.ipac.caltech.edu}.}

\tablenotetext{b}{Host-galaxy heliocentric radial velocities are taken
from NED; they are transformed to the Local Group and CMB
rest frames using the formulas also provided by NED;
\url{http://nedwww.ipac.caltech.edu/help/velc\_help.html\#notes}.}

\tablenotetext{c}{NED host-galaxy morphological type.}

\tablenotetext{d}{Positional offset relative to host-galaxy nucleus,
from the IAU List of Supernovae; 
\url{http://cfa-www.harvard.edu/iau/lists/Supernovae.html}.}

\tablenotetext{e}{Epoch of the first photometric observation,
relative to $B$ maximum light, in the SN rest frame.}

\tablenotetext{f}{Milky Way dust reddening, from Schlegel, Finkbeiner,
  \& Davis (1998).}

\tablerefs{(1) Leibundgut et al.~1991; 
(2) Hamuy et al.~1991;
(3) Buta \& Turner 1983;
(4) Tsvetkov 1982;
(5) Phillips et al.~1987;
(6) Wells et al.~1994;
(7) Lira et al.~1998;
(8) Hamuy et al.~1996b;
(9) Ford et al.~1993;
(10) Altavilla et al.~2004;
(11) Filippenko et al.~1992;
(12) Leibundgut et al.~1993;
(13) Turatto et al.~1996
(14) N. Suntzeff, personal communication;
(15) Riess et al.~1999;
(16) Richmond et al.~1995;
(17) Patat et al.~1996;
(18) Meikle et al.~1996;
(19) Riess et al.~2005;
(20) Salvo et al.~2001;
(21) Jha et al.~2006;
(22) Li et al.~1999;
(23) Turatto et al.~1998;
(24) Suntzeff et al.~1999;
(25) Jha et al.~1999a;
(26) Modjaz et al.~2001;
(27) W. Li, personal communication;
(28) Krisciunas et al.~2000;
(29) Strolger et al.~2002;
(30) Garnavich et al.~2004;
(31) Krisciunas et al.~2006;
(32) Krisciunas et al.~2001;
(33) Stritzinger et al.~2002;
(34) Krisciunas et al.~2004b;
(35) Valentini et al.~2003;
(36) Krisciunas et al.~2004a;
(37) Li et al.~2001;
(38) Candia et al.~2003;
(39) K. Mandel et al.~2007, in preparation;
(40) K. Krisciunas, personal communication;
(41) Krisciunas et al.~2003;
(42) Leonard et al.~2005;
(43) Szab\'o et al.~2003;
(44) Benetti et al.~2004;
(45) Li et al.~2003;
(46) Pignata et al.~2004}
\end{deluxetable}


\section{Groundwork}

Comparison of light curves from many SN Ia requires understanding and
correction for a number of effects to put the photometry on a common
basis. These include corrections from the observational photometric
system to standard passbands, correction for Galactic extinction,
correction for time dilation and the $K$-correction. We have updated a
number of these foundations in our development of MLCS2k2.

\subsection{$K$-correction \label{ch4-sec-kcorr}}

The stretching and shifting of spectra due to the cosmological
expansion leads to changes in measured flux observed through a fixed
detector passband as a function of redshift.  $K$-corrections for SN
Ia in $B$ and $V$ have been presented by Hamuy et al.~(1993), based on
spectra of three objects. Kim, Goobar, \& Perlmutter~(1996) provide
additional $K$-corrections in $R$, as well as developing a method of
``cross-filter'' $K$-corrections used at high redshift. These methods
use a fixed set of spectra to define the $K$-correction for all
objects, and do not include spectral energy distribution (SED)
variations arising from intrinsic differences among supernovae (either
in the detailed spectral features or in the continuum shape) or
changes in in the SED due to extinction.

Nugent, Kim, \& Perlmutter (2002) developed a method that accounts for
these variations with a simple, yet effective, ``trick'' in which both
the intrinsic and extinction-related SED variations are effected by
adjusting the SED by the $R_V = 3.1$ extinction law of Cardelli,
Clayton \& Mathis (1989; hereinafter CCM89) to match the color of the
SN as observed.\footnote{Here we use the extinction law to adjust the
SED redder \emph{or bluer}, to match the observed SN color.}  This
procedure is reasonably well motivated; at early times, SN Ia are in
the photospheric phase and the SED is continuum dominated, so that
adjustment of the SED by a relatively slowly varying function of
wavelength, like the CCM89 extinction law, to match the observed color
will do a good job of mimicking the true SED. At late times, much of
the color variation is due, in fact, to extinction and so the
adjustment is appropriate. Furthermore, the adjustment is done using a
color ``local'' to the spectral region being adjusted, minimizing any
adverse effects. The difference in the $K$-correction accounting for
these color variations and assuming a constant color for all objects
is typically small (generally $\lesssim 0.1$ mag), but
systematic. Nugent et al.~(2002) show the efficacy of this color-based
procedure.

Our concern in this paper is for accurate $K$-corrections at low
redshift ($z \lesssim 0.1$), so we restrict our attention to
$K$-corrections within each passband, not the cross-filter
$K$-correction. We have implemented the Nugent et al.~(2002) procedure
using a sample of 91 SN Ia spectra covering phases from $-$14 to $+$92
days past maximum light, compiled from a number of sources, including
archival IUE and HST data, Keck and Lick Observatory SN Ia spectra
from UC Berkeley (Filippenko 1997), and unpublished spectra from the
CfA SN Ia follow-up program (Matheson et al.~2007, in preparation). Of
these, 57 spectra extend far enough to the red to cover the $I$
passband, while 32 spectra cover the $U$ passband in the blue. The
number of spectra that cover the $U$-band decreases quickly as the
spectra are artificially redshifted to calculate the $K$-correction;
however, of the 133 SN Ia described here, the maximum redshift for
which there are $U$ observations is only $z \simeq 0.06$, meaning that
is the extent to which we need to calculate $K_{UU}$.

To derive the $K$-corrections, the basic procedure is as follows. For
each of \ubvri (we adopt the standard Bessell~1990 passbands), we
choose standard colors including that passband (specifically, for $U$:
\ub; $B$: \ub\ and \bv; $V$: \bv, \vr, and \vi; $R$: \vr\ and \ri; $I$:
\vi\ and \ri). Then for a particular combination of passband and color, we
calculate $K$-corrections in that passband, such that the spectra are
forced to take on a range of observed colors, using the CCM89 extinction
law to adjust the spectra to any particular color. Thus, we tabulate
the $K$-correction as a function of three parameters: epoch (i.e.,
days after maximum light), redshift and color. We measure the colors
in the observer frame, so that for example, to determine the
$K$-correction in the $V$-band for a supernova at maximum light at
redshift $z = 0.05$ with an observed \bv = 0.0 mag, we have taken our
maximum light spectra, redshifted them to $z = 0.05$, adjusted them
using the CCM89 extinction law to have \bv = 0.0 as measured with
synthetic photometry, and calculated the $K$-correction for that
adjusted spectrum. In this example, the answer is $K_{VV} = -0.04 \pm
0.02$ mag, where the uncertainty is measured from the scatter among
the individual spectra about a smooth curve. If the observed color had
been redder, e.g., \bv = 0.5 mag, then $K_{VV} = 0.06 \pm 0.01$ mag.

The advantage of calculating the $K$-correction as a function of
observer-frame color (as opposed to the SN rest-frame color) is that
no iteration is then required: the measured color directly determines
the SN SED in the observer frame. However, in other applications, for
example, if one wants to transform rest-frame templates (with known
rest-frame colors) to the observer-frame, it is more convenient to
tabulate the $K$-correction as a function of rest-frame color. For the
present sample, all SN have well-measured (to better than $\sim$0.1
mag) observer-frame colors, so we can apply the $K$-corrections to the
data directly, and expediently fit the SN in the rest frame.  Though
not used here, we are also developing an implementation of MLCS2k2 to
take rest-frame model light curves, convert them into the observer
frame, and then fit the observed data directly. That approach is more
appropriate, e.g., for high-redshift observations where observer-frame
color information is of lower quality and $K$-corrections based on the
model light-curve colors should be used.

\begin{figure}
\includegraphics[width=6in]{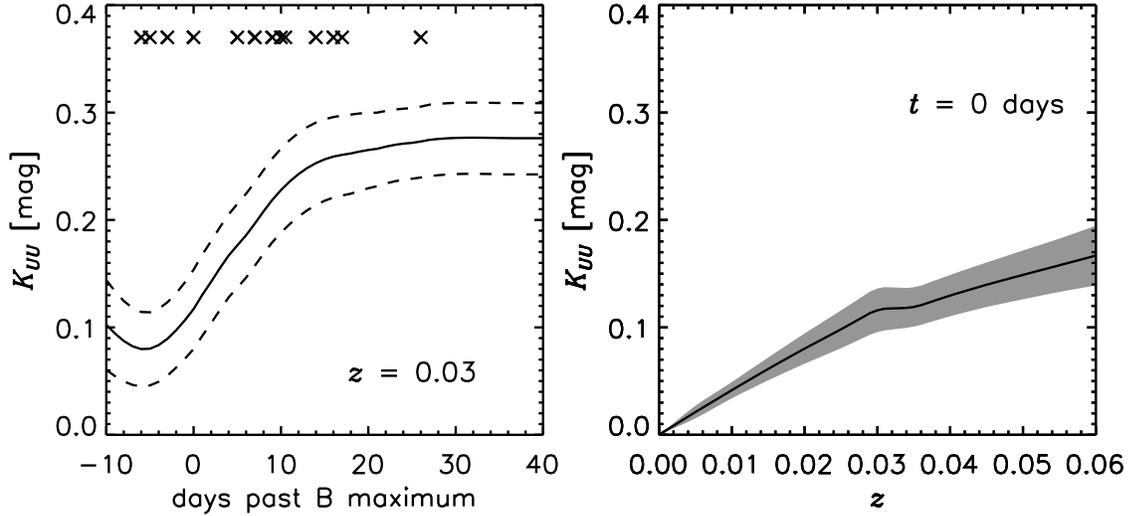}
\caption[$U$-band $K$-correction]{\singlespace $K$-corrections in the
  $U$-band. The left panel shows $K_{UU}$ as a function of supernova
  age at a redshift of $z = 0.03$. The solid line corresponds to an
  observed maximum-light \ub\ = $-0.35$ (corresponding to a SN
  rest-frame color at maximum light \ub\ = $-0.50$). The dashed lines
  above and below the solid line correspond to SN Ia that are 0.2 mag
  redder and bluer, respectively, in \ub\ at all epochs. The crosses 
  indicate the phases of the spectra used in the calculation. The right
  panel shows $K_{UU}$ at maximum light as a function of redshift,
  assuming a rest-frame color of \ub\ = $-0.50$. The shaded region
  indicates the $K$-correction uncertainty. The ``kinks'' in the right
  panel occur when individual spectra (out of 32 total) drop out of the
  $K$-correction calculation as their wavelength coverage ceases to
  encompass the entire $U$-band. \label{ch4-fig-kuu}}
\end{figure}

In Figure \ref{ch4-fig-kuu} we show the derived $K$-corrections in the
$U$-band. It is interesting to note that the $K$-corrections in $U$ at
maximum light are quite significant ($\sim$0.1 mag), even for very
modest redshifts. Ignoring the $K$-correction would lead to spurious
correlations of supernova parameters (including those that are
nominally distance-independent, such as decline rate or color) with
redshift and distance.

Nugent et al.~(2002) caution against using the CCM89 extinction law
adjustment procedure applied to normal SN Ia spectra in determining
$K$-corrections for SN 1991bg-like objects, whose strong Ti II
spectral features dominate the photometric colors (particularly in
$B$), even at early times. We have calculated $K$-corrections
independently for this class of SN Ia, a small minority of the entire
sample, using a separate spectral sample consisting of only
1991bg-like objects. The affected objects are SN 1986G, 1991bg, 1992K,
1992bo, 1993H, 1997cn, 1998bp, 1998de, 1999by, 1999da, and 1999gh.

\subsection{Extinction \label{ch4-sec-ex}}

Extinction by dust along the line of sight to a SN Ia is recognized by
its reddening effect on the SN colors. As discussed by Phillips et
al.~(1999) and Nugent et al.~(2002), the evolution of the SN Ia SED
over time leads to variations in the observed extinction in any given
passband. Furthermore the extinction itself alters the SED such that
the reddening is a nonlinear function of the total extinction. These
effects are small (typically at the level of a few hundredths of a
magnitude), but systematically affect the observed light curves in a
way that can impact the luminosity/light-curve shape relationship and
derived distances. In this section we describe in some detail the
methods we have used to account for host-galaxy extinction in MLCS2k2.

The dust extinction along a particular line of sight is typically
parameterized by the extinction in a given band $X$, $A_X$ and by the
amount of reddening, given by the color excess, typically $E(\bv)$.
Following Phillips et al.~(1999) and Nugent et al.~(2002), we
distinguish between the ``true'' reddening $E(\bv)_{\rm true}$ which
depends only on the dust itself (and is calculated from the effects of
that dust on idealized stellar spectra; CCM89), and the ``observed''
reddening $E(\bv)_{\rm obs} \equiv A_B - A_V$, which varies with time
as the supernova spectrum evolves. From these, we can construct the
ratio $R^{\rm true}_X \equiv A_X/E(\bv)_{\rm true}$, which varies in
time as the numerator varies, and $R^{\rm obs}_X \equiv
A_X/E(\bv)_{\rm obs}$, for which both the numerator and denominator
vary in time. We have calculated both of these ratios similarly to
Phillips et al.~(1999) and Nugent et al.~(2002), by simulating the
effects of extinction on the sample of 91 spectra described in \S
\ref{ch4-sec-kcorr}, using the $R_V = 3.1$ extinction law of
CCM89,\footnote{We have also explored using the extinction law
  presented by Fitzpatrick~(1999), with similar results.} as modified
by O'Donnell (1994), and synthetic photometry through the
Bessell~(1990) standard passbands.\footnote{Previous incarnations of
  MLCS used extinction coefficients tabulated for the Johnson $R$ and
  $I$ passbands, rather than the Kron/Cousins passbands. We have
  rectified this here.}

\begin{figure}
\begin{center}
\includegraphics[height=6.5in]{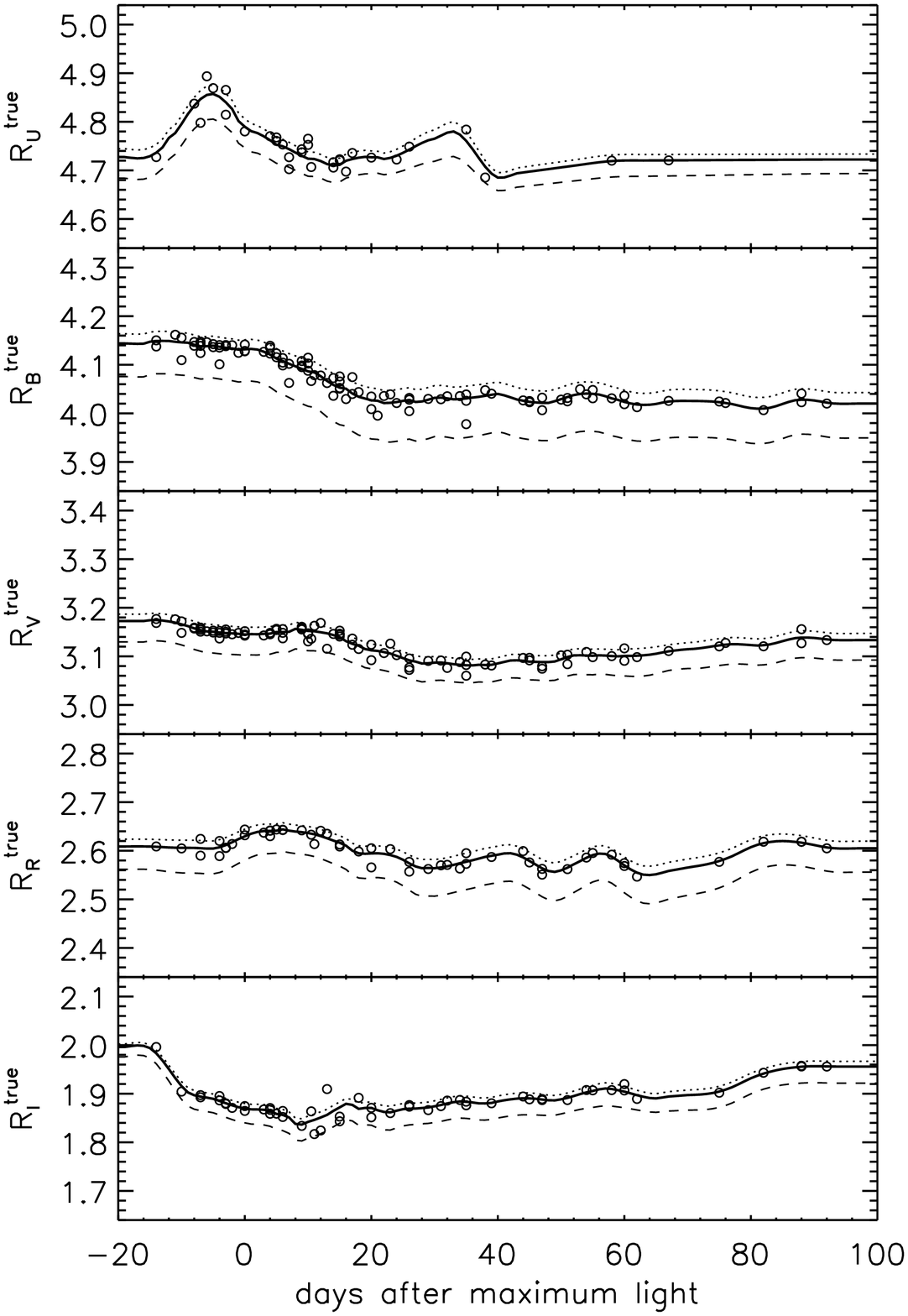}
\end{center}
\caption[$R^{\rm true}$]{\singlespace Variation of $R^{\rm true}_X
\equiv A_X/E(\bv)_{\rm true}$ in \ubvri as a function of supernova
phase. The open circles represent the measurement from the individual
spectra and the heavy solid line is the smoothed representation. The
dotted and dashed lines show the variation of $R^{\rm true}_X$ with
the total extinction (cf. Phillips et al.~1999, their Figure 2). The
dashed line corresponds to $E(\bv)_{\rm true} = +1.0$ mag relative to
the solid line (which was calculated for $E(\bv)_{\rm true}$
approaching zero). The dotted line corresponds to $E(\bv)_{\rm true} =
-0.5$ mag relative to the solid line (for illustration only, as it
would be unphysical unless the extinction zeropoint for the solid line
was significantly underestimated). \label{ch4-fig-rtrue}}
\end{figure}

\begin{figure}
\begin{center}
\includegraphics[height=6.5in]{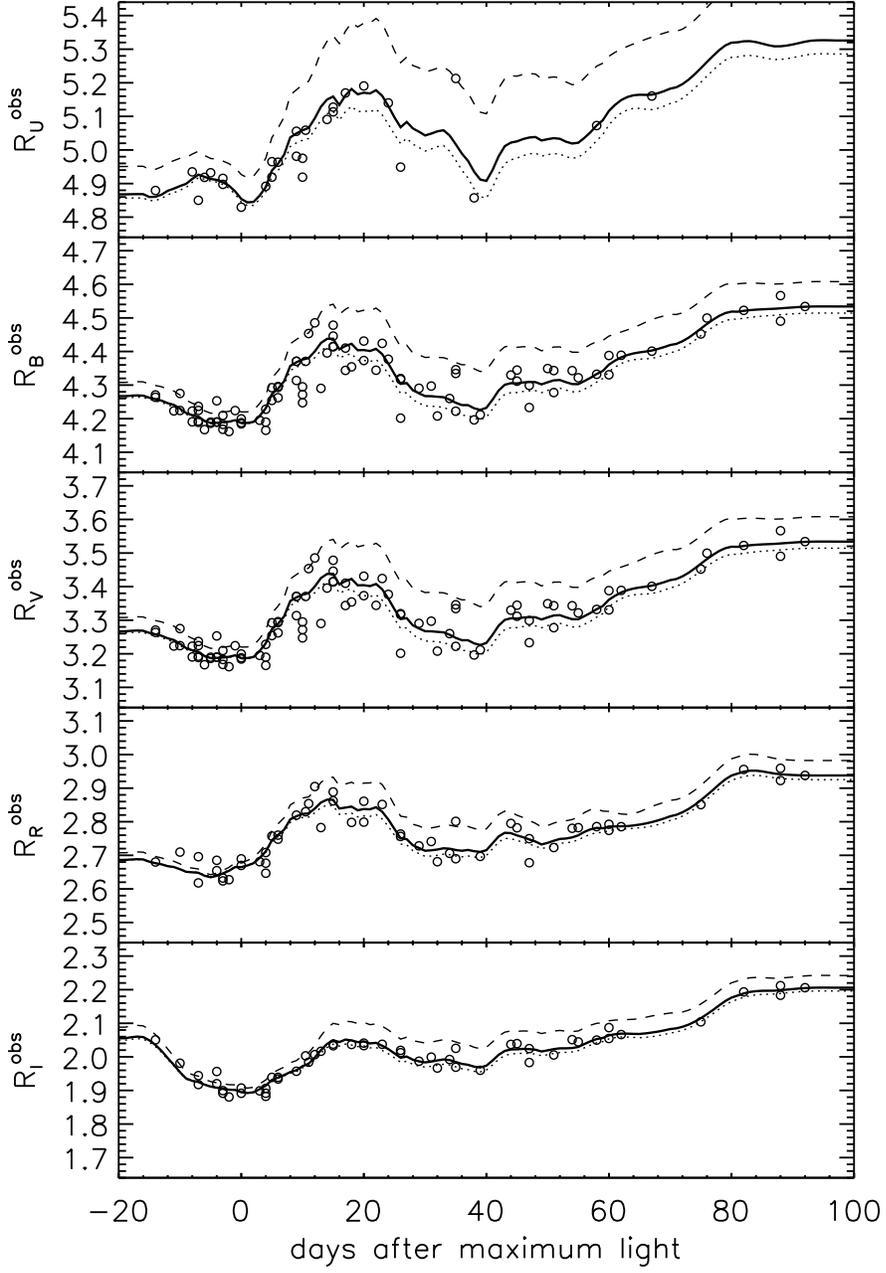}
\end{center}
\caption[$R^{\rm obs}$]{\singlespace Same as Figure
\ref{ch4-fig-rtrue}, except showing the variation of $R^{\rm obs}_X
\equiv A_X/E(\bv)_{\rm obs}$. Note that the y-axis ranges have been
increased in this figure, to encompass the larger variations in
$R^{\rm obs}_X$, and that $R^{\rm obs}_B = R^{\rm obs}_V + 1$ by
definition. \label{ch4-fig-robs}}
\end{figure}

We present our calculations of $R^{\rm true}_X$ and $R^{\rm obs}_X$ in
Figures \ref{ch4-fig-rtrue} and \ref{ch4-fig-robs}, respectively. We
find a good match in general to the results of Nugent et al.~(2002)
and Phillips et al.~(1999), which is not surprising, as we have many
of the input spectra in common. The figures show the time variation of
these two ratios as well as its variation over two magnitudes in
$E(\bv)_{\rm true}$. Both of these ratios have their uses, depending
on whether one knows the extinction \emph{a priori} (for instance,
using a Galactic reddening map) or is measuring it from the observed
SN colors. Because we have used the SED of SN Ia themselves to
calculate these quantities (and their time variation), it is
preferable to use these results rather than those from standard filter
tables (e.g., Table 6 of Schlegel, Finkbeiner \& Davis 1998).

So far we have restricted ourselves to the standard $R_V = 3.1$
extinction law of CCM89, but we would like our distance measuring
technique to allow for variations in the properties of the dust. CCM89
have shown that in the optical passbands, dust in our Galaxy follow
extinction laws that can be expressed in terms $R_V$.\footnote{We
adopt the notation that without a superscript ``true'' or ``obs'',
$R_V$ corresponds to the $R_V$ of CCM89, i.e., derived from
photoelectric measurements of Galactic stars.} Clearly, using
extinction laws with different values of $R_V$ will lead directly to
variations in $R^{\rm true}_X$ and $R^{\rm obs}_X$. The variation with
the total extinction itself adds further complications and makes using
these ratios cumbersome. For instance, calculating $R^{\rm obs}_I
\equiv A_I/(A_B - A_V)$ involves three terms, each varying
individually with epoch, total extinction and extinction law.

To make things simpler, we have chosen to work in the more natural
space of $A_X$ rather than using color excesses to parameterize the
extinction. This is convenient because we fit the light curves in each
passband directly, rather than fitting color curves, an approach that
does not require light curves to have multiple color measurements at
each epoch. Secondly, the extinction laws themselves are better
characterized in ratios of extinctions, e.g., $A_B/A_V$, rather than
in ratios of extinctions to color excesses.\footnote{This point is
made explicitly by CCM89 who note that ``There are some relationships
which emerge more clearly when $A(\lambda)/A(V)$ is considered than
when normalization by $E(\bv)$ is used.''}

\begin{figure}
\begin{center}
\includegraphics[height=6.5in]{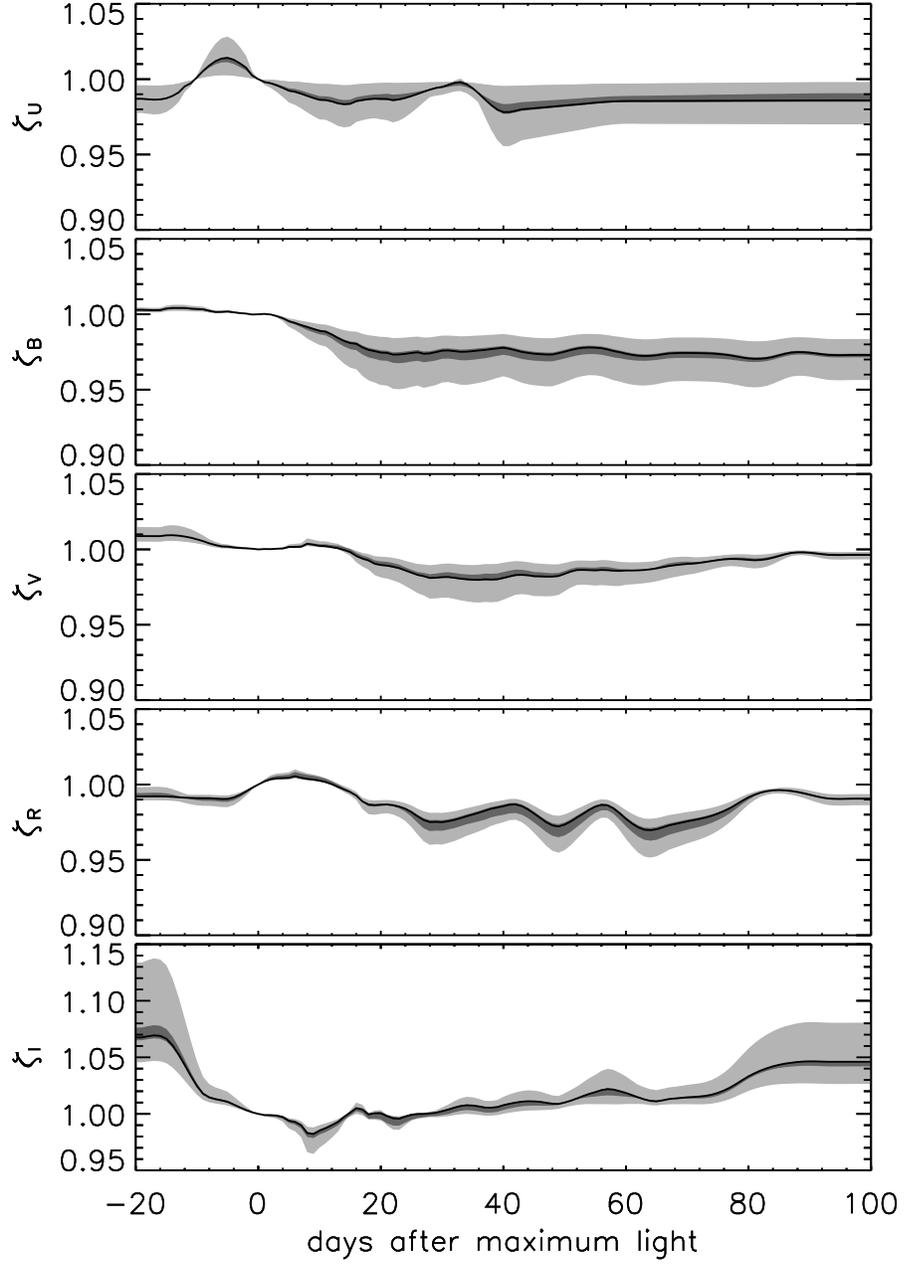}
\end{center}
\caption[$\vec{\zeta_X}$]{\singlespace Calculation of $\vec{\zeta_X}
\equiv \vec{A_X}/A^0_X$ in \ubvri. The dark solid line shows the result for
$R_V = 3.1$ and $E(\bv)_{\rm true}$ approaching zero. The dark gray shaded area
shows the range of variation in $\vec{\zeta_X}$ over 3 magnitudes in
$E(\bv)_{\rm true}$, while the light gray shaded area indicates the
variation in $\vec{\zeta_X}$ over that reddening range and over the
range $1.7 \leq R_V \leq 5.7$. \label{ch4-fig-zeta}}
\end{figure}

The first part of this framework is to separate out the time dependence
of the extinction. For each passband $X$, we define the quantity,
\begin{equation}
\vec{\zeta_X} \equiv \frac{\vec{A_X}}{A^0_X},
\end{equation}
where we denote quantities that are functions of SN phase with vector
arrows (i.e., $\vec{\zeta_X} = \vec{\zeta_X}(t)$), and $A^0_X$ is
defined as the extinction in passband $X$ at maximum light in
$B$. Thus, $\vec{\zeta_X}(t = 0) \equiv 1$ by definition (all times
are defined relative to maximum light in $B$). In Figure
\ref{ch4-fig-zeta} we show our calculation of $\vec{\zeta_X}$ in
\ubvri. The useful result is that $\vec{\zeta_X}$ captures all of
the time dependence in the extinction \emph{and} is insensitive to
both the total extinction $E(\bv)_{\rm true}$ and the extinction law
$R_V$. This point is also illustrated in Figure \ref{ch4-fig-zeta},
where the dark gray bands show the variation in $\vec{\zeta_X}$ over
the very wide range of 3 magnitudes of $E(\bv)_{\rm true}$ and the
light gray bands show the variation of a wide range of $R_V$ from 1.7
to 5.7 (including the full variation in $E(\bv)_{\rm true}$). While
the figure shows some differences in $\vec{\zeta_X}$ at levels greater
than a few percent (for instance, in $I$ at early times), these are
only realized for extreme values of both $R_V$ and $E(\bv)_{\rm
true}$. We can then confidently fix $\vec{\zeta_X}$ as shown by the
heavy solid lines in the figure, independent of the reddening law and
total extinction.

\begin{figure}
\includegraphics[width=6in]{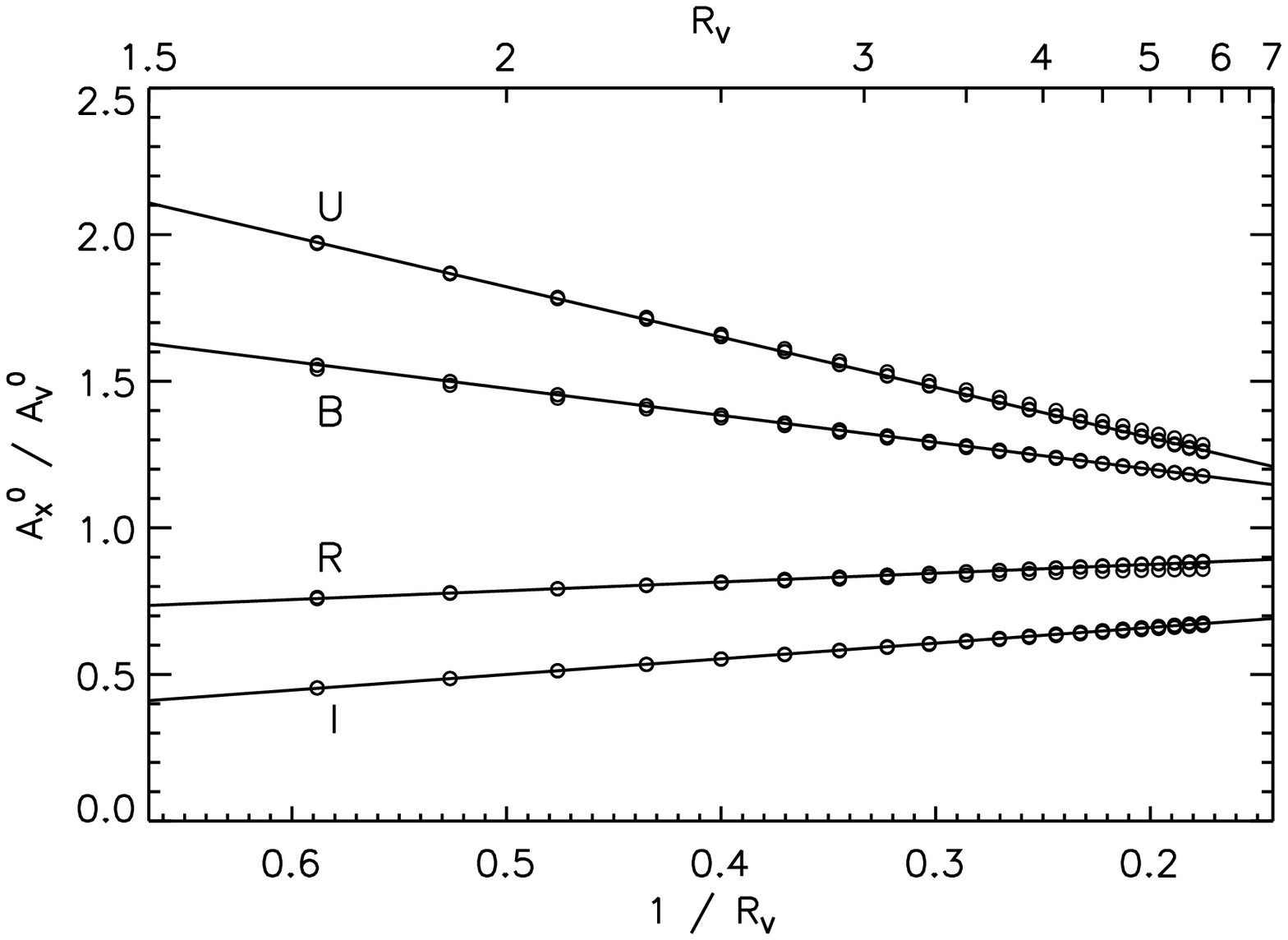}
\caption[Maximum Light Extinction Relations]{\singlespace Variation of
$A^0_X/A^0_V$ as a function of $R^{-1}_V$. The sold lines indicate the best
linear fit. The open circles show the calculated points; for each
passband and $R_V$ there are three open circles, with $E(\bv)_{\rm
true} \simeq -0.5$, $0.0$, and $+2.5$ mag, showing the small
differences in the relations as a function of the total
extinction. \label{ch4-fig-a0rv}}
\end{figure}

With the time dependence separated, we can relate the extinction in
any passband at any epoch to the maximum light extinction $A^0_X$ in
that passband. The maximum light extinctions in the different
passbands are interrelated and we can define the relations between
these (currently, five) parameters as a function of the total
extinction and the reddening law. CCM89 show that the ratio $A_X/A_V$
is a simple linear function of $(1/R_V)$. We have
calculated this relationship at maximum light explicitly using our
sample of spectra near that epoch, and fit for the coefficients
$\alpha_X$ and $\beta_X$, defined by
\begin{equation}
\frac{A^0_X}{A^0_V} \quad = \quad \alpha_X + \frac{\beta_X}{R_V}.
\end{equation}
The results are presented graphically in Figure \ref{ch4-fig-a0rv},
and listed in Table \ref{ch4-tab-a0rv} (which also shows explicitly
the relations for an $R_V = 3.1$ extinction law). Furthermore, the
figure illustrates that $A^0_X/A^0_V$ is a very weak function of
$E(\bv)_{\rm true}$ (in most cases the three open circles are
indistinguishable), and we are justified in ignoring the dependence on
the total extinction.

\begin{deluxetable}{cccc}
\tablecolumns{4}
\singlespace
\tablewidth{0pt}
\tablecaption{Maximum Light Extinction Coefficients \label{ch4-tab-a0rv}}
\tablehead{ & \multicolumn{3}{c}{$A^0_X/A^0_V = \alpha_X + \beta_X/R_V$} \\
\cline{2-4} \colhead{Passband ($X$)} & \colhead{$\alpha_X$} & 
\colhead{$\beta_X$} & \colhead{$\alpha_X + \beta_X/3.1$}}
\startdata
$U$ & 0.964 & \phs1.716 & 1.518 \\
$B$ & 1.016 & \phs0.919 & 1.313 \\
$V$ & 1.000 & \phs0.000 & 1.000 \\
$R$ & 0.935 & $-$0.300  & 0.839 \\
$I$ & 0.767 & $-$0.534  & 0.595 \\
\enddata
\end{deluxetable}


In the framework developed, then, we parameterize the extinction by
two numbers: $A^0_V$, corresponding to the extinction in the $V$
passband at maximum light, and $R_V$, describing the shape of the
extinction law. The fixed coefficients $\alpha_X$ and $\beta_X$
provide the maximum light extinctions in other passbands, and the
vectors $\vec{\zeta_X}$ contain the time variation. While the
parameterization in terms of $R^{\rm true}_X$ and $R^{\rm obs}_X$ is
still useful for certain tasks (see below), we switch to this new
framework in the MLCS2k2 fits.

\subsection{Separating Reddening and Intrinsic Color\label{ch4-sec-ex0}}

To define the training set that yields our desired template light
curves, we need an estimate of the extinction in the host galaxy of
each object. Various approaches are possible, e.g., assuming that SN
Ia in early-type host galaxies or the bluest SN Ia define an
extinction-free sample. Knowing the ``zeropoint'' of the extinction
(i.e., the true, unreddened color of a SN Ia) is not strictly
necessary if we are interested only in relative distances (or in tying
the SN Ia distances to another set, such as the Cepheid scale), but it
becomes essential if we want to use the positivity of the extinction
(dust cannot brighten a source or make it appear bluer) as a
constraint on our derived distances. We employ the observational
results of Lira~(1995), who noticed that the late time
($t \gtrsim +30$ days) \bv\ color evolution of SN Ia was remarkably
similar, regardless of light curve shape near maximum light. SN Ia
undergoing this transition to the nebular phase also show very similar
spectra (e.g., see Filippenko~1997). Phillips et al.~(1999) used this
``Lira Law,'' $(\bv)_0 = 0.725 - 0.0118(t_V - 60)$, giving the
intrinsic color in terms of the number of days past $V$ maximum light,
to measure extinctions and to constrain the \bv\ and \vi\ maximum
light colors of SN Ia (as a function of \dmf).  The Lira Law as quoted
was derived from a fit of late-time photometry of four objects (SN
1992A, 1992bc, 1992bo and 1994D), estimated to be free of host-galaxy
extinction. Phillips et al.~(1999) estimate that the intrinsic
dispersion about the relation is 0.05 mag.

\begin{figure}
\includegraphics[width=6in]{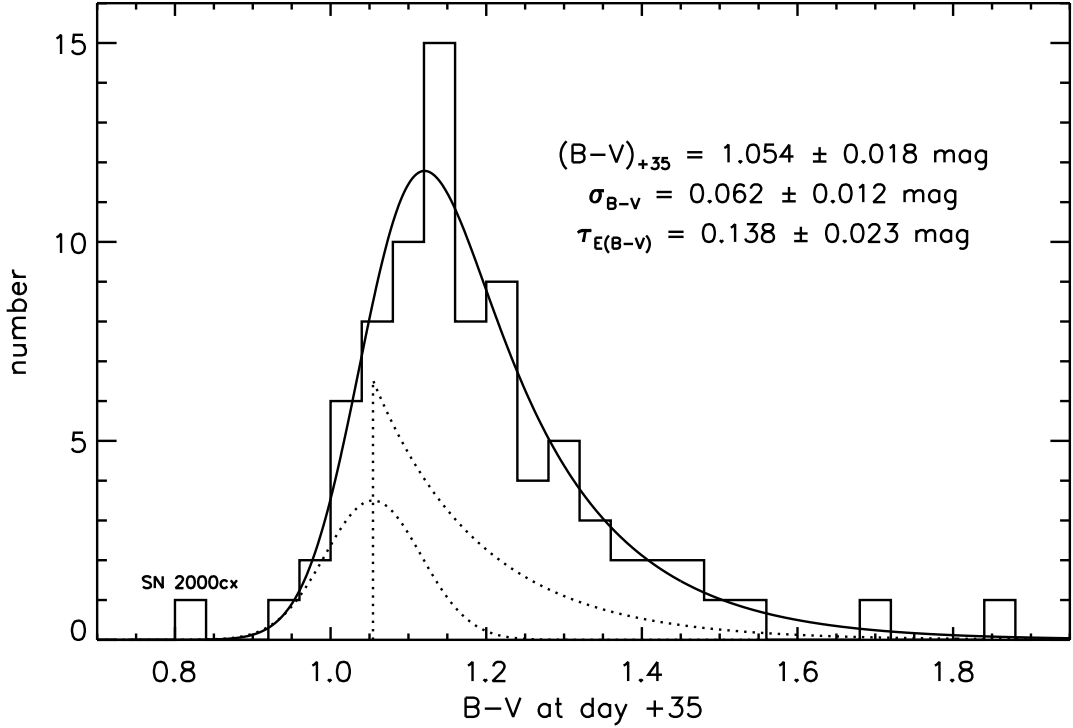}
\caption[Extinction Zeropoint]{\singlespace Histogram of 82 SN Ia
with well-measured late-time \bv\ color evolution. The data were
corrected for Galactic extinction and the $K$-correction, and
referenced to $+35$ days after $B$ maximum, adopting a late-time color
evolution slope of $-0.0118 \; {\rm mag \; day^{-1}}$. The
maximum-likelihood fit model is shown as the solid curve; it is the
convolution of the dotted curves shown (at an arbitrary scale for
clarity). SN 2000cx, a clear outlier, was not included in the
fit. \label{ch4-fig-zerocol}}
\end{figure}

We have attempted to check the Lira relation with another
approach. Rather than trying to choose an extinction-free subsample of
the data, we use all the data we can. Beginning with the sample of 133
SN Ia described above, we have corrected the photometry for Galactic
extinction, using the reddening maps of Schlegel et al.~(1998) and our
calculation of $R^{\rm true}_X$ above, with the assumption that the
Galactic component is described by an $R_V = 3.1$ extinction law. We
have also applied the $K$-correction (see \S \ref{ch4-sec-kcorr}) and
corrected for time dilation to bring the data to the SN rest frame. We
have then constructed \bv\ color curves for the sample, and attempted
to measure the late-time color evolution. We find that the Lira
late-time \bv\ slope of $-0.0118 \; {\rm mag \; day^{-1}}$ does an
excellent job of fitting the bulk of the observations.\footnote{Using
the 31 color curves of highest quality, we find a mean slope of
$-0.0123 \; {\rm mag \; day^{-1}}$ with a standard deviation of
$0.0011 \; {\rm mag \; day^{-1}}$.}  We thus fix this slope and fit a
straight line to all the \bv\ observations between 32 and 92 days past
$B$ maximum light to determine the intercept, $BV_{35}$ (which we
reference to a fiducial epoch of $t = +35$ days) and its observational
uncertainty $\sigma_{BV_{35}}$. Of the 133 objects listed in Table
\ref{ch4-tab-sninfo}, 28 have no useful late-time color information,
while an additional 23 have only poor data unsuitable for a good
measurement, due either to too few late time points or points with
large uncertainties. The $BV_{35}$ measurements for the remaining 82
objects are listed in Table \ref{ch4-tab-snlatecolor} and presented in
a histogram in Figure \ref{ch4-fig-zerocol}.


\begin{deluxetable}{lccclcc}
\tabletypesize{\scriptsize}
\tablecolumns{7}
\singlespace
\tablewidth{0pt}
\tablecaption{SN Ia Late-Time Colors \label{ch4-tab-snlatecolor}}
\tablehead{\colhead{SN Ia} & \colhead{$BV_{35}$} &
\colhead{$\sigma_{BV_{35}}$}  & \colhead{\phantom{Hello}} &
\colhead{SN Ia} & \colhead{$BV_{35}$} &
\colhead{$\sigma_{BV_{35}}$} \\
 & \colhead{mag} & \colhead{mag} & & & \colhead{mag} & \colhead{mag}}
\startdata
1972E  & 1.092 & 0.028 &  & 1997bp & 1.305 & 0.073 \\
1980N  & 1.139 & 0.027 &  & 1997bq & 1.216 & 0.026 \\
1981B  & 1.105 & 0.042 &  & 1997br & 1.362 & 0.030 \\
1986G  & 1.682 & 0.029 &  & 1997cw & 1.521 & 0.042 \\
1989B  & 1.438 & 0.040 &  & 1998V  & 1.123 & 0.031 \\
1990O  & 1.145 & 0.044 &  & 1998ab & 1.215 & 0.029 \\
1990T  & 1.174 & 0.027 &  & 1998aq & 1.099 & 0.026 \\
1990Y  & 1.326 & 0.033 &  & 1998bp & 1.127 & 0.029 \\
1991S  & 1.177 & 0.042 &  & 1998bu & 1.384 & 0.026 \\
1991T  & 1.223 & 0.028 &  & 1998de & 1.195 & 0.044 \\
1991ag & 1.121 & 0.029 &  & 1998dh & 1.214 & 0.032 \\
1991bg & 1.040 & 0.031 &  & 1998dk & 1.284 & 0.041 \\
1992A  & 1.036 & 0.029 &  & 1998dm & 1.461 & 0.031 \\
1992J  & 1.148 & 0.048 &  & 1998es & 1.166 & 0.042 \\
1992K  & 1.045 & 0.034 &  & 1999aa & 1.075 & 0.027 \\
1992P  & 1.152 & 0.031 &  & 1999ac & 1.142 & 0.029 \\
1992al & 1.091 & 0.026 &  & 1999aw & 0.943 & 0.049 \\
1992bc & 0.981 & 0.028 &  & 1999by & 1.067 & 0.027 \\
1992bg & 1.095 & 0.046 &  & 1999da & 1.308 & 0.074 \\
1992bl & 1.006 & 0.046 &  & 1999dq & 1.267 & 0.037 \\
1992bo & 1.036 & 0.072 &  & 1999ee & 1.415 & 0.026 \\
1993H  & 1.041 & 0.041 &  & 1999ek & 1.262 & 0.052 \\
1993L  & 1.283 & 0.045 &  & 1999gd & 1.490 & 0.078 \\
1993O  & 1.071 & 0.029 &  & 1999gh & 1.100 & 0.026 \\
1993ae & 1.022 & 0.029 &  & 1999gp & 1.166 & 0.035 \\
1993ag & 1.156 & 0.050 &  & 2000B  & 1.130 & 0.044 \\
1994D  & 0.970 & 0.034 &  & 2000bh & 1.176 & 0.026 \\
1994M  & 1.233 & 0.029 &  & 2000bk & 1.205 & 0.027 \\
1994Q  & 1.154 & 0.039 &  & 2000ca & 1.040 & 0.029 \\
1994ae & 1.055 & 0.027 &  & 2000cf & 1.167 & 0.026 \\
1995D  & 1.153 & 0.026 &  & 2000cx & 0.827 & 0.025 \\
1995E  & 1.857 & 0.089 &  & 2000dk & 1.007 & 0.082 \\
1995ac & 1.117 & 0.037 &  & 2000fa & 1.159 & 0.047 \\
1995ak & 1.331 & 0.033 &  & 2001V  & 1.144 & 0.026 \\
1995al & 1.225 & 0.032 &  & 2001bt & 1.314 & 0.035 \\
1995bd & 1.274 & 0.033 &  & 2001cn & 1.272 & 0.032 \\
1996C  & 1.107 & 0.049 &  & 2001cz & 1.221 & 0.032 \\
1996X  & 1.083 & 0.035 &  & 2001el & 1.358 & 0.027 \\
1996bl & 1.161 & 0.047 &  & 2002bo & 1.450 & 0.027 \\
1997E  & 1.111 & 0.031 &  & 2002er & 1.233 & 0.030 \\
1997Y  & 1.148 & 0.031 &  & 2003du & 1.050 & 0.026 \\
\enddata
\tablecomments{Determined from a linear fit to SN rest-frame 
\bv\ color measurements between 32 and 92 days past $B$ maximum 
light, with $(\bv)(t) = BV_{35} - 0.0118(t-35)$. See text for 
details.}
\end{deluxetable}


We assume that this distribution is a result of three independent
factors: first, observational uncertainty, second, an intrinsic \bv\
color with an unknown scatter, and third, reddening by dust in the
host galaxy. We then model the distribution of the intrinsic component
as a Gaussian with mean $(\bv)_{+35}$ and standard deviation
$\sigma_{B-V}$ (which subsumes the observational errors), and the
distribution of the reddening as an exponential with scale length
$\tau_{E(B-V)}$, such that the probability density of the reddening
peaks at zero, and falls to 1/$e$ of the peak at $\tau_{E(B-V)}$. The
probability distribution function of the sum of these two components
is just the convolution of the individual distributions, and we
perform a maximum-likelihood analysis using each point to determine
the best-fit model parameters. The results are also shown in Figure
\ref{ch4-fig-zerocol}, with the convolution overplotted on the
histogram and the components inset. The maximum-likelihood model
parameters are $(\bv)_{+35} = 1.054 \pm 0.018$ mag, $\sigma_{B-V} =
0.062 \pm 0.012$ mag, and $\tau_{E(B-V)} = 0.138 \pm 0.023$ mag, where
the uncertainties were determined by bootstrap resampling of the data
set.

There are a couple of cautionary points about this result. First, we
have excluded from the fit the very peculiar SN 2000cx (Li et
al.~2001), which is a clear outlier in the histogram and whose light
curve (but not peak luminosity) is significantly unlike other SN Ia
(Li et al.~2001; Jha et al.~2006). In addition, because the SN Ia we
study were discovered in a variety of supernova searches, they do not
comprise a complete or well-defined sample, and thus may be biased
with respect to the true distribution of SN Ia late-time color. In
particular, there may be a significant population of heavily obscured
SN Ia that are not present in the sample, leading to an underestimate
of $\tau_{E(B-V)}$, assuming an exponential distribution is valid at
all (Hatano, Branch, \& Deaton 1998). While our derived value for
$\tau_{E(B-V)}$ is valid for the sample we study and whose distances
we derive in this paper, other supernova surveys with flux or volume
limited samples, or searches in infrared, may find differing numbers
heavily extinguished SN Ia and different estimates of
$\tau_{E(B-V)}$.\footnote{If, for example, we truncate the observed
distribution to eliminate the reddest objects with $BV_{35} \ge
1.5$ mag, the derived scale length is $\tau = 0.104 \pm 0.021$ mag.}

Fortunately, we are chiefly interested in $(\bv)_{+35}$ and
$\sigma_{B-V}$, and it is unlikely our estimates of these are
significantly biased.  Our results with this maximum likelihood method
are in good accord with the Lira relation (which predicts $(\bv)_{+35}
\simeq 1.044$) and its estimated dispersion.  Our approach has the
advantage of using all normal SN Ia to determine the intrinsic color
zeropoint, rather than just the bluest objects.  As Figure
\ref{ch4-fig-zerocol} shows, in addition to having low reddening, the
bluest objects are also intrinsically bluer than the mean unreddened
color. Part of the ``intrinsic'' color scatter, $\sigma_{B-V}$ is due
to observational uncertainty in the color measurement. Based on our
color measurements, with an average uncertainty of 0.037 mag, we can
remove this component and estimate the true intrinsic color scatter
$\sigma_{(B-V){\rm int}} = 0.049$ mag.\footnote{Nobili et al.~(2003)
suggest that the distribution of \bv\ observations of unextinguished
SN Ia near 35 days past maximum can be explained entirely by
measurement errors, with no intrinsic component to the
dispersion. However, here we have combined late-time data from 32 to
92 days into one measurement using the Lira law, and this combination
reduces the observational uncertainty. Even with a conservative
estimate of a positive correlation among the late-time data for a
given SN, our results require an intrinsic late-time color dispersion
with a significance $\gtrsim$ 3-$\sigma$.}

Our model then establishes independent distributions for both the
intrinsic color of SN Ia and the reddening. Intrinsically, SN Ia at
35 days past maximum (and even later times, if corrected by the Lira
slope to the reference $+35$ day epoch) have a \bv\ color distribution
which can be described by a Gaussian with a mean of 1.054 mag and
standard deviation of 0.049 mag. In our sample, the distribution of
host-galaxy reddening by dust $E(\bv)$ is well described by an
exponential with a scale length of 0.138 mag. The observed color is
the sum of the intrinsic color and the reddening.

A straightforward application of Bayes's theorem yields the
prescription to turn the model around: determining estimates of the
reddening (and intrinsic color) from a measurement of the observed
color. For notational simplicity, we define $E \equiv E(\bv)$ as the
host-galaxy reddening, $L$ as the intrinsic late-time \bv\ color, and
$O$ as the observed late-time \bv\ color, with $O = E + L$. Bayes's
theorem says that the conditional probability $p(E \ \vert \ O)
\propto p(O \ \vert \ E) p(E)$, where $p(E)$ is just the reddening
distribution derived above, $p(E) \propto \exp(-E/0.138)$, for $E \geq
0$. The other term, the probability of measuring an observed color
\emph{given} the reddening, $p(O \ \vert \ E)$, is simply equivalent
to $p(L = O - E)$, the probability that the \emph{intrinsic} color
(which has a Gaussian distribution, as above) is that which added to
$E$ gives the observed color, $O$.

The conclusion is that for a SN Ia that has a measured late-time \bv\
color $BV_{35}$ with an observational uncertainty $\sigma_{BV_{35}}$
(with all quantities measured in magnitudes), the probability
distribution of the host-galaxy reddening $E(\bv)$, again called $E$
for notational simplicity, is given by
\begin{equation}
p(E \ \vert \ BV_{35},\sigma_{BV_{35}}) \; \propto \;
\cases{
\exp \left ( -\frac{\left(BV_{35} - E - 1.054\right)^2}
{2\left(\sigma_{BV_{35}}^2 + 0.049^2\right)} \right ) 
\exp \left (-\frac{E}{0.138} \right )
& if $E \geq 0$, \cr
\cr
\, 0 & if $E < 0$.\cr
}
\label{eqn:E}
\end{equation}

This is a generalization of the Bayesian filter of Riess, Press, \&
Kirshner (1996a), where our model now includes the uncertainty caused
by the intrinsic scatter in SN Ia late-time color in addition to the
observational uncertainty. We use equation \ref{eqn:E} with the data
in Table \ref{ch4-tab-snlatecolor} to get the initial host-galaxy
reddening estimates required in training MLCS2k2.

The intrinsic scatter in the late-time colors (0.049 mag) dominates
the uncertainty in the mean unreddened color from the maximum
likelihood fit (0.018 mag), and thus we strongly urge that
``negative'' extinction be disallowed in any model. The idea of
negative extinctions (because of a color zeropoint that is estimated
redward of the true color zeropoint) is a convenient fiction, but only
if the intrinsic color scatter is smaller than the zeropoint
uncertainty. Here we have the opposite case; blue SN Ia (with $BV_{35}
= 1.0$ mag, for example) are blue because of the intrinsic color
variation, not because they are the only ``truly'' unextinguished
objects. Thus, ``correcting'' them with a negative extinction is a
mistake. As long as there is no systematic error in the mean
unreddened color approaching the level of the intrinsic scatter, the
mean of equation \ref{eqn:E}, $\int_{0}^{\infty} p(E \ \vert \
BV_{35}, \sigma_{BV_{35}}) E dE$, provides an \emph{unbiased} estimate
of the host-galaxy reddening.

\section{Model \label{ch4-sec-model}}

We are now in a position to define our light curve model. For each
passband, $X$, we fit the observed light curves (corrected for
Galactic extinction, $K$-correction, and time dilation), $\vec{m_X}$
as follows (arrowed quantities span the SN rest-frame phase):
\begin{equation}
\vec{m_X}(t - t_0) = \vec{M^0_X} + \mu_0 + \vec{\zeta_X} \left(\alpha_X +
\beta_X/R_V\right) A^0_V + \vec{P_X} \Delta + \vec{Q_X} \Delta^2,
\end{equation}
where $t_0$ is the epoch of maximum light in $B$, $\vec{M^0_X}$ are
the absolute magnitudes of the fiducial SN Ia, $\mu_0$ is the true
distance modulus, $R_V$ and $A_V^0$ are the host-galaxy extinction
parameters (\S \ref{ch4-sec-ex}), $\Delta$ is the
luminosity/light-curve shape parameter, and $\vec{P_X}$ and
$\vec{Q_X}$ are vectors describing the change in light curve shape as
a (quadratic) function of $\Delta$. There are five ``free'' parameters
in the model: $t_0$, $\mu_0$, $\Delta$, $A^0_V$, and $R_V$. As in
previous versions of MLCS, we solve for the optimal vectors
$\vec{M^0_X}$, $\vec{P_X}$, and $\vec{Q_X}$ using a training set for
which we estimate initial values of the free parameters based on
relative distances from the Hubble Law.

Given a training set and solution for the optimal vectors, we
construct an empirical model covariance matrix $S$ that incorporates
the variance and covariance in the residuals of the training set data
from the model (minus the variance and covariance in the training set
data itself). Following Riess et al.~(1998a), the diagonal elements of
the $S$ matrix are derived from the variance about the model, while
the off-diagonal elements are estimated from two-point correlations
(in the same passband at different epochs, in different passbands at
the same epoch and in different passbands at different epochs).

Armed with the template vectors and the model covariance matrix, we
can apply the model. Along with the light curve observations
$\vec{m_X}$, conscientious observers provide a covariance matrix of
``noise'', $N$, albeit typically only a diagonal one consisting of the
variance of each data point (even though SN light curve data can be
highly correlated, for instance in the fact that all the light curve
points are usually referenced to the same few field comparison
stars). We correct these data for Galactic extinction, and incorporate
the uncertainty estimate of Schlegel et al.~(1998), whereby
$\sigma\left(E\left(\bv\right)_{\rm true}\right) \simeq 0.16 \times
E(\bv)_{\rm true}$, updating both the diagonal and off-diagonal
elements of $N$ (the uncertainties in the Galactic extinction
correction are highly correlated, but generally small). We also apply
the $K$-correction (\S \ref{ch4-sec-kcorr}), and incorporate the
$K$-correction uncertainty in the diagonal elements of N
(unfortunately we do not have enough data to estimate the
$K$-correction correlations) and correct for time dilation.

We find the best-fit model parameters via $\chi^2$ minimization, with 
\begin{equation}
\chi^2 = \vec{r}^{\; T}C^{-1}\vec{r},
\end{equation}
where $\vec{r}$ is the vector of residuals (in all bands) for a given
set of model parameters, and $C = S + N$. We use a downhill simplex
method (amoeba; Press et al.~1992) to perform the minimization (though
linear algebra suffices in solving for $\mu_0$ and $A^0_V$, if the
other model parameters are fixed) and determine their best-fit
parameters. Because the $S$ matrix is empirically determined from the
training set, application of the model to objects in the training set
will necessarily have a minimum reduced $\chi^2_\nu$ ($\equiv \chi^2
$ per degree of freedom) close to unity, but in applying the model to
other objects, the minimum $\chi^2_\nu$ still yields useful
goodness-of-fit information (indicating how similar the light curves
of the new object are to those in the training set, given the nature
of the model).

We have also incorporated the ability to add priors on any of the
model parameters directly into the fit. The previous versions of MLCS,
for instance, used a Bayesian filter requiring the extinction to be
non-negative, but this was enforced after the best fits were
determined. The advantage of that approach was convenience and
expediency, but at the expense of ignoring correlations among the
parameters. Those versions of MLCS used $\mu_V$, the distance modulus
uncorrected for extinction, as the basic distance model parameter
rather than $\mu_0 = \mu_V - A_V$, in order to minimize this
undesirable correlation, thus making the Bayesian filter simple and
effective. However, there are situations in which one might like to
include other prior information into the fit. For example, a
sparsely-sampled light curve may not allow for a good determination of
the time of maximum light, but we may have other information about the
time of maximum from spectroscopy (Riess et al.~1997a). In the extreme
limit of the ``snapshot'' method, (Riess et al.~1998b), spectroscopy
could be used to determine prior constraints on both $t_0$ and
$\Delta$. Alternately, we may have observations in additional,
non-modeled passbands (such as in the near-infrared), which impose
prior constraints on $A^0_V$ or $R_V$. Incorporating priors,
$\hat{p}$, into the fit is straightforward, we simply adjust the
$\chi^2$ to
\begin{equation}
\chi^2 \; = \; \vec{r}^{\; T}C^{-1}\vec{r} - 2 \ln
\hat{p}(t_0,\mu_0,\Delta,A^0_V,R_V),
\label{ch4-eqn-chisq}
\end{equation}
taking care to reinterpret this ``$\chi^2$'' in the proper way when
estimating the goodness of the model fit (which depends only on the
first term, as opposed to the parameter uncertainties). This procedure
allows the parameters to be ``naturally'' filtered during the fit.

\section{Training}

The training set is the most critical part of the analysis; it
requires objects with accurate estimates of the model parameters, in
order to construct the template vectors and covariance matrix. As in
Riess et al.~(1998a), we use the Hubble law to determine precise
relative distances, thus useful objects in the training set need to
have recession velocities that are dominated by cosmological
expansion, not peculiar motions.  The objects also need to have
well-measured light curves at maximum light so that $t_0$ and $V_{\rm
max}$ (to be used in determining $\Delta$, as below) can be reliably
determined. Finally, we require accurate host-galaxy extinction
estimates for the training set objects. This estimate comes from
well-sampled $B$ and $V$ light curves at the epochs past $+35$ days
when the intrinsic color is not a strong function of luminosity, as
described in \S \ref{ch4-sec-ex0}. Ideally, we would also like
estimates of the host-galaxy extinction law $R_V$ for each SN, but
these are not easily determined \emph{a priori}, and values based on
the SN light curves themselves would require knowledge of the
intrinsic colors we are trying to determine! To avoid this conundrum,
then, we restrict the initial training sample to objects not very
significantly reddened (as determined from the \bv\ color at $\sim$35
days past maximum), so that variation in $R_V$ does not have a large
effect, allowing us to fix $R_V$ = 3.1 for the initial training set.

Our initial training set consists of 37 SN Ia, with $cz \geq 2500 \;
\kms$ (measured in the CMB frame) and well-sampled light
curves beginning earlier than 10 days past maximum light: SN 
1990O,
1990af,
1992P,
1992ae,
1992al,
1992bc,
1992bg,
1992bl,
1992bo,
1992bp,
1992br,
1993H,
1993O,
1993ag,
1994M,
1995ac,
1996C,
1996bl,
1997E,
1997Y,
1997bq,
1998V,
1998ab,
1998bp,
1998de,
1998es,
1999aa,
1999ac,
1999da,
1999gp,
2000ca,
2000cf,
2000dk,
2000fa,
2001V,
2001cz, and
2002er.
For each SN, we initially calculate $\Delta$ as the difference between
$V$ magnitude at maximum light (corrected for host-galaxy extinction)
and the Hubble line given the host-galaxy redshift. We assign an
uncertainty in $\Delta$ from the quadrature sum of the uncertainty in
the direct fit of $V_{\rm max}$, the extinction uncertainty and a
distance uncertainty based on a peculiar velocity uncertainty of
$\sigma = 300 \; \kms$. These initial guesses of $\Delta$ span a range
of over two magnitudes, and the sample includes a wide range of over-
and under-luminous objects.

From the initial guesses of the model parameters and uncertainties, we
derive the best fit vectors, $\vec{M^0_X}$, $\vec{P_X}$, and
$\vec{Q_X}$, as well as the model covariance matrix $S$, all of which
are sampled daily over the range $-10 \leq t \leq +90$ days past $B$
maximum, where the data are constraining. With these vectors, we can
apply this initial model to our larger sample of supernovae; deriving
new estimates of $\Delta$ and $A_V^0$ while constraining the distance
moduli to their Hubble Law estimates (within the uncertainties listed
above). Iterating this procedure leads to convergence on a consistent
set of model parameters, template vectors and a model covariance
matrix.\footnote{In principle, this procedure could artificially
decrease the scatter in our distance measurements, if the contribution
of individual SN Ia to the templates ``circularly'' affected their
fits when the model is applied. In practice, however, the templates
are derived from a large enough number of objects that this effect is
negligible. We have further tested this by applying the model derived
only from the initial training set to the remaining objects, as well
as by partitioning the full sample and iteratively training the model
on one half of the data, and then applying it to the other half. In
all cases the scatter in the Hubble diagram of the objects not used in
the training set is significantly decreased, typically from $\sim$0.5
mag to $\lesssim$0.2 mag. We thus choose to retain the largest
training set sample we can to most robustly determine the templates
and, importantly, their associated model uncertainty.}

\begin{figure}
\includegraphics[width=6in]{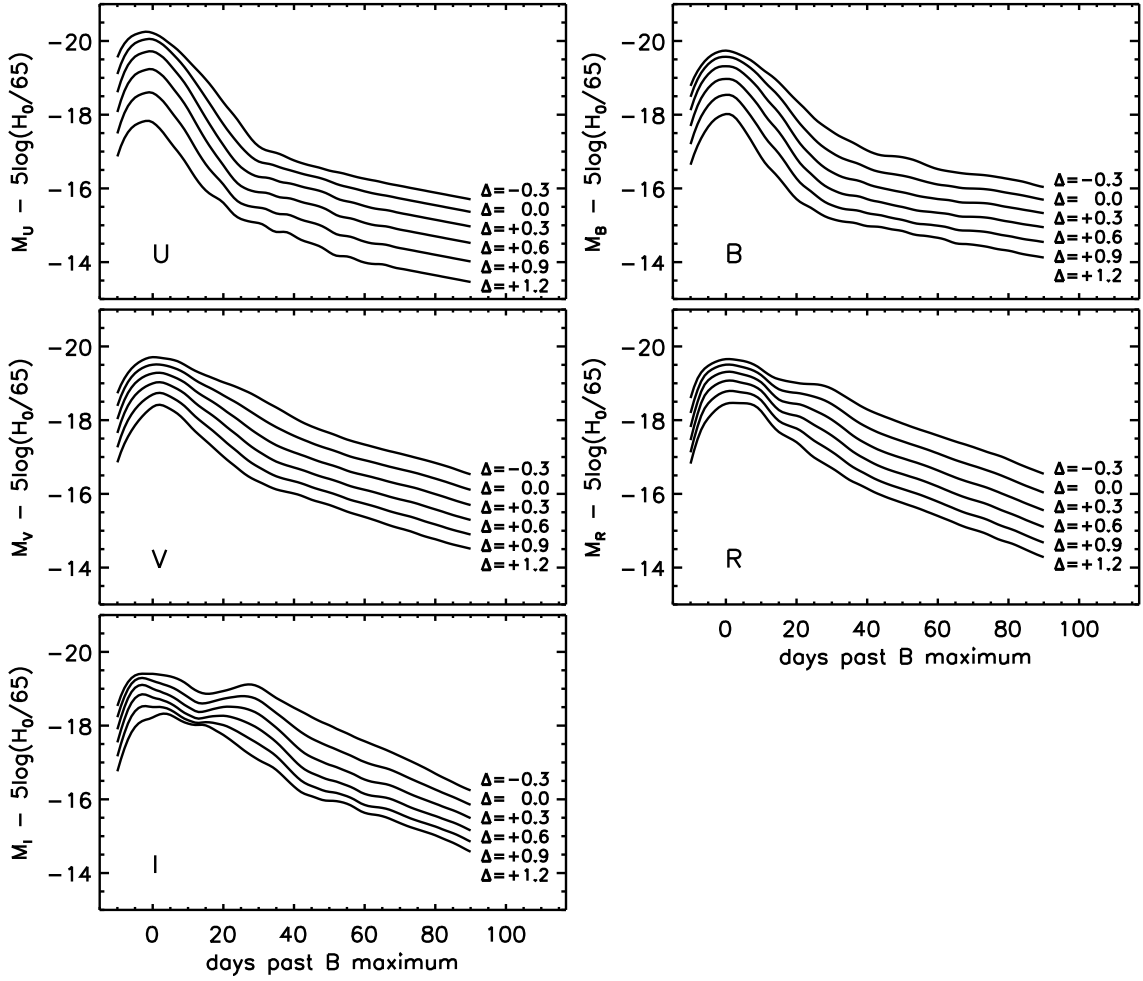}
\caption[Light curve templates]{\singlespace MLCS2k2 intrinsic \ubvri
light curve templates, $\vec{M_X} = \vec{M^0_X} + \vec{P_X}\Delta +
\vec{Q_X}\Delta^2$, shown over a range of luminosity and light-curve
shape from $\Delta = -0.3$ (brighter) to $\Delta = +1.2$
(fainter). \label{ch4-fig-templates}}
\end{figure}

In Figure \ref{ch4-fig-templates}, we present the one-parameter family
of unreddened template light curves in \ubvri over a wide range of
luminosity, based on the final template vectors derived in the
training process. The characteristic result that intrinsically
brighter SN Ia (low $\Delta$) have broader light curves is clearly
established in all passbands. We note that while $\Delta$ is
originally estimated by the relative V-band magnitude difference at
maximum light, during the iterative training the meaning of $\Delta$
changes. It should be viewed simply as a unitless fit parameter that
describes the family of light curves, and can differ significantly
from its original value for a given supernova. In particular, it
should not be subtracted directly from $V_{\rm max}$ to obtain a
corrected peak magnitude, though this has been a convenient usage in
the past.\footnote{The reason for this is the increased dispersion and
shallower slope in SN Ia luminosity at the slowly-declining end of the
distribution, which has become clear only with the growing samples of
SN Ia now being used to calibrate the luminosity/light-curve-shape
relationship. The iterative training for MLCS2k2 associates objects
with similar light curve shapes, which leads the model vectors away
from their starting point at $\vec{P_V}(t=0) = 1$ and $\vec{Q_V}(t=0)
= 0$ to convergence at $\vec{P_V}(t=0) = 0.736$ and $\vec{Q_V}(t=0) =
0.182$.} The fiducial MLCS2k2 $\Delta = 0$ model light curve has
\dmfb = 1.07 (read directly off the $B$-band template) and
stretch parameter, $s_B = 0.96$ (fit as described by Jha et
al.~2006). The intrinsic absolute $V$ magnitude at the time of $B$
maximum light is given by
\begin{equation}
M_V(t=0) = -19.504 + 0.736\Delta + 0.182\Delta^2 +
5 \log \left(\frac{H_0}{65 \; \kmsmpc}\right) 
\quad {\rm mag.}
\label{ch4-eqn-absvmax}
\end{equation}

\begin{figure}
\begin{center}
\includegraphics[height=6in]{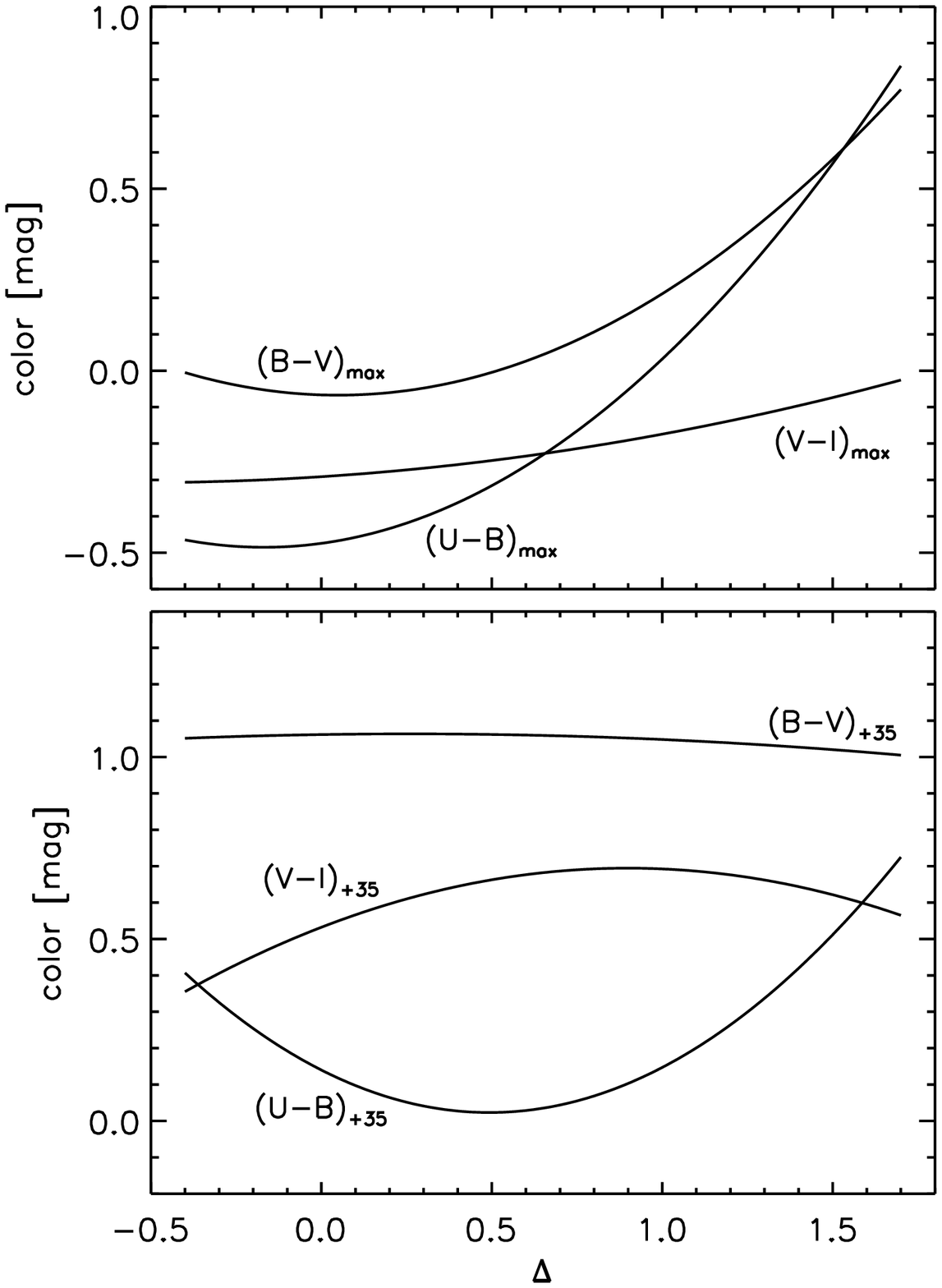}
\end{center}
\caption[MLCS2k2 colors]{\singlespace MLCS2k2 maximum light (top
panel) and late time (+35 days, bottom panel) intrinsic \ub, \bv, and
\vi\ colors as a function of decreasing intrinsic luminosity
(increasing $\Delta$). \label{ch4-fig-colors}}
\end{figure}

The MLCS2k2 relations also allow us to determine the intrinsic colors
of SN Ia with varying luminosity at any epoch. In Figure
\ref{ch4-fig-colors}, we show a sample of the color relations in \ub,
\bv, and \vi, at maximum light and 35 days after maximum. Our fiducial
$\Delta = 0$ SN Ia has $\ub\ = -0.47$ mag, $\bv\ = -0.07$ mag, $\vr\ =
0.00$ mag, and $\vi\ = -0.29$ mag, at the time of $B$ maximum light.
The intrinsic dispersion in these colors is captured in the model
covariance matrix.  Because the intrinsic colors are determined by the
model at any epoch, all of the observations can be used to constrain
the extinction and $\Delta$, weighted by the observational and model
uncertainties. This is a more flexible approach than just using
late-time or maximum-light relations to determine the extinction
(e.g., Phillips et al.~1999), or using colors at one specific epoch to
determine the intrinsic luminosity and extinction (e.g., at $t = +12$
days, as suggested by Wang et al.~2005).

\section{Application \label{ch4-sec-app}}

We have applied the model to our full sample of 133 SN Ia, employing
priors, $\hat{p}(t_0,\mu_0,\Delta,A^0_V,R_V)$, via equation
\ref{ch4-eqn-chisq}. Of course, in applying the model to a SN Ia light
curve we impose no prior constraints on the distance modulus,
$\mu_0$. We require $t_0$ to be within $\pm$3 days of the input
estimate used to derive the time-dependent Galactic extinction and
K-corrections; in cases where the fit $t_0$ is outside this range or
otherwise inconsistent with the input value, we start the fit over
using the updated estimate for $t_0$. We use a uniform prior on
$\Delta$ over the range $-0.3 \leq \Delta \leq 1.6$, with a Gaussian
rolloff (with $\sigma = 0.1$) on either side of that range; this
restricts the fits to the range of $\Delta$ encompassed by the
training set (roughly between $-$0.4 and 1.7) and for which the model
is valid. We do not see objects in our sample ``piling up'' at these
boundaries, but it will be important to check that application of
MLCS2k2 to a new object does not require a fit that extrapolates
beyond the training set.

Additionally, we constrain the maximum light host-galaxy extinction
parameter $A_V^0$ to be non-negative, as well as incorporate the
results of our determination of the SN Ia intrinsic color and
reddening distributions (\S \ref{ch4-sec-ex0}). There we determined
that the host-galaxy reddening distribution was well described by an
exponential with a scale length of $\tau_{E(B-V)} = 0.138$ mag.
Because our MLCS2k2 model parameter is $A_V^0$, the maximum light
extinction in $V$, we need to convert the results based on the
late-time \bv\ color, multiplying by $R_V^{\rm obs}$ (to change from
$E(\bv)$ to $A_V$) and dividing by $\zeta_V$. With average late-time
values of these quantities, we derive our final prior on the maximum
light host-galaxy extinction, $\hat{p}(A_V^0) \propto
\exp(-A_V^0/0.457 \ {\rm mag})$ for $A_V^0 \geq 0$. Our results show
little sensitivity to the exact value for the exponential scale in the
denominator.

Finally, we use a prior constraint on $R_V$ estimated from its
distribution over multiple lines of sight to stars in the Galaxy
(CCM89). For $\hat{p}(R_V)$, we adopt a functional form that is
Gaussian in $R_V^{-1}$ with a mean $\langle R_V \rangle = 3.1$ and
standard deviation $(\langle R_V^2 \rangle - \langle R_V
\rangle^2)^{1/2} = 0.4$.  The choice of a prior on $R_V$ has a
significant effect on the best-fit parameters of heavily extinguished
SN Ia, and we discuss this issue more fully below (\S
\ref{ch4-sec-rv}).

To fit the MLCS2k2 model, we first calculate $\chi^2$ (equation
\ref{ch4-eqn-chisq}) over a four-dimensional grid of ($t_0$, $\mu_0$,
$\Delta$, $A_V^0$), with a fixed $R_V = 3.1$. Because variation in the
reddening law has only a very small effect on the measured distance if
the extinction is low, we expand this to a five-dimensional grid
(including $R_V$ as a fit parameter) just for those objects which show
$A_V^0 > 0.5$ mag in the initial fit. Though mapping $\chi^2$ over a
large grid is computationally expensive, it allows us to fully explore
our non-linear model, and makes evident, for example, objects which
could be well fit in separate regions of parameter space or otherwise
have a complicated $\chi^2$ surface (which can occur especially for SN
Ia with sparse photometry).

\begin{figure}
\includegraphics[width=6in]{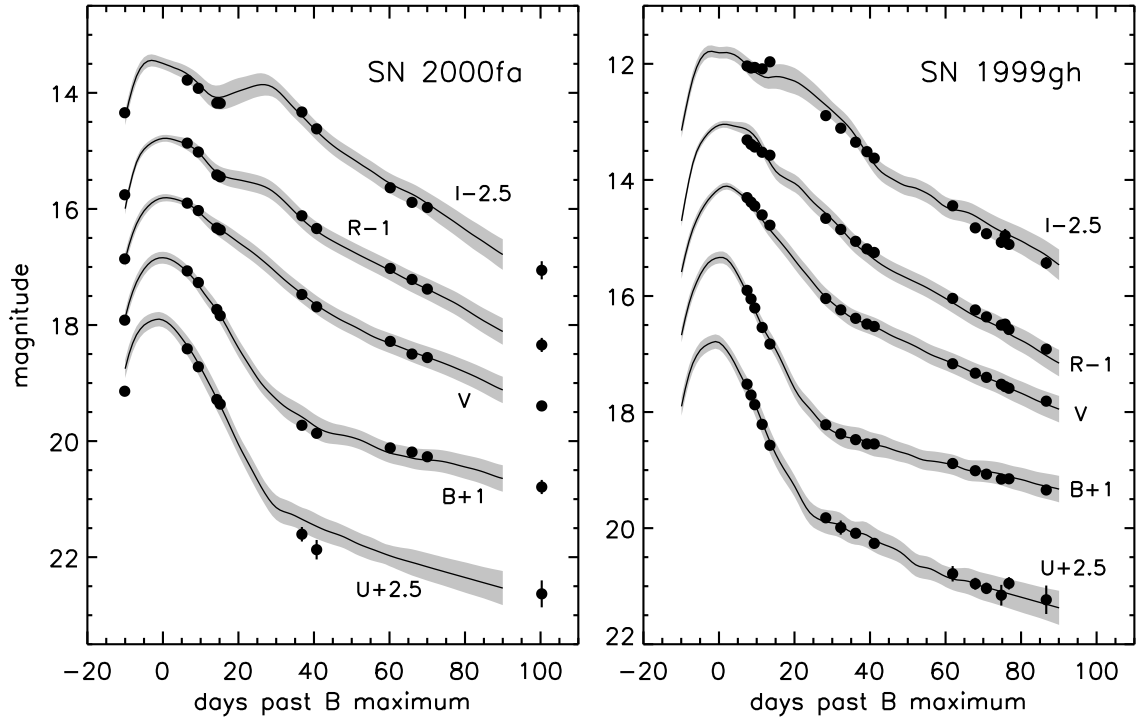}
\caption[Sample MLCS2k2 fit]{\singlespace MLCS2k2 fits of SN 2000fa
and SN 1999gh. The SN rest-frame \ubvri photometry (filled circles) is
shown with the best-fit model light curves (solid lines), as well as
the model uncertainty (shaded regions) derived from the diagonal
elements of the model covariance matrix. The supernovae are well fit,
even with gaps in the light curves and a lack of observations at
maximum light.\label{ch4-fig-sampfit}}
\end{figure}

In Figure \ref{ch4-fig-sampfit} we show example light curve fits for
SN 2000fa and SN 1999gh (Jha et al.~2006), whose light curves are
typical in the nearby SN Ia sample. SN 2000fa has a slightly broader
light curve than average and is moderately extinguished by dust in its
host galaxy, while SN 1999gh is a fast-declining, subluminous SN Ia
with little host extinction. The shaded regions in the figure show the
model uncertainty (more precisely, the square root of the diagonal
elements of $S$) which generally dominates the uncertainty in the
data.

\begin{figure}
\begin{center}
\includegraphics[height=7.7in]{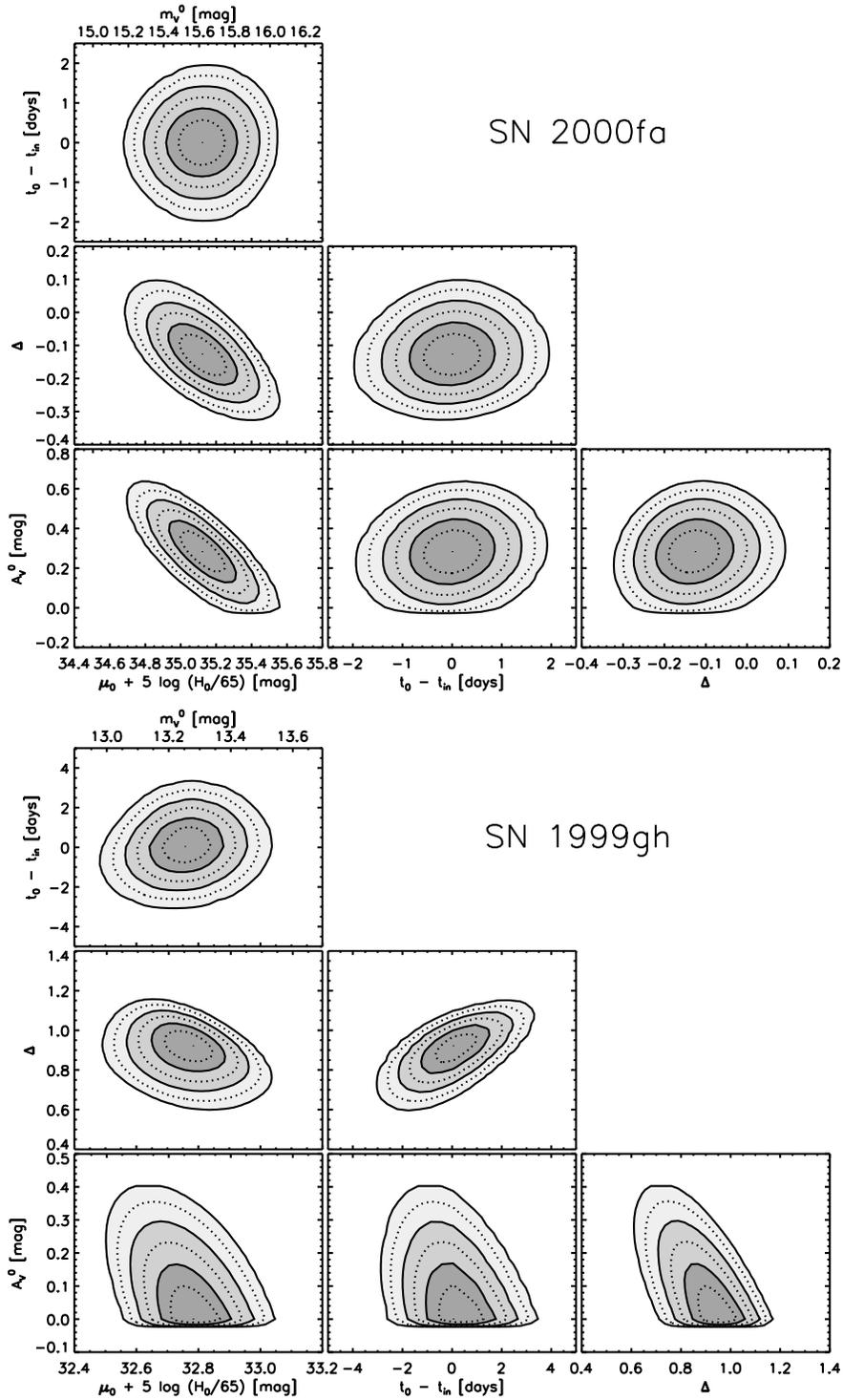}
\end{center}
\caption[2d Contours]{\singlespace Two-dimensional MLCS2k2 probability
densities for the fits to SN 2000fa and SN 1999gh. Each panel shows
the four-dimensional probability density marginalized over the two
remaining parameters. The solid lines and shaded regions indicate 1,
2, and 3-$\sigma$ confidence regions (more precisely, 68.3\%, 95.4\%,
and 99.7\% enclosed probability regions), while the dotted lines show
\onehalf-$\sigma$ contours. \label{ch4-fig-grids}}
\end{figure}

\begin{figure}
\begin{center}
\includegraphics[height=7.7in]{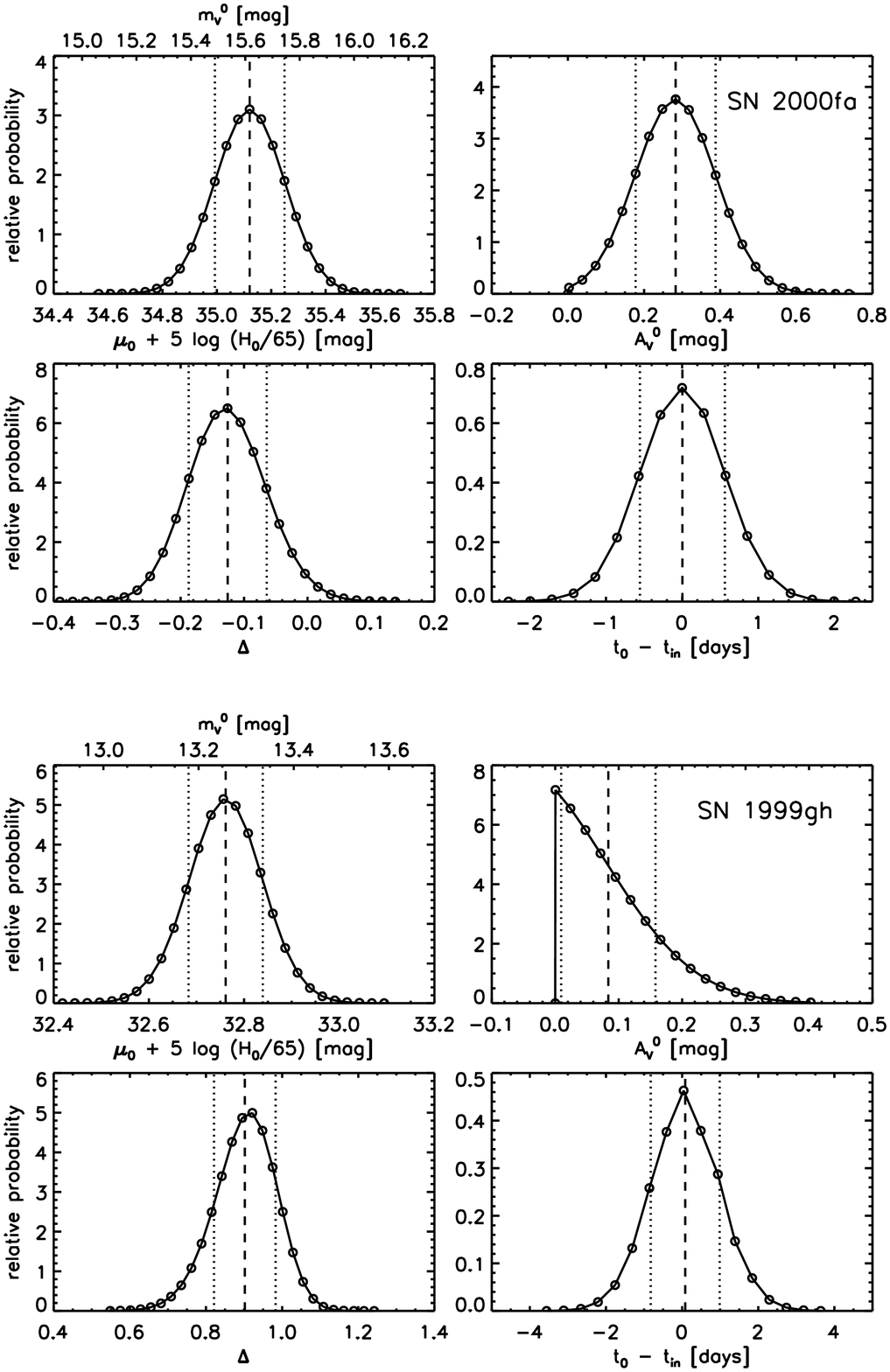}
\end{center}
\caption[1d Plots]{\singlespace MLCS2k2 probability densities for the
fits to SN 2000fa and SN 1999gh. Each panel shows the probability
distribution for each free parameter, marginalizing over the three
remaining parameters. The empty circles show the grid points at
which the fits were calculated. The distribution mean is denoted by
the dashed line, while the dotted lines show one standard deviation
about the mean.\label{ch4-fig-oneds}}
\end{figure}

From the likelihood grid we calculate $p(t_0, \mu_0, \Delta, A_V^0,
R_V) = \exp(-\chi^2/2)$, the posterior probability density function
(pdf). We marginalize this distribution to determine the pdfs of the
parameters of interest. Though the full surface is difficult to
visualize or present, we show all of the two-dimensional
marginalizations of the probability densities for SN 2000fa and SN
1999gh in Figure \ref{ch4-fig-grids}, clearly illustrating bivariate
correlations among the model parameters. Finally, in Figure
\ref{ch4-fig-oneds} we show the one-dimensional pdfs for each of the
model parameters. Our best estimate of each parameter is given by the
\emph{mean} of these distributions (not the mode/peak), and their
standard deviations give an estimate of the uncertainty.  Note that in
some cases these distributions are asymmetric or otherwise
non-Gaussian (for example, $p(A_V^0)$ for SN 1999gh), so the mean and
standard deviation do not always provide a complete description. In
typical cases, however, the distribution of $\mu_0$, the parameter of
primary interest, can be well approximated by a Gaussian. While
derivative calculations based on these parameters (fitting the Hubble
diagram, for example) should formally employ the full pdf, we have
found that after combination of just a few objects, the formally
correct (but computationally intensive) method yields results
indistinguishable from those based on the means and standard
deviations.

We present the full results of the MLCS2k2 fits to our SN Ia sample in
Table \ref{ch4-tab-mlcs}. The tabulated values correspond to the means
and standard deviations of the posterior pdf for each parameter taken
\emph{individually}, i.e. marginalizing over all the other
parameters. As such the model derived from setting each parameter to
its mean value does not correspond exactly to the true best fit (at
which the pdf is maximized), though the difference is typically
small. Since we are generally interested in the individual parameter
distributions (and the distance modulus in particular), it is more
useful to present the fit results in this form.

Two objects, SN 2000cx and SN 2002cx, cannot be fit by MLCS2k2. Li et
al.~(2001, 2003) discuss the peculiarity of these objects, whose light
curves are distinct from the vast majority of SN Ia. In addition, SN
1998D and SN 1999cw have poor fits, with very broad or multi-peaked
pdfs. Both of these objects have a paucity of data (no more than three
photometric points in any filter), and more importantly, the first
observations for each object occurred well after maximum light ($+$32
and $+$24 days, respectively; see Table \ref{ch4-tab-sninfo}). This
allows for a wide range of models to fit the data well, and the
uncertain results are not very informative, but we present them for
completeness.

Finally, analysis of the distance modulus residuals (see \S
\ref{ch4-sec-hflow} and \ref{ch4-sec-bubble}) leads us to conclude
that the $\mu_0$ (and $m_V^0$) uncertainties listed in Table
\ref{ch4-tab-mlcs} are slightly underestimated. This is likely due to
our less-than-perfect knowledge of the model covariance matrix $S$. In
particular, our analysis shows that the objects with the smallest
distance uncertainties are the ones that show the strongest evidence
for unmodeled variance, suggesting there is a systematic floor to the
distance uncertainty (as opposed to a multiplicative factor to adjust
all the uncertainties). We thus recommend that the $\mu_0$ and $m_V^0$
uncertainties in Table \ref{ch4-tab-mlcs} be increased by adding
$\sigma_{\rm add} =$ 0.08 mag in quadrature to yield a final estimate
of $\sigma(\mu_0)$ and $\sigma(m_V^0)$, and we adopt this augmented
uncertainty for all subsequent results. \footnote{Note that this
number shows only the deficiency of our implementation; in a perfect
version of a model like MLCS2k2, the average distance uncertainties
would match the scatter in the residuals without augmentation. This is
because the intrinsic variation of SN Ia around the model is
encapsulated into the model covariance matrix, just not perfectly in
our case. Other distance fitters that do not include any model
uncertainty often add the standard deviation of the residuals in
quadrature to individual distance errors; this would not be correct
for MLCS2k2.}


\begin{deluxetable}{lccccccc}
\tabletypesize{\scriptsize}
\tablecolumns{8}
\singlespace
\tablewidth{0pt}
\tablecaption{MLCS2k2 Light Curve Fits \label{ch4-tab-mlcs}}
\tablehead{ \colhead{SN Ia} & \colhead{$t_0 - 2400000$} &
 \colhead{$\mu_0 + 5 \log h_{65}$} & \colhead{$\Delta$} &
 \colhead{$A_V^0$} & \colhead{$R_V$} & \colhead{$m_V^0$} &
 \colhead{Notes} \\ & \colhead{HJD} & \colhead{mag} & &
 \colhead{mag} & & \colhead{mag} & }
\startdata
1972E  &  41446.64 $\pm$ 0.96  &  27.829 $\pm$ 0.110  &  $-$0.145 $\pm$ 0.062  &  0.068 $\pm$ 0.059  &        3.1        &   8.324 $\pm$ 0.110 & \\
1980N  &  44585.60 $\pm$ 0.63  &  31.602 $\pm$ 0.099  &  $+$0.032 $\pm$ 0.060  &  0.282 $\pm$ 0.082  &        3.1        &  12.098 $\pm$ 0.099 & \\
1981B  &  44670.95 $\pm$ 0.73  &  31.104 $\pm$ 0.140  &  $-$0.081 $\pm$ 0.060  &  0.364 $\pm$ 0.121  &        3.1        &  11.599 $\pm$ 0.140 & \\
1981D  &  44680.21 $\pm$ 0.53  &  31.009 $\pm$ 0.228  &  $+$0.257 $\pm$ 0.204  &  0.631 $\pm$ 0.270  &  3.07 $\pm$ 0.37  &  11.505 $\pm$ 0.228 & \\
1986G  &  46561.36 $\pm$ 0.24  &  27.460 $\pm$ 0.203  &  $+$1.054 $\pm$ 0.065  &  2.227 $\pm$ 0.224  &  2.87 $\pm$ 0.27  &   7.956 $\pm$ 0.203 & \\
1989B  &  47564.32 $\pm$ 0.59  &  30.040 $\pm$ 0.144  &  $+$0.035 $\pm$ 0.087  &  1.330 $\pm$ 0.144  &  2.86 $\pm$ 0.29  &  10.536 $\pm$ 0.144 & \\
1990N  &  48081.90 $\pm$ 0.17  &  32.153 $\pm$ 0.090  &  $-$0.242 $\pm$ 0.046  &  0.174 $\pm$ 0.084  &        3.1        &  12.648 $\pm$ 0.090 & \\
1990O  &  48076.12 $\pm$ 1.02  &  35.814 $\pm$ 0.095  &  $-$0.150 $\pm$ 0.066  &  0.071 $\pm$ 0.063  &        3.1        &  16.310 $\pm$ 0.095 & HF \\
1990T  &  48083.06 $\pm$ 1.16  &  36.555 $\pm$ 0.169  &  $-$0.111 $\pm$ 0.095  &  0.200 $\pm$ 0.120  &        3.1        &  17.050 $\pm$ 0.169 & HF \\
1990Y  &  48115.94 $\pm$ 1.28  &  36.221 $\pm$ 0.248  &  $+$0.036 $\pm$ 0.161  &  0.497 $\pm$ 0.150  &  3.03 $\pm$ 0.36  &  16.717 $\pm$ 0.248 & HF \\
1990af &  48195.86 $\pm$ 0.44  &  36.714 $\pm$ 0.160  &  $+$0.519 $\pm$ 0.113  &  0.144 $\pm$ 0.134  &        3.1        &  17.209 $\pm$ 0.160 & HF \\
1991M  &  48334.98 $\pm$ 1.41  &  33.516 $\pm$ 0.179  &  $+$0.191 $\pm$ 0.112  &  0.269 $\pm$ 0.174  &        3.1        &  14.012 $\pm$ 0.179 & \\
1991S  &  48348.17 $\pm$ 1.18  &  37.274 $\pm$ 0.131  &  $-$0.147 $\pm$ 0.074  &  0.129 $\pm$ 0.101  &        3.1        &  17.770 $\pm$ 0.131 & HF \\
1991T  &  48374.03 $\pm$ 0.14  &  30.776 $\pm$ 0.080  &  $-$0.220 $\pm$ 0.031  &  0.340 $\pm$ 0.070  &        3.1        &  11.271 $\pm$ 0.080 & \\
1991U  &  48356.18 $\pm$ 1.38  &  35.612 $\pm$ 0.167  &  $-$0.096 $\pm$ 0.086  &  0.379 $\pm$ 0.187  &        3.1        &  16.108 $\pm$ 0.167 & HF \\
1991ag &  48414.23 $\pm$ 1.45  &  34.007 $\pm$ 0.097  &  $-$0.134 $\pm$ 0.061  &  0.060 $\pm$ 0.056  &        3.1        &  14.503 $\pm$ 0.097 & HF \\
1991bg &  48603.12 $\pm$ 0.31  &  31.418 $\pm$ 0.101  &  $+$1.404 $\pm$ 0.047  &  0.611 $\pm$ 0.123  &  3.21 $\pm$ 0.41  &  11.913 $\pm$ 0.101 & \\
1992A  &  48640.63 $\pm$ 0.19  &  31.654 $\pm$ 0.067  &  $+$0.447 $\pm$ 0.054  &  0.038 $\pm$ 0.032  &        3.1        &  12.150 $\pm$ 0.067 & \\
1992G  &  48669.96 $\pm$ 1.06  &  32.584 $\pm$ 0.141  &  $-$0.198 $\pm$ 0.047  &  0.499 $\pm$ 0.136  &  3.07 $\pm$ 0.37  &  13.079 $\pm$ 0.141 & \\
1992J  &  48671.64 $\pm$ 1.60  &  36.453 $\pm$ 0.265  &  $+$0.445 $\pm$ 0.236  &  0.259 $\pm$ 0.185  &        3.1        &  16.949 $\pm$ 0.265 & HF \\
1992K  &  48674.37 $\pm$ 1.76  &  33.207 $\pm$ 0.153  &  $+$1.378 $\pm$ 0.116  &  0.223 $\pm$ 0.144  &        3.1        &  13.702 $\pm$ 0.153 & HF \\
1992P  &  48719.20 $\pm$ 0.86  &  35.585 $\pm$ 0.124  &  $-$0.142 $\pm$ 0.082  &  0.136 $\pm$ 0.092  &        3.1        &  16.081 $\pm$ 0.124 & HF \\
1992ae &  48803.18 $\pm$ 1.34  &  37.731 $\pm$ 0.167  &  $+$0.051 $\pm$ 0.102  &  0.163 $\pm$ 0.141  &        3.1        &  18.227 $\pm$ 0.167 & HF \\
1992ag &  48806.95 $\pm$ 0.81  &  35.247 $\pm$ 0.151  &  $+$0.027 $\pm$ 0.077  &  0.424 $\pm$ 0.134  &        3.1        &  15.743 $\pm$ 0.151 & HF \\
1992al &  48837.86 $\pm$ 0.49  &  34.125 $\pm$ 0.065  &  $-$0.144 $\pm$ 0.035  &  0.045 $\pm$ 0.038  &        3.1        &  14.621 $\pm$ 0.065 & HF \\
1992aq &  48833.33 $\pm$ 1.18  &  38.651 $\pm$ 0.117  &  $+$0.193 $\pm$ 0.116  &  0.065 $\pm$ 0.053  &        3.1        &  19.147 $\pm$ 0.117 & HF \\
1992au &  48830.62 $\pm$ 1.55  &  37.379 $\pm$ 0.216  &  $+$0.292 $\pm$ 0.196  &  0.115 $\pm$ 0.097  &        3.1        &  17.875 $\pm$ 0.216 & HF \\
1992bc &  48912.06 $\pm$ 0.17  &  34.859 $\pm$ 0.042  &  $-$0.236 $\pm$ 0.030  &  0.017 $\pm$ 0.014  &        3.1        &  15.354 $\pm$ 0.042 & HF \\
1992bg &  48915.39 $\pm$ 1.23  &  36.126 $\pm$ 0.130  &  $+$0.008 $\pm$ 0.076  &  0.130 $\pm$ 0.091  &        3.1        &  16.622 $\pm$ 0.130 & HF \\
1992bh &  48920.43 $\pm$ 0.97  &  36.899 $\pm$ 0.133  &  $-$0.051 $\pm$ 0.085  &  0.204 $\pm$ 0.114  &        3.1        &  17.395 $\pm$ 0.133 & HF \\
1992bk &  48938.95 $\pm$ 1.59  &  37.184 $\pm$ 0.170  &  $+$0.414 $\pm$ 0.157  &  0.097 $\pm$ 0.082  &        3.1        &  17.679 $\pm$ 0.170 & HF \\
1992bl &  48946.33 $\pm$ 1.14  &  36.497 $\pm$ 0.123  &  $+$0.353 $\pm$ 0.117  &  0.061 $\pm$ 0.060  &        3.1        &  16.993 $\pm$ 0.123 & HF \\
1992bo &  48986.26 $\pm$ 0.16  &  34.742 $\pm$ 0.091  &  $+$0.595 $\pm$ 0.072  &  0.051 $\pm$ 0.049  &        3.1        &  15.237 $\pm$ 0.091 & HF \\
1992bp &  48980.41 $\pm$ 0.74  &  37.793 $\pm$ 0.106  &  $+$0.095 $\pm$ 0.092  &  0.060 $\pm$ 0.056  &        3.1        &  18.289 $\pm$ 0.106 & HF \\
1992br &  48984.63 $\pm$ 1.34  &  37.977 $\pm$ 0.185  &  $+$0.888 $\pm$ 0.173  &  0.125 $\pm$ 0.115  &        3.1        &  18.472 $\pm$ 0.185 & HF \\
1992bs &  48985.05 $\pm$ 1.30  &  37.636 $\pm$ 0.161  &  $-$0.031 $\pm$ 0.089  &  0.172 $\pm$ 0.137  &        3.1        &  18.132 $\pm$ 0.161 & HF \\
1993B  &  49003.99 $\pm$ 1.35  &  37.779 $\pm$ 0.148  &  $-$0.115 $\pm$ 0.076  &  0.198 $\pm$ 0.119  &        3.1        &  18.274 $\pm$ 0.148 & HF \\
1993H  &  49068.92 $\pm$ 0.46  &  35.109 $\pm$ 0.095  &  $+$0.904 $\pm$ 0.076  &  0.139 $\pm$ 0.086  &        3.1        &  15.605 $\pm$ 0.095 & HF \\
1993L  &  49095.40 $\pm$ 1.51  &  31.986 $\pm$ 0.305  &  $+$0.041 $\pm$ 0.198  &  0.722 $\pm$ 0.172  &  3.04 $\pm$ 0.36  &  12.482 $\pm$ 0.305 & \\
1993O  &  49134.40 $\pm$ 0.43  &  37.140 $\pm$ 0.095  &  $+$0.025 $\pm$ 0.070  &  0.077 $\pm$ 0.059  &        3.1        &  17.636 $\pm$ 0.095 & HF \\
1993ac &  49269.70 $\pm$ 1.19  &  36.881 $\pm$ 0.189  &  $-$0.054 $\pm$ 0.108  &  0.377 $\pm$ 0.191  &        3.1        &  17.377 $\pm$ 0.189 & HF \\
1993ae &  49288.58 $\pm$ 1.19  &  34.468 $\pm$ 0.187  &  $+$0.401 $\pm$ 0.157  &  0.055 $\pm$ 0.055  &        3.1        &  14.963 $\pm$ 0.187 & HF \\
1993ag &  49316.67 $\pm$ 0.68  &  37.041 $\pm$ 0.131  &  $+$0.119 $\pm$ 0.089  &  0.166 $\pm$ 0.097  &        3.1        &  17.537 $\pm$ 0.131 & HF \\
1993ah &  49302.52 $\pm$ 1.33  &  35.640 $\pm$ 0.200  &  $+$0.147 $\pm$ 0.164  &  0.136 $\pm$ 0.121  &        3.1        &  16.135 $\pm$ 0.200 & HF \\
1994D  &  49432.47 $\pm$ 0.10  &  31.187 $\pm$ 0.067  &  $+$0.142 $\pm$ 0.035  &  0.109 $\pm$ 0.047  &        3.1        &  11.682 $\pm$ 0.067 & \\
1994M  &  49473.61 $\pm$ 0.90  &  35.244 $\pm$ 0.121  &  $+$0.261 $\pm$ 0.079  &  0.226 $\pm$ 0.115  &        3.1        &  15.739 $\pm$ 0.121 & HF \\
1994Q  &  49496.72 $\pm$ 1.09  &  35.759 $\pm$ 0.141  &  $-$0.140 $\pm$ 0.075  &  0.192 $\pm$ 0.121  &        3.1        &  16.255 $\pm$ 0.141 & HF \\
1994S  &  49518.28 $\pm$ 0.50  &  34.356 $\pm$ 0.077  &  $-$0.084 $\pm$ 0.074  &  0.054 $\pm$ 0.045  &        3.1        &  14.852 $\pm$ 0.077 & HF \\
1994T  &  49514.54 $\pm$ 0.52  &  36.012 $\pm$ 0.112  &  $+$0.731 $\pm$ 0.103  &  0.093 $\pm$ 0.075  &        3.1        &  16.507 $\pm$ 0.112 & HF \\
1994ae &  49684.65 $\pm$ 0.15  &  32.563 $\pm$ 0.065  &  $-$0.191 $\pm$ 0.036  &  0.070 $\pm$ 0.047  &        3.1        &  13.058 $\pm$ 0.065 & \\
1995D  &  49768.60 $\pm$ 0.44  &  32.813 $\pm$ 0.074  &  $-$0.186 $\pm$ 0.041  &  0.081 $\pm$ 0.060  &        3.1        &  13.308 $\pm$ 0.074 & \\
1995E  &  49774.67 $\pm$ 0.54  &  33.198 $\pm$ 0.178  &  $+$0.006 $\pm$ 0.057  &  2.241 $\pm$ 0.176  &  2.85 $\pm$ 0.28  &  13.694 $\pm$ 0.178 & \\
1995ac &  49992.99 $\pm$ 0.40  &  36.511 $\pm$ 0.108  &  $-$0.231 $\pm$ 0.047  &  0.252 $\pm$ 0.107  &        3.1        &  17.007 $\pm$ 0.108 & HF \\
1995ak &  50021.16 $\pm$ 0.93  &  34.741 $\pm$ 0.138  &  $+$0.122 $\pm$ 0.065  &  0.613 $\pm$ 0.131  &  3.12 $\pm$ 0.39  &  15.237 $\pm$ 0.138 & HF \\
1995al &  50028.96 $\pm$ 0.44  &  32.714 $\pm$ 0.081  &  $-$0.282 $\pm$ 0.035  &  0.177 $\pm$ 0.065  &        3.1        &  13.209 $\pm$ 0.081 & \\
1995bd &  50086.31 $\pm$ 0.24  &  34.006 $\pm$ 0.125  &  $-$0.239 $\pm$ 0.051  &  0.523 $\pm$ 0.241  &  3.01 $\pm$ 0.35  &  14.502 $\pm$ 0.125 & HF \\
1996C  &  50128.42 $\pm$ 0.90  &  35.947 $\pm$ 0.104  &  $-$0.104 $\pm$ 0.055  &  0.136 $\pm$ 0.091  &        3.1        &  16.443 $\pm$ 0.104 & HF \\
1996X  &  50190.85 $\pm$ 0.33  &  32.432 $\pm$ 0.065  &  $+$0.095 $\pm$ 0.048  &  0.061 $\pm$ 0.044  &        3.1        &  12.927 $\pm$ 0.065 & \\
1996Z  &  50215.24 $\pm$ 1.46  &  32.760 $\pm$ 0.238  &  $+$0.112 $\pm$ 0.210  &  0.609 $\pm$ 0.316  &  3.05 $\pm$ 0.36  &  13.256 $\pm$ 0.238 & HF \\
1996ab &  50224.69 $\pm$ 1.07  &  38.934 $\pm$ 0.146  &  $+$0.164 $\pm$ 0.127  &  0.098 $\pm$ 0.091  &        3.1        &  19.430 $\pm$ 0.146 & HF \\
1996ai &  50255.21 $\pm$ 0.81  &  31.151 $\pm$ 0.187  &  $-$0.098 $\pm$ 0.068  &  3.662 $\pm$ 0.185  &  2.09 $\pm$ 0.12  &  11.647 $\pm$ 0.187 & \\
1996bk &  50368.46 $\pm$ 1.02  &  32.195 $\pm$ 0.142  &  $+$1.011 $\pm$ 0.124  &  0.711 $\pm$ 0.184  &  3.10 $\pm$ 0.38  &  12.691 $\pm$ 0.142 & \\
1996bl &  50376.32 $\pm$ 0.58  &  36.096 $\pm$ 0.113  &  $-$0.116 $\pm$ 0.064  &  0.208 $\pm$ 0.115  &        3.1        &  16.592 $\pm$ 0.113 & HF \\
1996bo &  50386.93 $\pm$ 0.30  &  34.043 $\pm$ 0.138  &  $+$0.057 $\pm$ 0.070  &  0.938 $\pm$ 0.138  &  3.02 $\pm$ 0.35  &  14.538 $\pm$ 0.138 & HF \\
1996bv &  50403.72 $\pm$ 1.27  &  34.213 $\pm$ 0.157  &  $-$0.206 $\pm$ 0.065  &  0.547 $\pm$ 0.159  &  3.07 $\pm$ 0.37  &  14.709 $\pm$ 0.157 & HF \\
1997E  &  50467.58 $\pm$ 0.41  &  34.105 $\pm$ 0.091  &  $+$0.341 $\pm$ 0.084  &  0.213 $\pm$ 0.104  &        3.1        &  14.600 $\pm$ 0.091 & HF \\
1997Y  &  50486.88 $\pm$ 1.37  &  34.563 $\pm$ 0.104  &  $+$0.060 $\pm$ 0.078  &  0.196 $\pm$ 0.087  &        3.1        &  15.058 $\pm$ 0.104 & HF \\
1997bp &  50548.94 $\pm$ 0.52  &  32.895 $\pm$ 0.092  &  $-$0.180 $\pm$ 0.049  &  0.537 $\pm$ 0.086  &  2.88 $\pm$ 0.31  &  13.391 $\pm$ 0.092 & HF \\
1997bq &  50557.99 $\pm$ 0.18  &  33.458 $\pm$ 0.109  &  $-$0.095 $\pm$ 0.063  &  0.513 $\pm$ 0.095  &  3.03 $\pm$ 0.35  &  13.953 $\pm$ 0.109 & HF \\
1997br &  50559.26 $\pm$ 0.23  &  32.248 $\pm$ 0.104  &  $-$0.303 $\pm$ 0.031  &  0.804 $\pm$ 0.112  &  3.03 $\pm$ 0.35  &  12.743 $\pm$ 0.104 & \\
1997cn &  50586.64 $\pm$ 0.77  &  34.532 $\pm$ 0.070  &  $+$1.381 $\pm$ 0.046  &  0.071 $\pm$ 0.059  &        3.1        &  15.027 $\pm$ 0.070 & HF \\
1997cw &  50630.75 $\pm$ 0.98  &  34.075 $\pm$ 0.151  &  $-$0.179 $\pm$ 0.064  &  1.092 $\pm$ 0.145  &  2.97 $\pm$ 0.33  &  14.570 $\pm$ 0.151 & HF \\
1997dg &  50720.05 $\pm$ 0.84  &  36.149 $\pm$ 0.114  &  $-$0.006 $\pm$ 0.076  &  0.201 $\pm$ 0.102  &        3.1        &  16.645 $\pm$ 0.114 & HF \\
1997do &  50766.21 $\pm$ 0.45  &  33.601 $\pm$ 0.118  &  $-$0.146 $\pm$ 0.080  &  0.312 $\pm$ 0.102  &        3.1        &  14.097 $\pm$ 0.118 & HF \\
1997dt &  50785.60 $\pm$ 0.31  &  32.723 $\pm$ 0.191  &  $-$0.126 $\pm$ 0.070  &  1.849 $\pm$ 0.198  &  3.03 $\pm$ 0.34  &  13.219 $\pm$ 0.191 & \\
1998D  &  50841.07 $\pm$ 2.02  &  34.089 $\pm$ 0.627  &  $+$0.571 $\pm$ 0.482  &  0.346 $\pm$ 0.186  &        3.1        &  14.585 $\pm$ 0.627 & poor fit \\
1998V  &  50891.27 $\pm$ 0.84  &  34.395 $\pm$ 0.102  &  $-$0.055 $\pm$ 0.064  &  0.209 $\pm$ 0.115  &        3.1        &  14.891 $\pm$ 0.102 & HF \\
1998ab &  50914.43 $\pm$ 0.25  &  35.213 $\pm$ 0.100  &  $-$0.118 $\pm$ 0.051  &  0.394 $\pm$ 0.082  &        3.1        &  15.709 $\pm$ 0.100 & HF \\
1998aq &  50930.80 $\pm$ 0.13  &  31.974 $\pm$ 0.046  &  $-$0.063 $\pm$ 0.036  &  0.024 $\pm$ 0.019  &        3.1        &  12.470 $\pm$ 0.046 & \\
1998bp &  50936.39 $\pm$ 0.33  &  33.304 $\pm$ 0.075  &  $+$1.114 $\pm$ 0.057  &  0.188 $\pm$ 0.100  &        3.1        &  13.800 $\pm$ 0.075 & HF \\
1998bu &  50952.40 $\pm$ 0.23  &  30.283 $\pm$ 0.117  &  $-$0.015 $\pm$ 0.038  &  1.055 $\pm$ 0.114  &  3.13 $\pm$ 0.36  &  10.778 $\pm$ 0.117 & \\
1998co &  50987.76 $\pm$ 1.32  &  34.431 $\pm$ 0.126  &  $+$0.464 $\pm$ 0.196  &  0.301 $\pm$ 0.183  &        3.1        &  14.926 $\pm$ 0.126 & HF \\
1998de &  51026.69 $\pm$ 0.17  &  34.400 $\pm$ 0.082  &  $+$1.448 $\pm$ 0.036  &  0.398 $\pm$ 0.101  &        3.1        &  14.896 $\pm$ 0.082 & HF \\
1998dh &  51029.83 $\pm$ 0.22  &  32.846 $\pm$ 0.087  &  $-$0.051 $\pm$ 0.042  &  0.471 $\pm$ 0.061  &  2.76 $\pm$ 0.27  &  13.342 $\pm$ 0.087 & \\
1998dk &  51056.58 $\pm$ 1.48  &  33.802 $\pm$ 0.219  &  $-$0.127 $\pm$ 0.134  &  0.508 $\pm$ 0.152  &  3.10 $\pm$ 0.38  &  14.297 $\pm$ 0.219 & HF \\
1998dm &  51062.03 $\pm$ 1.16  &  33.067 $\pm$ 0.178  &  $-$0.192 $\pm$ 0.073  &  1.045 $\pm$ 0.160  &  3.08 $\pm$ 0.37  &  13.563 $\pm$ 0.178 & \\
1998dx &  51071.32 $\pm$ 0.80  &  36.917 $\pm$ 0.099  &  $+$0.190 $\pm$ 0.099  &  0.086 $\pm$ 0.065  &        3.1        &  17.413 $\pm$ 0.099 & HF \\
1998ec &  51088.41 $\pm$ 1.07  &  35.082 $\pm$ 0.161  &  $-$0.169 $\pm$ 0.077  &  0.569 $\pm$ 0.125  &  3.01 $\pm$ 0.35  &  15.578 $\pm$ 0.161 & HF \\
1998ef &  51113.99 $\pm$ 0.22  &  34.142 $\pm$ 0.105  &  $+$0.269 $\pm$ 0.083  &  0.046 $\pm$ 0.044  &        3.1        &  14.638 $\pm$ 0.105 & HF \\
1998eg &  51110.69 $\pm$ 1.23  &  35.353 $\pm$ 0.120  &  $+$0.014 $\pm$ 0.118  &  0.224 $\pm$ 0.118  &        3.1        &  15.849 $\pm$ 0.120 & HF \\
1998es &  51141.89 $\pm$ 0.15  &  33.263 $\pm$ 0.078  &  $-$0.287 $\pm$ 0.030  &  0.227 $\pm$ 0.073  &        3.1        &  13.758 $\pm$ 0.078 & HF \\
1999X  &  51203.48 $\pm$ 1.04  &  35.546 $\pm$ 0.132  &  $-$0.181 $\pm$ 0.075  &  0.139 $\pm$ 0.095  &        3.1        &  16.041 $\pm$ 0.132 & HF \\
1999aa &  51231.97 $\pm$ 0.15  &  34.468 $\pm$ 0.042  &  $-$0.271 $\pm$ 0.028  &  0.020 $\pm$ 0.018  &        3.1        &  14.963 $\pm$ 0.042 & HF \\
1999ac &  51250.60 $\pm$ 0.17  &  33.334 $\pm$ 0.081  &  $-$0.086 $\pm$ 0.039  &  0.308 $\pm$ 0.072  &        3.1        &  13.830 $\pm$ 0.081 & HF \\
1999aw &  51253.96 $\pm$ 0.29  &  36.551 $\pm$ 0.043  &  $-$0.369 $\pm$ 0.032  &  0.019 $\pm$ 0.018  &        3.1        &  17.047 $\pm$ 0.043 & HF \\
1999by &  51309.50 $\pm$ 0.14  &  31.158 $\pm$ 0.073  &  $+$1.348 $\pm$ 0.030  &  0.170 $\pm$ 0.083  &        3.1        &  11.653 $\pm$ 0.073 & \\
1999cc &  51315.62 $\pm$ 0.46  &  35.859 $\pm$ 0.092  &  $+$0.337 $\pm$ 0.097  &  0.148 $\pm$ 0.088  &        3.1        &  16.355 $\pm$ 0.092 & HF \\
1999cl &  51342.28 $\pm$ 0.26  &  30.604 $\pm$ 0.163  &  $+$0.031 $\pm$ 0.087  &  2.666 $\pm$ 0.160  &  2.22 $\pm$ 0.15  &  11.099 $\pm$ 0.163 & \\
1999cp &  51363.22 $\pm$ 0.26  &  33.520 $\pm$ 0.098  &  $-$0.103 $\pm$ 0.102  &  0.072 $\pm$ 0.062  &        3.1        &  14.015 $\pm$ 0.098 & HF \\
1999cw &  51352.49 $\pm$ 1.47  &  33.363 $\pm$ 0.141  &  $-$0.306 $\pm$ 0.067  &  0.161 $\pm$ 0.073  &        3.1        &  13.859 $\pm$ 0.141 & poor fit \\
1999da &  51370.57 $\pm$ 0.17  &  33.905 $\pm$ 0.101  &  $+$1.451 $\pm$ 0.038  &  0.233 $\pm$ 0.115  &        3.1        &  14.400 $\pm$ 0.101 & HF \\
1999dk &  51413.92 $\pm$ 0.78  &  34.273 $\pm$ 0.098  &  $-$0.233 $\pm$ 0.047  &  0.201 $\pm$ 0.094  &        3.1        &  14.769 $\pm$ 0.098 & HF \\
1999dq &  51435.70 $\pm$ 0.15  &  33.668 $\pm$ 0.067  &  $-$0.338 $\pm$ 0.024  &  0.369 $\pm$ 0.077  &        3.1        &  14.163 $\pm$ 0.067 & HF \\
1999ee &  51469.29 $\pm$ 0.14  &  33.472 $\pm$ 0.088  &  $-$0.208 $\pm$ 0.030  &  0.797 $\pm$ 0.085  &  2.76 $\pm$ 0.27  &  13.968 $\pm$ 0.088 & HF \\
1999ef &  51457.78 $\pm$ 1.27  &  36.657 $\pm$ 0.111  &  $-$0.057 $\pm$ 0.077  &  0.051 $\pm$ 0.044  &        3.1        &  17.153 $\pm$ 0.111 & HF \\
1999ej &  51482.37 $\pm$ 0.85  &  34.459 $\pm$ 0.123  &  $+$0.302 $\pm$ 0.088  &  0.166 $\pm$ 0.099  &        3.1        &  14.954 $\pm$ 0.123 & \\
1999ek &  51481.65 $\pm$ 0.50  &  34.360 $\pm$ 0.111  &  $+$0.050 $\pm$ 0.067  &  0.475 $\pm$ 0.251  &        3.1        &  14.856 $\pm$ 0.111 & HF \\
1999gd &  51518.40 $\pm$ 1.14  &  34.606 $\pm$ 0.167  &  $+$0.024 $\pm$ 0.073  &  1.381 $\pm$ 0.156  &  2.99 $\pm$ 0.33  &  15.101 $\pm$ 0.167 & HF \\
1999gh &  51513.53 $\pm$ 0.91  &  32.761 $\pm$ 0.078  &  $+$0.902 $\pm$ 0.081  &  0.084 $\pm$ 0.074  &        3.1        &  13.257 $\pm$ 0.078 & HF \\
1999gp &  51550.30 $\pm$ 0.17  &  35.622 $\pm$ 0.061  &  $-$0.323 $\pm$ 0.030  &  0.076 $\pm$ 0.051  &        3.1        &  16.118 $\pm$ 0.061 & HF \\
2000B  &  51562.85 $\pm$ 1.30  &  34.628 $\pm$ 0.216  &  $+$0.253 $\pm$ 0.190  &  0.286 $\pm$ 0.136  &        3.1        &  15.123 $\pm$ 0.216 & HF \\
2000E  &  51576.80 $\pm$ 0.35  &  31.721 $\pm$ 0.107  &  $-$0.277 $\pm$ 0.039  &  0.609 $\pm$ 0.189  &  3.01 $\pm$ 0.35  &  12.216 $\pm$ 0.107 & \\
2000bh &  51636.31 $\pm$ 1.47  &  35.309 $\pm$ 0.131  &  $-$0.026 $\pm$ 0.067  &  0.138 $\pm$ 0.093  &        3.1        &  15.805 $\pm$ 0.131 & HF \\
2000bk &  51646.02 $\pm$ 1.08  &  35.390 $\pm$ 0.156  &  $+$0.613 $\pm$ 0.129  &  0.290 $\pm$ 0.148  &        3.1        &  15.885 $\pm$ 0.156 & HF \\
2000ca &  51666.12 $\pm$ 0.46  &  35.250 $\pm$ 0.060  &  $-$0.125 $\pm$ 0.048  &  0.030 $\pm$ 0.024  &        3.1        &  15.746 $\pm$ 0.060 & HF \\
2000ce &  51667.45 $\pm$ 1.01  &  34.376 $\pm$ 0.179  &  $-$0.149 $\pm$ 0.060  &  1.761 $\pm$ 0.179  &  2.91 $\pm$ 0.30  &  14.872 $\pm$ 0.179 & HF \\
2000cf &  51672.33 $\pm$ 0.83  &  36.382 $\pm$ 0.104  &  $-$0.006 $\pm$ 0.065  &  0.194 $\pm$ 0.097  &        3.1        &  16.878 $\pm$ 0.104 & HF \\
2000cn &  51707.83 $\pm$ 0.16  &  35.153 $\pm$ 0.086  &  $+$0.716 $\pm$ 0.076  &  0.171 $\pm$ 0.109  &        3.1        &  15.648 $\pm$ 0.086 & HF \\
2000cx &  \nodata & \nodata & \nodata & \nodata & \nodata & \nodata & bad fit \\
2000dk &  51812.47 $\pm$ 0.28  &  34.408 $\pm$ 0.077  &  $+$0.578 $\pm$ 0.067  &  0.033 $\pm$ 0.032  &        3.1        &  14.904 $\pm$ 0.077 & HF \\
2000fa &  51892.25 $\pm$ 0.57  &  35.121 $\pm$ 0.128  &  $-$0.126 $\pm$ 0.061  &  0.283 $\pm$ 0.105  &        3.1        &  15.616 $\pm$ 0.128 & HF \\
2001V  &  51973.28 $\pm$ 0.16  &  34.187 $\pm$ 0.065  &  $-$0.280 $\pm$ 0.032  &  0.092 $\pm$ 0.052  &        3.1        &  14.683 $\pm$ 0.065 & HF \\
2001ay &  52023.96 $\pm$ 0.78  &  35.926 $\pm$ 0.102  &  $-$0.405 $\pm$ 0.045  &  0.374 $\pm$ 0.098  &        3.1        &  16.422 $\pm$ 0.102 & HF \\
2001ba &  52034.18 $\pm$ 0.54  &  35.889 $\pm$ 0.075  &  $-$0.093 $\pm$ 0.056  &  0.041 $\pm$ 0.037  &        3.1        &  16.385 $\pm$ 0.075 & HF \\
2001bt &  52063.91 $\pm$ 0.21  &  33.952 $\pm$ 0.114  &  $+$0.071 $\pm$ 0.052  &  0.616 $\pm$ 0.111  &  3.00 $\pm$ 0.35  &  14.447 $\pm$ 0.114 & HF \\
2001cn &  52071.10 $\pm$ 0.74  &  34.130 $\pm$ 0.106  &  $+$0.007 $\pm$ 0.052  &  0.447 $\pm$ 0.100  &        3.1        &  14.625 $\pm$ 0.106 & HF \\
2001cz &  52103.89 $\pm$ 0.31  &  34.294 $\pm$ 0.111  &  $-$0.078 $\pm$ 0.054  &  0.254 $\pm$ 0.116  &        3.1        &  14.790 $\pm$ 0.111 & HF \\
2001el &  52182.55 $\pm$ 0.20  &  31.544 $\pm$ 0.077  &  $-$0.193 $\pm$ 0.032  &  0.696 $\pm$ 0.059  &  2.40 $\pm$ 0.19  &  12.040 $\pm$ 0.077 & \\
2002bf &  52337.87 $\pm$ 0.50  &  35.480 $\pm$ 0.134  &  $-$0.111 $\pm$ 0.065  &  0.254 $\pm$ 0.131  &        3.1        &  15.975 $\pm$ 0.134 & HF \\
2002bo &  52356.89 $\pm$ 0.14  &  31.945 $\pm$ 0.105  &  $-$0.048 $\pm$ 0.041  &  1.212 $\pm$ 0.099  &  2.57 $\pm$ 0.22  &  12.441 $\pm$ 0.105 & \\
2002cx &  \nodata & \nodata & \nodata & \nodata & \nodata & \nodata & bad fit \\
2002er &  52524.84 $\pm$ 0.15  &  33.034 $\pm$ 0.085  &  $+$0.170 $\pm$ 0.057  &  0.470 $\pm$ 0.112  &        3.1        &  13.530 $\pm$ 0.085 & HF \\
2003du &  52765.62 $\pm$ 0.50  &  33.189 $\pm$ 0.049  &  $-$0.247 $\pm$ 0.028  &  0.037 $\pm$ 0.026  &        3.1        &  13.685 $\pm$ 0.049 & \\
\enddata
\tablecomments{Values listed are means and standard deviations for
each of the parameters, as determined from their one-dimensional
probability distributions marginalized over the other parameters. We
use the standard notation $h_{65} \equiv H_0/(65 \; \kmsmpc)$. The
extinction parameters $A_V^0$ and $R_V$ correspond to host-galaxy
extinction only; Galactic reddening is listed in Table
\ref{ch4-tab-sninfo}. The corrected apparent peak magnitude $m_V^0$
differs from the distance modulus $\mu_0$ by an additive constant,
$m_V^0 \equiv \mu_0 + M_V^0$.}
\end{deluxetable}


\section{Discussion}

\subsection{Hubble Flow Sample \label{ch4-sec-hflow}}

Constructing a Hubble diagram of nearby SN Ia (Hubble 1929; Kirshner
2004) requires a Hubble flow sample, for which the measured recession
velocities are dominated by the cosmological redshift (as opposed to
peculiar motions).  We consider objects with $cz \geq 2500 \; \kms$ in
the CMB rest-frame to be in the Hubble flow, yielding 101 objects out
of our original 133. Additionally, we require good distances for these
SN Ia, eliminating SN 2002cx which could not be fit by
MLCS2k2. Similarly, we further cull the Hubble Flow sample to exclude
objects whose first observation occurs more than 20 days past maximum
light ($t_1 > 20$ d; see Table \ref{ch4-tab-sninfo}), eliminating two
SN Ia (SN 1998D and SN 1999cw). Excessive host-galaxy extinction also
leads to uncertain distances and we exclude objects with mean $A_V^0
>$ 2.0 mag (another two objects: SN 1995E and SN 1999cl). We explore
the consequences of this relatively permissive extinction cut below in
\S \ref{ch4-sec-rv}.

We constructed a Hubble diagram of the 96 SN Ia that passed these
cuts, and noticed a significant outlier: SN 1999ej in NGC 495
(Friedman, King, \& Li 1999; $cz_{\rm CMB} = 3831 \; \kms$) gave a
residual of $\sim$0.6 mag relative to the tight locus defined by the
other objects. Spectroscopy of SN 1999ej showed it to be a normal SN
Ia (Jha et al.~1999b) and the photometric data (Jha et al.~2006),
though not plentiful with only 4 fit points each in \ubvri,
nonetheless provide a good MLCS2k2 distance with no indication of
peculiarity. The solution to this puzzle is that NGC 495 is in a
particularly dense region of the group/poor cluster Zw 0107+3212
(whose brightest cluster galaxy is the nearby NGC 507). The cluster
has a mean redshift $c\bar{z}_{\rm CMB} \simeq 4690 \; \kms$ and
estimated velocity dispersion $\sigma$ ranging from 440--590 $\kms$
(Ramella et al.~2002; Miller et al.~2002). Using this cluster mean
redshift instead for SN 1999ej makes it fully consistent with the rest
of the sample, suggesting the observed recession velocity of NGC 495
is strongly affected by its environment, but we have chosen simply to
exclude SN 1999ej from the Hubble flow sample. From a NED search, we
did not find any independent evidence for large peculiar velocities
among the other Hubble flow objects.

\begin{figure}
\includegraphics[width=6in]{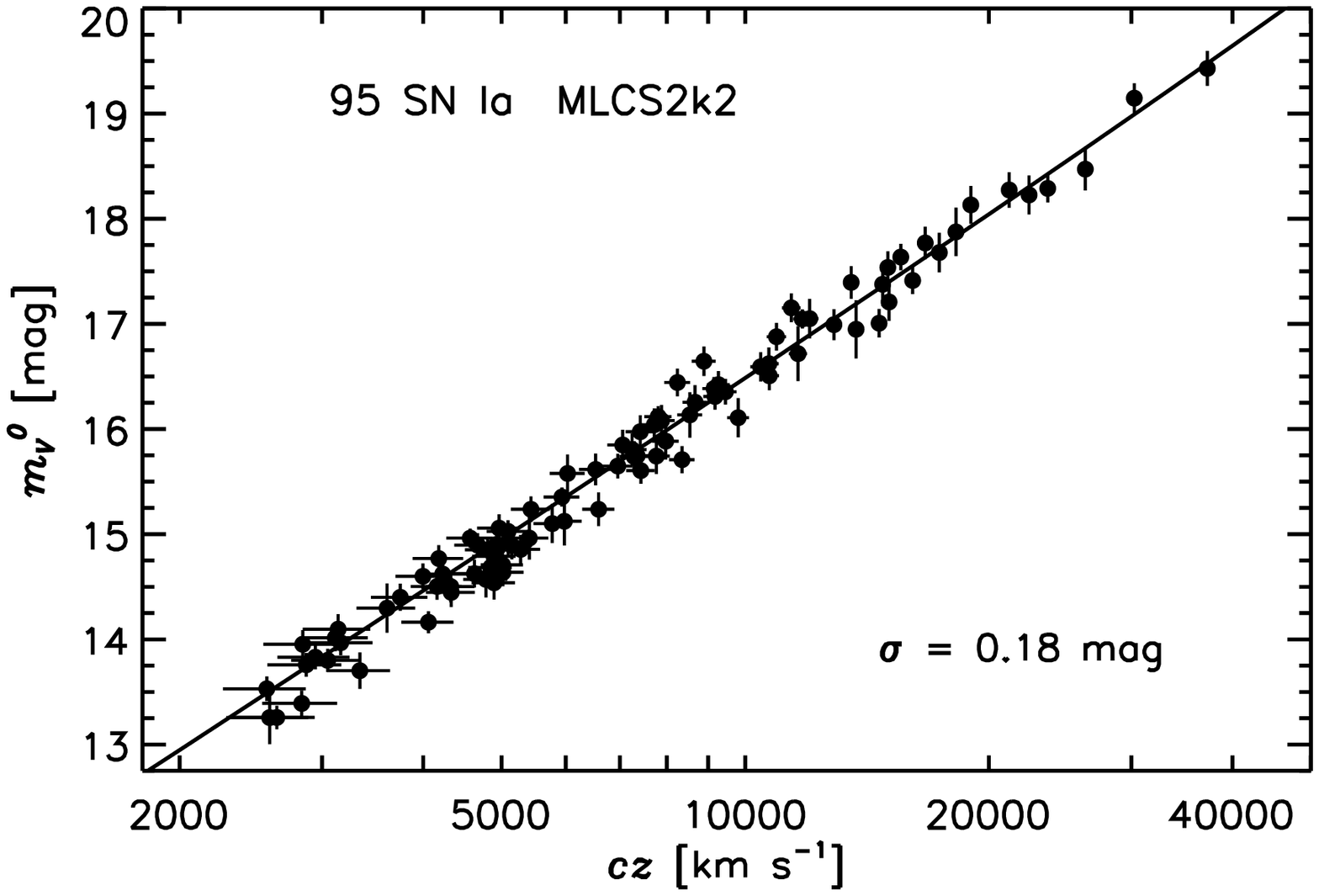}
\caption[Diagram Hubble]{\singlespace Apparent magnitude-redshift
relation for our Hubble flow sample of 95 SN Ia, using MLCS2k2 to
correct for host-galaxy extinction and intrinsic luminosity
differences, and redshifts in the CMB frame. The shape of the
solid-line is fixed by the inverse-square law and (for high $cz$) the
adopted $\Omega_M = 0.3$, $\Omega_\Lambda = 0.7$ cosmology; the
intercept is the only free parameter and it is determined to
$\pm$0.018 mag. \label{ch4-fig-hdiag}}
\end{figure}

\begin{figure}
\includegraphics[width=6in]{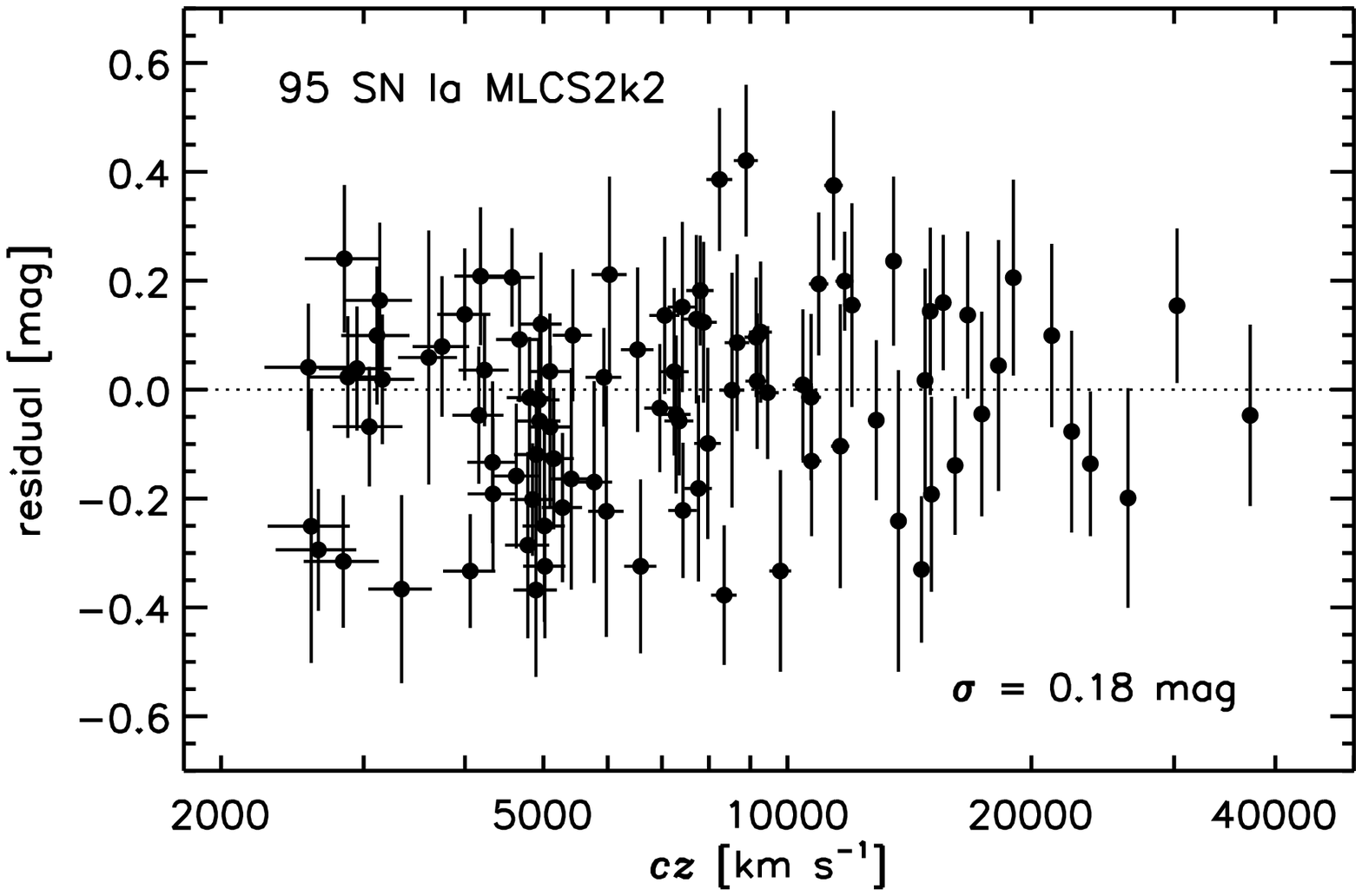}
\caption[Residuals]{\singlespace Magnitude residuals of the Hubble
flow sample after MLCS2k2 correction, relative to the best-fit Hubble
line shown in Figure \ref{ch4-fig-hdiag}. \label{ch4-fig-resid}}
\end{figure}

Our final Hubble flow sample is the largest compiled to date for SN Ia
in the nearby Universe with homogeneous distances.  It consists of 95
SN Ia (denoted HF in Table \ref{ch4-tab-mlcs}), including 35 out of 44
of the objects presented by Jha et al.~(2006), and more than doubles
the MLCS Hubble flow sample of Jha et al.~(1999a).

The Hubble diagram for this sample is shown in Figure
\ref{ch4-fig-hdiag}, where we have fit for only one free parameter,
the intercept.\footnote{We have adopted an $\Omega_{\rm M} = 0.3$,
$\Omega_\Lambda = 0.7$ cosmology, which is significant for the more
distant objects, and assumed a peculiar velocity uncertainty of $\pm
300 \; \kms$, important for the nearby objects. Including in
quadrature the error floor (see \S \ref{ch4-sec-app}) of $\sigma_{\rm
add} =$ 0.08 mag yields a reduced $\chi^2$ of 1.06 for 94 degrees of
freedom. This combination of the peculiar velocity uncertainty and
$\sigma_{\rm add}$ results in a good fit across the whole Hubble flow
sample as well as selected subsamples.} This intercept is
scale-dependent, but can be expressed several different ways. One
relates it to the absolute magnitude of the fiducial SN Ia, and
provides the derivation for the zeropoint in equation
\ref{ch4-eqn-absvmax} (which in turn defines the relation between
$\mu_0$ and $m_V^0$):
\begin{equation}
M_V^0 - 5 \log H_0 = m_V^0 - 25 - 5 \log \left( c\left(1+z\right) 
\int_{0}^{z} \left[ \Omega_{\rm M} \left(1+z'\right)^3 +
\Omega_\Lambda \right]^{-1/2} dz' \right),
\end{equation}
valid for a flat, $\Omega_{\rm M} + \Omega_\Lambda = 1$ Universe. The
weighted fit to the full Hubble flow sample gives $M_V^0 - 5 \log
h_{65} = -19.504 \, \pm \, 0.018$ mag (statistical uncertainty
only). Alternately we can follow the notation of Jha et al.~(1999a) and
calculate the ``intercept of the ridge line,'' $a_V$,
defined as\footnote{In Jha et al.~(1999a) the intercept of the ridge line
was given as $a_V = \log cz - 0.2 m_V^0$. Here we explicitly show the
dependence on cosmological parameters.} 
\begin{equation}
a_V = \log \left( c\left(1+z\right) \int_{0}^{z} \left[
  \Omega_{\rm M} \left(1+z'\right)^3 + \Omega_\Lambda \right]^{-1/2} dz'
\right) - 0.2 m_V^0,
\end{equation}
again assuming an $\Omega_{\rm M} + \Omega_\Lambda = 1$ Universe.  The
full Hubble flow sample gives a best-fit value and formal uncertainty
of $a_V = 0.7139 \, \pm \, 0.0037$. Unfortunately we cannot directly
compare these intermediate results with the values given by Jha et
al.~(1999a), because MLCS2k2 uses new template vectors with new
magnitude and $\Delta$ zeropoints, which uniformly shifts all
distances, $a_V$, and $M^0_V$.

The scatter about the Hubble line in Figures \ref{ch4-fig-hdiag} and
\ref{ch4-fig-resid} is $\sigma = 0.18$ mag ($\sim$8\% in
distance). Part of the scatter in the Hubble diagram comes from our
inexact knowledge of cosmological redshifts because of galactic
peculiar velocities. Our estimate of the peculiar velocity uncertainty
of $\pm$ 300 $\kms$ corresponds to an uncertainty of 0.26 mag for the
nearest objects in the Hubble flow sample ($cz \simeq 2500 \; \kms$),
which drops to under 0.02 mag for the most distant object ($cz = 37239
\; \kms$). The mean peculiar velocity uncertainty for the full sample
is 0.11 mag\footnote{Because of the inverse dependence on the
redshift, this is not the same as the peculiar velocity uncertainty at
the mean redshift, $c\bar{z}_{\rm CMB} = 8796 \; \kms$, where $\pm 300
\; \kms$ corresponds to $\pm$0.07 mag.}, and subtracting this in
quadrature from the overall dispersion implies that the
\emph{intrinsic} dispersion in SN Ia distances is at most 0.14 mag
($\lesssim$ 7\% in distance) for samples similar to the one presented
here.

Our Hubble flow sample is the largest ever fit self-consistently with
one technique, and comprises data from diverse sources. In addition,
we have attempted to limit the number of objects rejected from the
sample. We believe the measured dispersion is more reflective of
the true dispersion among SN Ia as they are currently being
observed. Further restricting samples to the best objects (e.g., those
with low reddening or close to fiducial luminosity/light-curve shape)
clearly provides an avenue to reducing the dispersion, but it is
important that these restrictions be carefully defined to avoid
biasing applications of these samples to measuring cosmological
parameters such as $H_0$, $\Omega_\Lambda$, or $w$.  For example, Guy
et al.~(2005) show that the quoted dispersion of 0.08 mag and 0.07 mag
for the color-selected samples presented by Wang et al.~(2003) and
Wang et al.~(2005), respectively, increase to 0.15 mag and 0.18 mag,
when less stringent color cuts are applied. Comparison of MLCS2k2 to
other SN Ia distance-fitting techniques based on the scatter in
respective Hubble diagrams requires supernova samples that are
comparable (and ideally identical) in size and scope.

\begin{figure}
\includegraphics[width=6in]{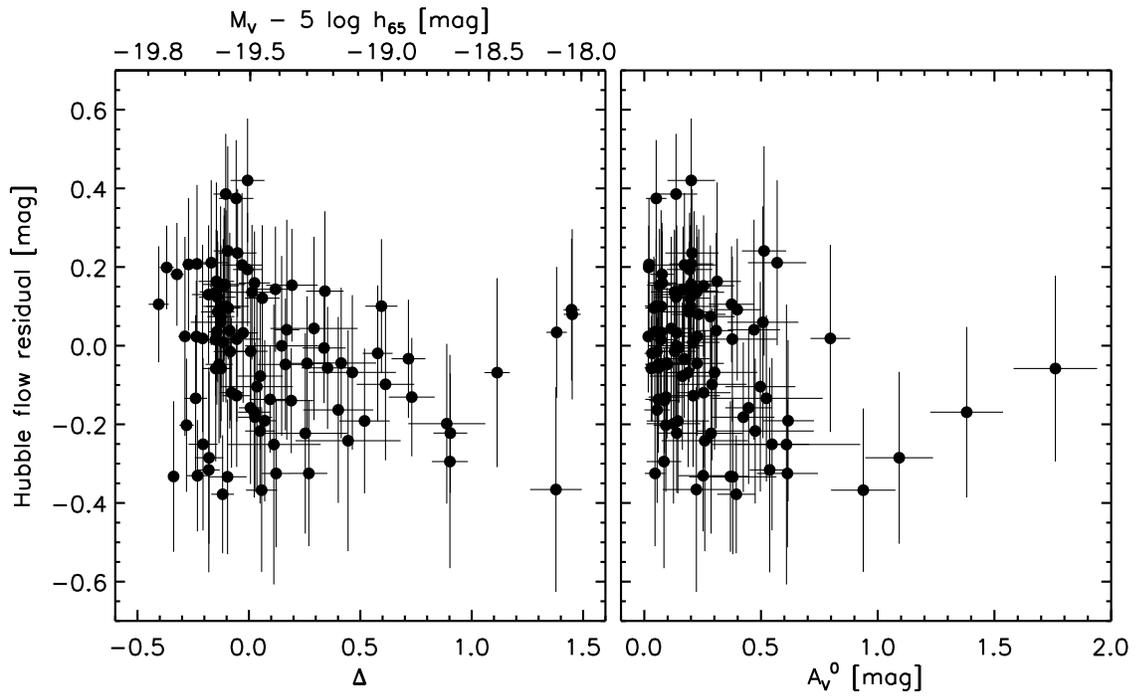}
\caption[Residuals correlations]{\singlespace MLCS2k2 Hubble flow
sample residuals versus mean values of the light-curve shape parameter
$\Delta$ and host-galaxy extinction $A_V^0$. The conversion from
$\Delta$ to the peak $V$ absolute magnitude, given on the top axis
of the left panel, is based on equation \ref{ch4-eqn-absvmax}.
\label{ch4-fig-resdav}}
\end{figure}

We display the distribution of the Hubble flow residuals versus the
measured $\Delta$ and $A_V^0$ in Figure \ref{ch4-fig-resdav}. No
obvious trends are present, though the scatter of the points decreases
with increasing $\Delta$, meaning the lower luminosity SN Ia have a
tighter correlation around the luminosity/light-curve shape
relationship. If we arbitrarily divide the sample into three bins, with slow
decliners ($\Delta < -0.15$), normal SN Ia ($-0.15 \leq \Delta \leq
0.3$), and fast decliners ($\Delta > 0.3$), the residual dispersions
are $\sigma =$ 0.21 mag ($N = 17$ objects), 0.18 mag ($N = 57$), and
0.14 mag ($N = 21$), respectively. 

\begin{figure}
\includegraphics[height=7in]{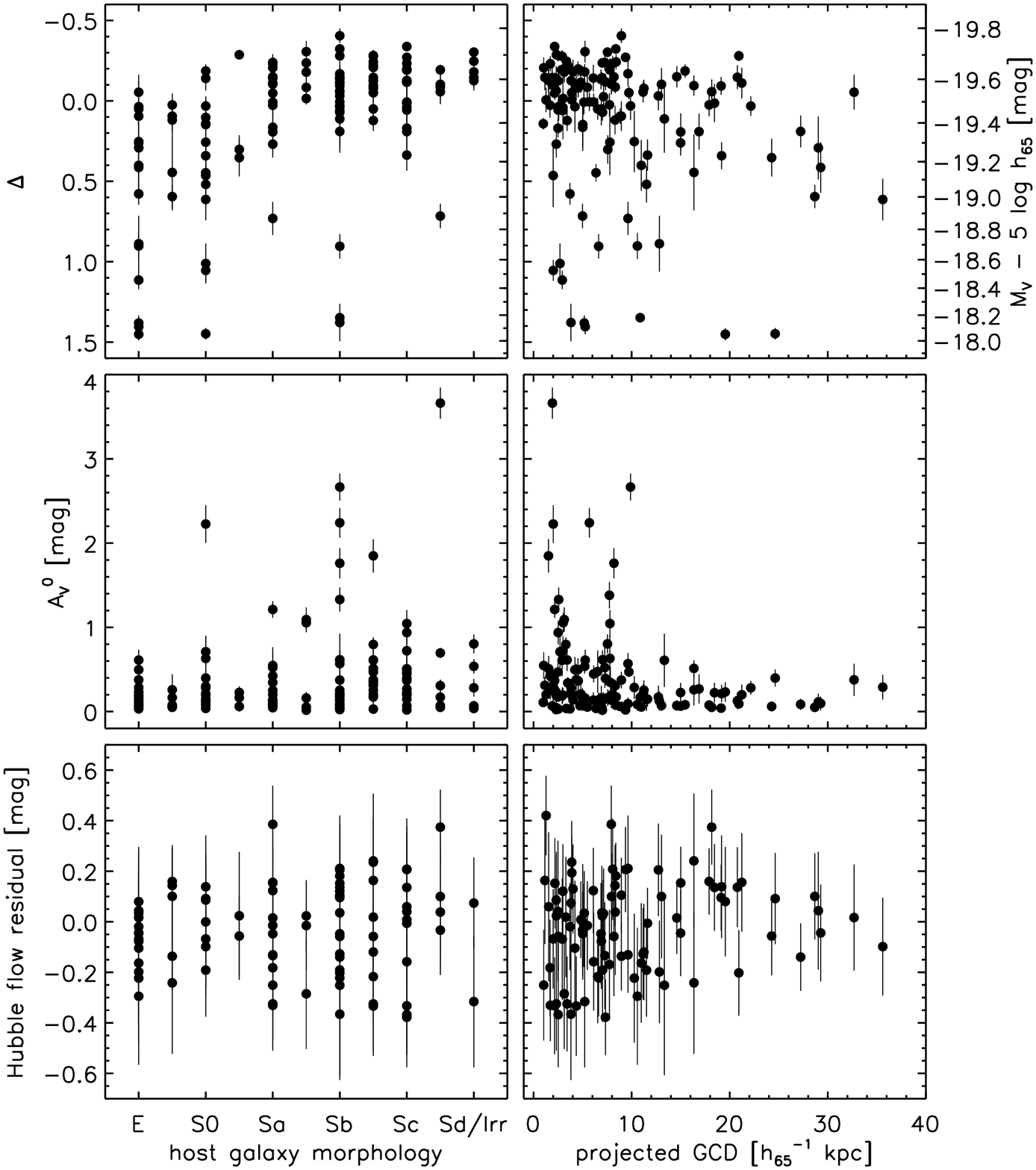}
\caption[Galaxy Correlations]{\singlespace Correlations with
host-galaxy morphology and projected galactocentric distance
(GCD). The top panels show mean values of the light-curve shape
parameter $\Delta$ for the full sample, and the middle panels show
the distribution of host-galaxy extinction $A_V^0$. The bottom panels show
the residuals relative to the best fit Hubble line for the Hubble flow
sample only (after MLCS2k2 correction for luminosity differences and
extinction). Projected GCDs are calculated using the angular offsets
presented in Table \ref{ch4-tab-sninfo}, with angular diameter
distances ($\Omega_{\rm M} = 0.3$, $\Omega_\Lambda = 0.7$) calculated
from the redshift for objects with $cz_{\rm CMB} \geq 2500 \; \kms$
and from the MLCS2k2 supernova distances for objects with lower
recession velocities.
\label{ch4-fig-gal} }
\end{figure}

In Figure \ref{ch4-fig-gal} we show correlations in the MLCS2k2 fit
parameters ($\Delta$ and $A_V^0$), as well as the residuals in the
Hubble flow sample, with host-galaxy morphological type and projected
separation. These confirm well-known trends for SN Ia (e.g., Hamuy et
al.~2000, and references therein): slow-declining (low $\Delta$,
high luminosity) SN Ia and heavily extinguished SN Ia are not
generally found in early-type galaxies or in the outskirts of their
hosts. Note there are some exceptions, such as SN 1990Y, 1991bg, and
1993ac, all fit with mean $A_V^0 \geq 0.3$ mag in elliptical hosts
(though in the case of the very subluminous SN 1991bg, there is a
significant correlation between $\Delta$ and $A_V^0$). The lower panels
of Figure \ref{ch4-fig-gal} show the residuals after MLCS2k2
calibration of the Hubble flow sample; note the large reduction in the
vertical scale, showing the efficacy of corrections for intrinsic
luminosity differences and extinction. The Hubble flow SN Ia do not
show any trend in their residuals with respect to host galaxy
morphology or projected galactocentric distance, though perhaps the
residual scatter is decreased at large separations. There is also a
hint that SN Ia in elliptical hosts have slightly negative residuals
after MLCS2k2 correction (meaning they are corrected to be too
bright/too nearby); however the weighted average residual is only
$-0.06 \pm 0.06$ mag, consistent with zero. Grouping together E and
E/S0 hosts (which show a similar paucity of slowly-declining SN Ia in
the upper left panel) yields a weighted mean residual of $-0.02 \pm
0.04$ mag. The observed scatter in the residuals of the early-type
hosts is less than the overall sample, with $\sigma = 0.11$ mag for
elliptical hosts only, and $\sigma = 0.13$ mag for E, E/S0, and S0
hosts; a large fraction of this scatter may be from peculiar
velocities, which should contribute $\sim$0.09 mag to the dispersion
for these galaxies.

Gallagher et al.~(2005) have approached these issues in more detail,
with integrated spectra of the host galaxies of many of these SN Ia,
allowing them to correlate SN properties (before and after MLCS2k2
correction) with additional parameters such as host metallicity, star
formation rate, and star formation history. Their results suggest no
clear correlations with the MLCS2k2 Hubble flow residuals, though
there is marginal evidence for a relation between the residuals and
host metallicity. 

\subsection{A Hubble Bubble? \label{ch4-sec-bubble}}

Zehavi et al.~(1998) presented evidence for a large local void based
on SN Ia distances which suggested a monopole in the peculiar velocity
field. They found that the Hubble constant estimated from SN Ia
within 70 $h^{-1}$ Mpc was 6.5\% $\pm$ 2.2\% higher than $H_0$ measured
from SN Ia outside this region, assuming a flat $\Omega_{\rm M} = 1$
Universe. The significance of this void decreases in the current
concordance cosmology ($\Omega_{\rm M} = 0.3$, $\Omega_\Lambda = 0.7$)
to 4.5\% $\pm$ 2.1\%, but it is still worthwhile to test whether the
larger SN Ia sample and updated distances presented here provide
evidence for or against a void.

Following Zehavi et al.~(1998), it is convenient to work with the
supernova distances in units of $\kms$, making them independent of the
distance scale. The ambiguity in the zeropoint of the Hubble diagram
disappears if we use the quantity $H_0 d_{\rm SN}$, which can be
calculated
\begin{equation}
H_0 d_{\rm SN} = 65 \left[ 10^{0.2\left(\mu_{65} - 25\right)} \right] 
\quad \kms,
\end{equation}
where we define $\mu_{65} \equiv \mu_0 + 5 \log h_{65}$, tabulated in
the third column of Table \ref{ch4-tab-mlcs}. These can be compared to
scale-independent luminosity distances: for an object at a
cosmological redshift $z$ in a flat Universe,
\begin{equation}
H_0 d_{\rm L}(z) = c \left(1+z\right) \int_{0}^{z} \left[
  \Omega_{\rm M} \left(1+z'\right)^3 + \Omega_\Lambda \right]^{-1/2} dz',
\end{equation}
and the difference $u = H_0 d_{\rm L}(z) - H_0 d_{\rm SN}$ is the host
galaxy peculiar velocity. The deviation from the Hubble law is given
by $\delta H/H = u/H_0 d_{\rm SN}$.

\begin{figure}
\begin{center}
\includegraphics[height=6.5in]{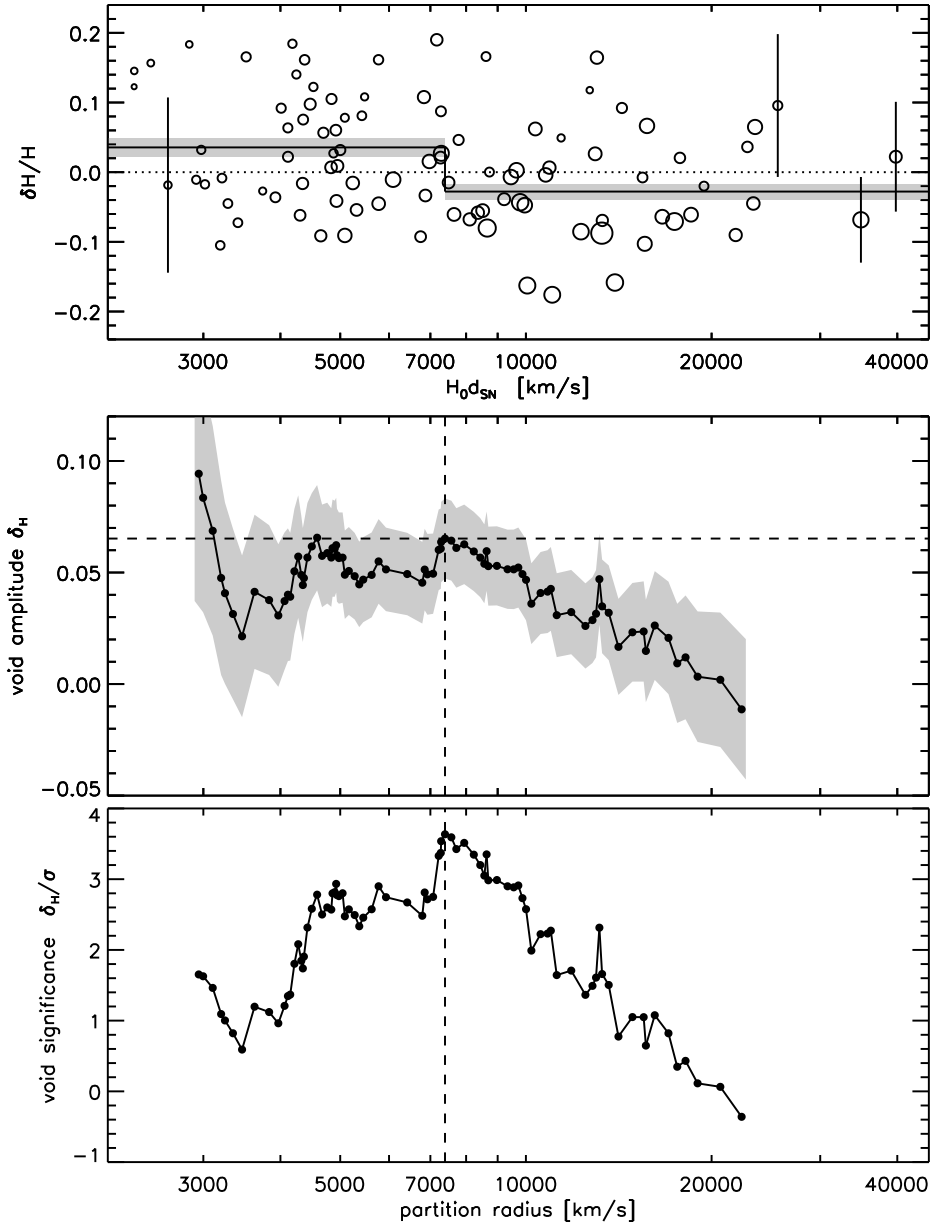}
\end{center}
\caption[Hubble Bubble]{\singlespace Indications of a local void. The
top panel shows the deviation from the Hubble law for each object in
the full Hubble flow sample. The radius of each circle is inversely
proportional to the uncertainty, with several representative points
showing the error bars explicitly to calibrate the symbol size. The
solid line shows the best-fit Hubble constants in the two-zone model
with the most significant void; the shaded regions give the 1-$\sigma$
uncertainty on each of these values. The middle panel shows the void
amplitude, $\delta_H \equiv H_{\rm inner}/H_{\rm outer} - 1$, as a
function of the radius at which the sample is partitioned, with the
shaded region illustrating the 1-$\sigma$ uncertainty. The lower panel
shows the significance of the void, i.e. the void amplitude divided by
its uncertainty. The most significant void occurs when the sample is
partitioned at $H_0 d_{\rm SN} \simeq 7400 \; \kms$, with $\delta_H =
6.5 \pm 1.8$\%.
\label{ch4-fig-bubble}}
\end{figure}

The top panel of Figure \ref{ch4-fig-bubble} shows $\delta H/H$ for
the 95 SN Ia in the full Hubble flow sample, where we have used our
standard assumptions: a peculiar velocity uncertainty of $\pm$300
$\kms$, $\sigma_{\rm add} = 0.08$ mag, and a $\Omega_{\rm M} = 0.3$,
$\Omega_\Lambda = 0.7$ cosmology. We then partition the full sample
into two parts at each value of $H_0 d_{\rm SN}$ (such that there are
at least six objects in the smaller subset) and calculate the best-fit
Hubble constants $H_{\rm inner}$ and $H_{\rm outer}$, and their
corresponding uncertainties (see equations 2 and 3 of Zehavi et
al.~1998). We define the void amplitude, $\delta_H \equiv (H_{\rm
inner} - H_{\rm outer})/H_{\rm outer}$, and display this quantity,
with its uncertainty, as a function of the partition radius in the
middle panel of Figure \ref{ch4-fig-bubble}.  We normalize $\delta_H$
by its uncertainty in the lower panel to illustrate the void
significance.

The full Hubble flow sample clearly shows a void signature; the most
significant void is derived when the sample is partitioned at $H_0
d_{\rm SN} \simeq 7400 \; \kms$ (between SN 2000ca and SN 2000bh at
$H_0 d_{\rm SN} =$ 7293 and 7494 $\kms$, respectively), with $\delta_H
= 6.5 \pm 1.8$\%, similar in both location and amplitude to the Zehavi
et al.~(1998) result. We have performed a number of statistical tests
to assess the significance of the result. We created $3 \times 10^5$
Monte Carlo realizations of the data set with Hubble law deviations
drawn from a Gaussian distribution with a standard deviation equal to
the uncertainty in each data point, and fit for the most significant
void or overdensity at each partition radius. Only 0.2\% of the time
was there a void as significant as the one seen in the actual data (at
any location).  Because this result depends upon our assumed error
distribution, we also created synthetic data sets with the Hubble law
deviations randomly resampled from the deviations in the full data set
(both with and without resampling of their errors). We have also
performed a full bootstrap resampling analysis to determine the
distribution of the void amplitude and significance around their best
fit values. In all of these tests a void with the significance as in
the actual data was seen at most 1.2\% of the time, and typically much
less often, depending on the details of the synthetic sample,
suggesting the result is valid at the 2.5 to 3.5-$\sigma$ level of
confidence.

The result also seems robust to jackknife tests; eliminating any one,
two, or even three points from the sample does not result in a very
large change in the void characteristics. With the three largest
outliers removed (without justification; these objects which are not
peculiar in any way), the void amplitude only decreases to $5.2 \pm
1.8\%$. We have also confirmed that the result persists using the 51
SN Ia from 1997 onwards only, an independent sample from Zehavi et
al.~(1998). The newer data show an even stronger result, $\delta_H =
9.1 \pm 2.6\%$ (at the same location), but there are only a handful of
recent SN Ia more distant than 10,000 $\kms$ and the most robust
results come from the full sample. Indications of this void can also
be seen in other distance estimates to SN Ia (though with largely
overlapping samples), including the ``gold'' sample of Riess et
al.~(2004), which used a slightly earlier implementation of MLCS2k2
(Jha 2002), and \dmf\ distances presented by Prieto, Rest, \& Suntzeff
(2006), suggesting it is not an artifact limited to our particular
analysis.

The significance of the void at 7400 $\kms$ is high partly because it
is near the middle of the sample, such that $H_{\rm inner}$ and
$H_{\rm outer}$ (and thus, their ratio) are most precisely
measured. The data suggest that void of similar amplitude may be
present if the sample is partitioned near $H_0 d_{\rm SN} \simeq 4800
\; \kms$. Indeed, if we attempt to fit a three-zone model, the data
support a model with $\delta_H \simeq 8\%$ closer than 4600 $\kms$,
and $\delta_H \simeq 5\%$ for $4600 \; \kms \lesssim H_0 d_{\rm SN}
\lesssim 7400 \; \kms$, both relative to the outer region beyond 7400
$\kms$. However, another three-zone scenario with nearly the same
likelihood has $\delta_H \simeq 5\%$ nearer than 7400 $\kms$, and an
\emph{overdense} region with $\delta_H \simeq -2\%$ at $7400 \; \kms
\lesssim H_0 d_{\rm SN} \lesssim 14000 \; \kms$ relative to the more
distant Hubble flow, which is similar to the three-zone infall region
model seen by Zehavi et al.~(1998). Because there are fewer points in
each region, the uncertainties in these numbers are increased to
$\sim$2.5\%. Moreover, the data do not favor a three zone model over a
two zone model (a $\chi^2$ decrease of less than 1 per new model
parameter), unlike the situation for a two zone model over a one zone
model (a $\chi^2$ decrease of more than 7 per new parameter).

While it seems likely this void signature is present in the SN data,
is it really present in the Universe? The void boundary does occur at
a distance comparable to large mass concentrations in the local
Universe, including the Great Wall and the Southern Wall (Geller 1997
and references therein). A void with $\delta_H = 6.5\%$ on this scale
would imply an underdensity $\delta \rho/\rho \simeq$ 20 to 40\% for
$\Omega_{\rm M} = 0.3$, depending on our location within the void
(Lahav et al.~1991; Turner, Cen, \& Ostriker~1992). Such a large-scale
density contrast, while not ruled out, is relatively unlikely in
current $\Lambda$CDM models; the fraction of mass residing in such a
void ranges from $10^{-4}$ to a few percent, depending on the exact
size and underdensity (Furlanetto \& Piran~2006).  Furthermore, tests
of the Hubble bubble using other distance measures have not
corroborated the SN Ia result. Using Tully-Fisher distances to galaxy
clusters, Giovanelli et al.~(1999) find $\delta_H = 1.0 \pm 2.2\%$ at
7000 $\kms$, while Hudson et al.~(2004) derive $\delta_H = 2.3 \pm
1.9\%$ via Fundamental Plane cluster distances, both consistent with
no void. Each method is subject to various systematic effects that
could lead to spurious results. For the SN, these include
K-corrections (but which are unlikely to cause a $\sim$0.1 mag error
at $z \simeq 0.03$), the effect of higher-order multipoles on the
monopole signature (but the sky distribution of the Hubble flow
objects generally shows the void in all directions), or a possible
photometric offset between the Cal\'an/Tololo sample (accounting for
most of the more distant objects) and more recent samples such as CfA
I and II (making up most of the more nearby objects; Prieto, Rest, \&
Suntzeff 2006). More data (throughout the nearby redshift range, $z
\lesssim 0.15$) and a more thorough analysis are needed to
definitively resolve this open issue.

Regardless of whether the Hubble bubble is due to a real local void in
the Universe or an artifact of SN Ia distances, the feature is present
in the Hubble flow SN sample, and this has important implications for
using SN Ia as tools for precision cosmology.\footnote{This is not to
say the reason for the discrepancy is not important. If the SN Ia are
revealing a true void, one would presumably proceed to measure the
global Hubble constant and other cosmological parameters simply using
a Hubble flow sample beyond the void region (with $z \gtrsim$ 0.025;
Riess et al.~2004). However, if the indications for a void are due to
an unknown systematic error, the effects on the cosmological utility
of SN Ia could be much more severe.} The Hubble flow sample is
critical to both measurements of $H_0$ (a differential measurement
between the Hubble flow objects and nearby Cepheid-calibrated SN Ia)
and $\Omega_\Lambda$ or $w$ (a differential measurement between the
Hubble flow objects and high-redshift SN Ia), and both applications
are affected by this result.  The effect on $H_0$ is readily apparent;
our derived $H_0$ will be 6.5\% higher using a Hubble flow sample with
$H_0 d_{\rm SN} \lesssim 7400 \; \kms$ compared to a Hubble flow
sample with $H_0 d_{\rm SN} \gtrsim 7400 \; \kms$. This is mitigated
if we use the full Hubble flow sample, but that still yields a small
but not insignificant effect: the full sample $H_0$ is larger by 3\%
compared to just the subsample beyond 7400 $\kms$ (which gives an
intercept of the ridge line, $a_V = 0.7015 \, \pm \, 0.0049$).

\begin{figure}
\includegraphics[width=6in]{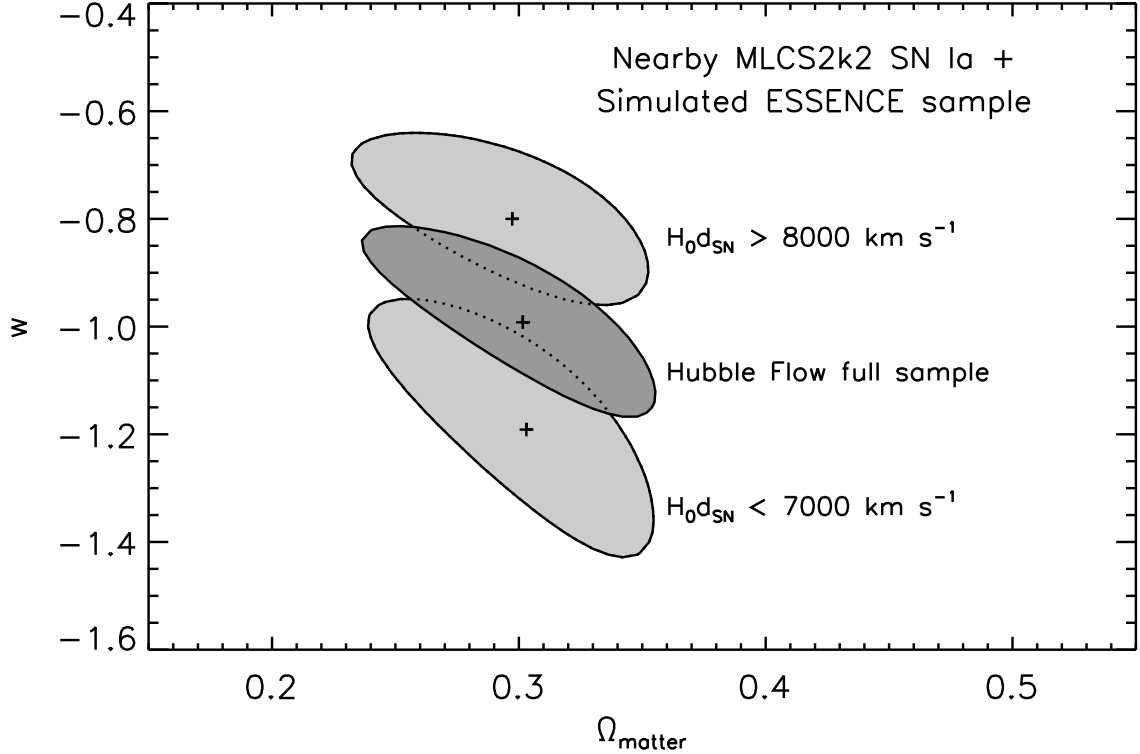}
\caption[Effect on w]{\singlespace Effect of a local void on
constraints of the dark energy equation of state, using a simulated
sample of 200 SN Ia with 0.3 $\leq z \leq$ 0.8 as expected from the
ESSENCE survey (Miknaitis et al.~2007). We perform a cosmological fit
using the same simulated high-redshift sample plus three nearby
samples: (1) the full MLCS2k2 Hubble flow sample with 95 SN Ia, (2)
the nearby Hubble flow objects with $H_0 d_{\rm SN} < 7000 \; \kms$
(48 SN Ia) and (3) the distant Hubble flow objects with $H_0 d_{\rm
SN} > 8000 \; \kms$ (40 SN Ia). The figure shows 68.3\% (1-$\sigma$)
confidence regions (shaded), and the mean values of $\Omega_{\rm M}$
and $w$ (crosses). The input cosmology for the simulated objects is
$\Omega_{\rm M} = 0.3$, $\Omega_\Lambda = 0.7$ ($w = -1$), with the
distance scale set by the full Hubble flow sample. In the cosmological
fit, we assume a flat Universe and a prior on $\Omega_{\rm M} = 0.30
\pm 0.04$.  The different low-redshift samples have a strong effect on
the estimation of $w$, with $w = -0.99 \pm 0.12$ for the full Hubble
flow sample, $w = -1.19 \pm 0.17$ for the nearby Hubble flow sample
and $w = -0.80 \pm 0.11$ for the distant Hubble flow
sample. \label{ch4-fig-essence}}
\end{figure}

For high-redshift applications, the void signature can be much more
important. Whereas the choice of using the full Hubble flow sample or
one just with objects beyond the void region (with $z \gtrsim 0.025$)
does not significantly affect the conclusion from SN Ia that we live
in an accelerating Universe (Riess et al.~2004), such a choice has a
large effect in current efforts to measure the equation of state of
the dark energy. In Figure \ref{ch4-fig-essence} we show constraints
on $w$ from a simulation in which the current Hubble flow sample is
used in conjunction with an artificial data set of 200 SN Ia with $0.3
\leq z \leq 0.8$, as expected from the ongoing ESSENCE survey
(Miknaitis et al.~2007). The results show a $\sim$20\% difference in
the derived value of $w$ using the full Hubble flow sample compared to
one only including objects with $H_0 d_{\rm SN} \geq 8000 \; \kms$ ---
twice the target statistical uncertainty of the survey. For future
surveys with thousands of high redshift SN Ia, this single systematic
uncertainty in the Hubble flow sample could easily dwarf all other
sources of uncertainty combined. Clearly, then, precision cosmology
with SN Ia will require an investment in \emph{nearby} SN Ia (both in
data and analysis) comparable to the immense efforts ongoing and
envisioned at high redshift.

\subsection{The Local Group Motion \label{ch4-sec-flows}}

\begin{figure}
\begin{center}
\includegraphics[width=6in]{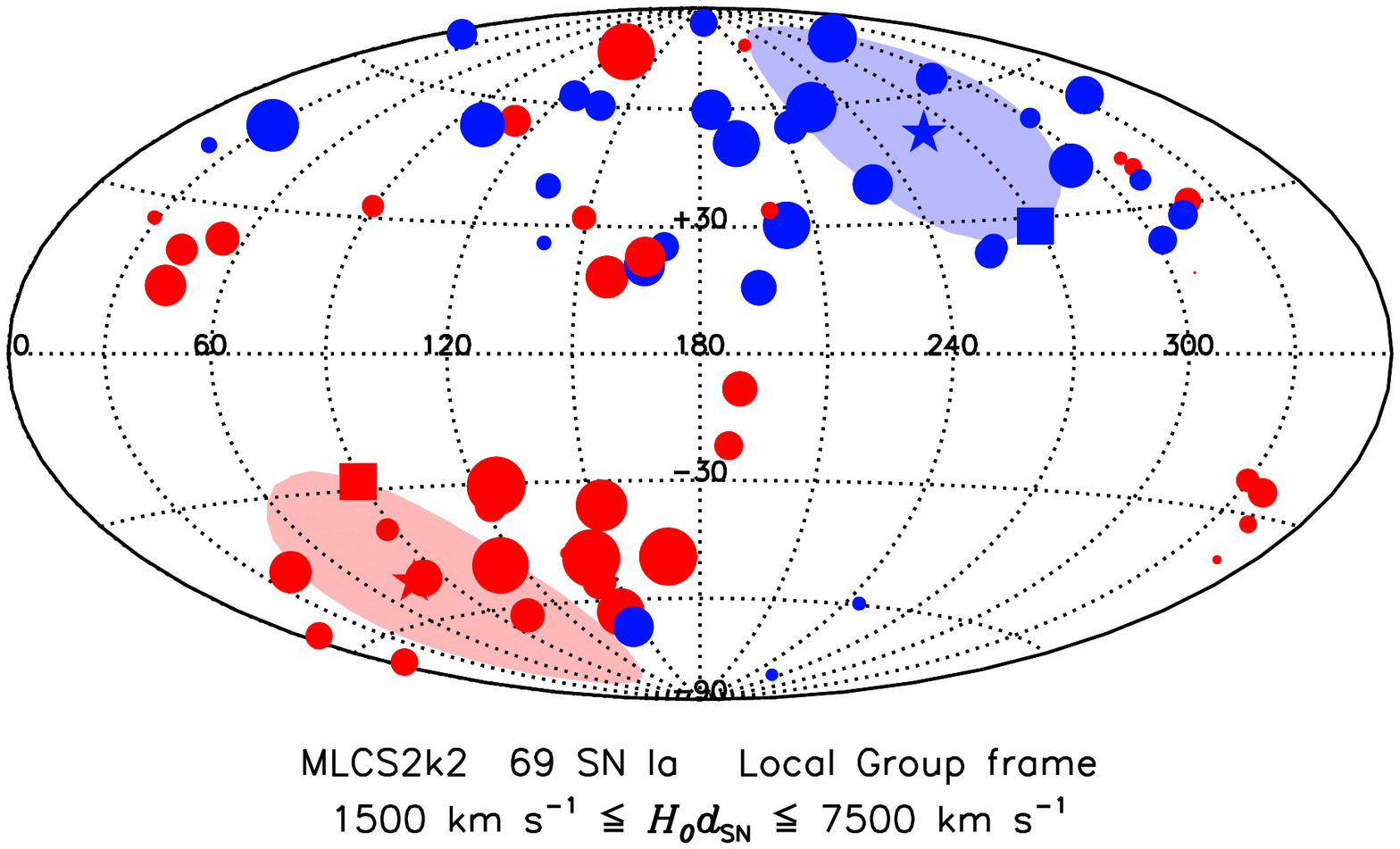}
\end{center}
\caption[Peculiar Velocities]{\singlespace Peculiar velocities of 69
SN Ia with $1500 \; \kms \leq H_0 d_{\rm SN} \leq 7500 \; \kms$ in the
rest-frame of the Local Group, plotted in Galactic coordinates (with
central meridian $\ell = 180\arcdeg$). The blue circles indicate SN Ia
in galaxies with negative peculiar velocities (i.e., approaching us
relative to the cosmic expansion). The red circles indicate positive
peculiar velocities, objects from which we are moving away.  The area
of the symbols is proportional to the amplitude of the peculiar
velocity. The squares mark the location of the CMB dipole (Fixsen et
al.~1996) and are drawn to scale relative to the circles, with the
Local Group moving towards (269\arcdeg, $+$28\arcdeg) at $635 \;
\kms$. The stars show the position and amplitude of the best-fit
simple dipole model to the SN data, with the shaded area indicating
locations within the formal 2-$\sigma$ confidence
region. \label{ch4-fig-flow}}
\end{figure}

The precision of SN Ia distances makes them well suited to measure
peculiar velocities of nearby galaxies (e.g., Riess, Press, \&
Kirshner~1995b).  Many new SN Ia are being discovered at redshifts
conducive to studying the local flow field ($z \lesssim 0.03$). In
Figure \ref{ch4-fig-flow}, we show a clear detection of the motion of
the Local Group relative to the frame defined by nearby SN Ia. We plot
the host galaxy peculiar velocities (in the Local Group frame, $u_{\rm
LG} = H_0 d_{\rm L}(z_{\rm LG}) - H_0 d_{\rm SN}$) of 69 SN Ia with
$1500 \; \kms \leq H_0 d_{\rm SN} \leq 7500 \; \kms$ in Galactic
coordinates (excluding the cluster member SN 1999ej and SN 2000cx
which could not be fit by MLCS2k2).  The distance range was chosen to
exclude Virgo cluster galaxies at the lower end, and be within the
void signature discussed above at the upper end (so that monopole
deviations are minimized, in addition to excluding objects with
increasingly uncertain absolute peculiar velocities).

A dipole signature is clearly present in the data; a simple (naive)
dipole fit to these SN data indicates the Local Group is moving at
$541 \pm 75 \; \kms$ towards $(l,b)$ = $(258\arcdeg \pm 18\arcdeg$,
$+51\arcdeg \pm 12\arcdeg)$. This is consistent with the position and
amplitude of the CMB dipole, $635 \; \kms$ towards $(269\arcdeg$,
$+28\arcdeg)$, at approximately the 2-$\sigma$ confidence level. In
addition, the paucity of SN Ia discovered at low Galactic latitude
skews the best fit dipole away from the Galactic plane, suggesting
that the SN Ia peculiar velocities are providing evidence for
convergence of the local flow field to the CMB dipole at roughly the
depths probed by these objects.

The full MLCS2k2 sample is quite amenable to more sophisticated
analysis of the flow field (Riess et al.~1997b), such as recently
performed by Radburn-Smith, Lucey, \& Hudson (2004), who constrained
$\beta \equiv \Omega_{\rm M}^{0.6}/b = 0.55 \pm 0.06$ comparing the
nearby SN sample of Tonry et al.~(2003) with the IRAS PSCz galaxy
survey (Saunders et al.~2000). The ever-increasing nearby SN sample
should also allow for merging and extension of elaborate local flow
models, for example, such as provided by surface-brightness
fluctuation distances (Tonry et al.~2000). For maximum utility, we
strongly encourage nearby SN discovery efforts in the southern
hemisphere (which would help fill the empty region in the lower right
quadrant of Figure \ref{ch4-fig-flow}).

\subsection{Extragalactic Extinction Laws \label{ch4-sec-rv}}

Extinguished SN Ia, though less than ideal distance indicators,
nevertheless provide one of the few ways to constrain the extinction
law along individual lines of sight through distant galaxies (see also
e.g., Mu\~noz et al.~2004, but note McGough et al.~2005).  The
traditional method of using SN Ia to determine the extinction and
reddening properties of extragalactic dust (typically parameterized by
$R_V$) has been to plot the magnitude residuals (assuming a relative
distance, such as from the Hubble law) in a given passband and epoch
(for example, in $B$ or $V$ at maximum light) versus the SN color at
some epoch (for instance, \bv\ also at maximum light). Because SN Ia
are not perfect standard candles, this simple method fails: the
magnitude residuals and the intrinsic colors are both functions of
intrinsic luminosity. In the calibrated candle framework, the
light-curve shape provides the intrinsic luminosity (through a
parameterization such as \dmf). However, SN Ia with different light
curve shapes have varying luminosities \emph{and} colors, so
determination of $R_V$ depends critically on disentangling the
intrinsic variation from the contributions of dust.\footnote{Tripp \&
Branch (1999) constructed a two parameter model for SN Ia distances
based on fitting SN Ia peak magnitudes as a linear function of \dmf\
and peak \bv\ that explicitly forgoes any attempt at the difficult
problem of separating these two causes for faint, red SN Ia. This
model is particularly sensitive to the makeup of the sample, and
requires care in its application. The same applies to the model of
Parodi et al.~(2000).}

By choosing objects in early-type host galaxies, or those that are not
located in spiral arms, and using the observations at times when the
intrinsic colors are nearly independent of luminosity (e.g., the Lira
law for the late-time evolution of \bv; \S \ref{ch4-sec-ex0}), it is
possible to derive and correct for the relation between intrinsic
luminosity and color and construct a sample of magnitude residuals and
color excesses which are dominated by the effect of host-galaxy
dust. From this procedure Phillips et al.~(1999) found $R_V = 2.6 \pm
0.4$, lower than but roughly consistent with the canonical $R_V = 3.1$
for Galactic dust. Similarly, Altavilla et al.~(2004) find $R_V = 2.5$
and Reindl et al.~(2005) give $R_V = 2.65 \pm 0.15$, using generally
similar (though different in detail) methods of separating the
intrinsic and dust effects on luminosity and color. Using a previous
incarnation of MLCS (albeit one with too strong a dependence of
intrinsic color on light curve shape) Riess et al.~(1996b) found $R_V
= 2.55 \pm 0.30$.

Here we present a different method of determining $R_V$, individually
for each SN Ia, using MLCS2k2. This alternative is discussed by Riess
et al.~(1996b), and does not require an independent estimate of the SN
Ia distance. Rather, we include $R_V$ directly as an MLCS2k2 fit
parameter, and make use of the relations between the extinction in
different passbands (see Figure \ref{ch4-fig-a0rv} and Table
\ref{ch4-tab-a0rv}). With only two observed passbands, such as $B$ and
$V$ (and thus, only one observed color), the three parameters $\mu_0$,
$A_V^0$, and $R_V$ are degenerate, so we require observations in at
least three passbands to constrain $R_V$ directly.\footnote{Another
approach that would work for Hubble flow SN Ia with only two observed
passbands in the MLCS2k2 framework would be to use the Hubble law to
put a prior on $\mu_0$, and then fit for $A_V^0$ and $R_V$. This would
essentially be the same as using magnitude residuals with respect to
the Hubble line as discussed above. However, almost all of the nearby
SN Ia presented here have observations in three or more filters, and
so we can use the direct method.}  Because the distance modulus shifts
magnitudes in all passbands uniformly, this approach is effectively
the same as constraining $R_V$ from multiple color excesses (e.g.,
$E(\vi)/E(\bv)$; Riess et al.~1996b).

An advantage of this method is that we do not require independent
distances; therefore, extinguished SN Ia with recession velocities too
small to give reliable Hubble flow distances can still be used to
constrain $R_V$. In addition, because MLCS2k2 explicitly utilizes a
model covariance matrix, the constraints on $R_V$ properly account for
correlated uncertainties in the data and the model, as well as the
effect of the intrinsic dispersion in the luminosities and
colors. Furthermore, MLCS2k2 requires the data in all passbands (for
one SN) be fit by a consistent $R_V$, and required constraints such as
$R_B = R_V + 1$ are automatically enforced (not necessarily the case
in the alternate method where residuals in $B$ and $V$ are
independently regressed against the color excess; Phillips et
al.~1999; Reindl et al.~2005).

\begin{figure}
\begin{center}
\includegraphics[height=6.5in]{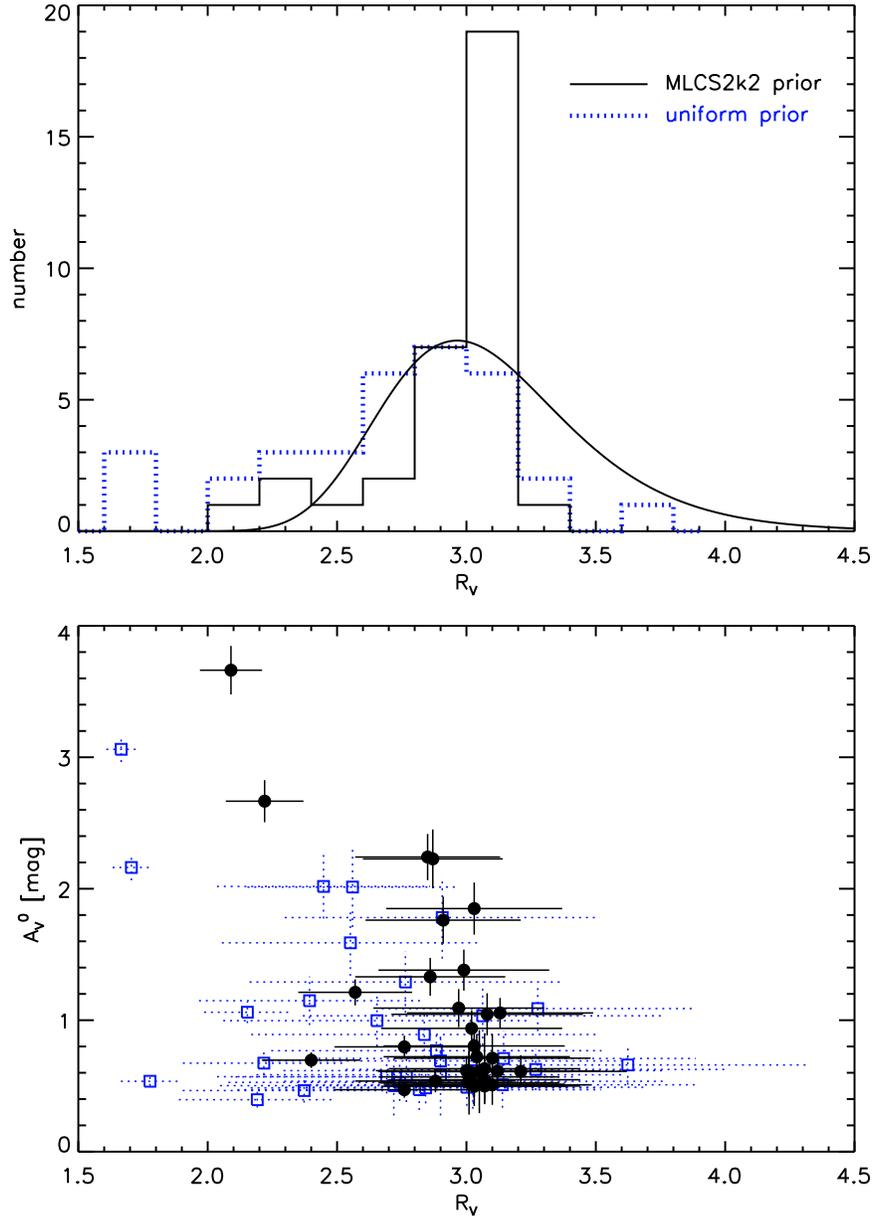}
\end{center}
\caption[Extinction Law]{\singlespace Distribution of the host galaxy
extinction law parameter, $R_V$, and its correlation with the host
galaxy extinction $A_V^0$. The solid curve in the upper panel shows
the adopted MLCS2k2 prior, $\hat{p}(R_V)$, which is a Gaussian in
$R_V^{-1}$, and has a mean $R_V = 3.1$ with a standard deviation of
0.4.  The solid histogram shows the distribution of the mean $R_V$ for
33 extinguished SN Ia, using MLCS2k2 and $\hat{p}(R_V)$, listed in
Table \ref{ch4-tab-mlcs}. The dotted histogram (blue in the
electronic edition) shows the distribution if a uniform prior for $R_V
\geq 1.6$ is adopted instead. The filled circles in lower panel show
the mean $R_V$ and $A_V^0$ using the MLCS2k2 prior for these 33
objects, while the empty squares with dotted error bars (blue in the
electronic version) correspond to results based on the uniform prior.
\label{ch4-fig-rv}}
\end{figure}

The main disadvantage to this approach is a lack of sensitivity. The
intrinsic variation in SN Ia colors, observational uncertainties, and
the subtle relative color differences (in the optical bands) with
changes in the extinction law combine to make the constraints on $R_V$
relatively weak for individual SN Ia. Unfortunately, the $U$ band,
which has the most sensitivity to varying $R_V$, also has the most
intrinsic dispersion (and this is clearly not related to $R_V$; Jha
et al.~2006). In the upper panel of Figure \ref{ch4-fig-rv}, we plot a
histogram of the mean values of $R_V$ for the sufficiently
extinguished SN Ia tabulated in Table \ref{ch4-tab-mlcs}, and show
the relation between $R_V$ and $A_V^0$ for these objects in the lower
panel.

On first inspection, the data show a clustering near $R_V \simeq 3.1$,
with a handful of objects with smaller values and hardly any with
significantly larger values. However, because of the lack of
sensitivity to $R_V$ even for these moderately extinguished objects,
the results turn out to be quite dependent on the choice of prior,
$\hat{p}(R_V)$. As described in \S \ref{ch4-sec-app}, we have used a
prior on $R_V$ based on its distribution along Galactic lines of sight
from CCM89 which is a Gaussian in $R_V^{-1}$ and is shown as the solid
curve in the upper panel of Figure \ref{ch4-fig-rv}. When data are
increasingly informative, the posterior pdf is decreasingly sensitive
to the choice of prior, but in this case the data only weakly
constrain $R_V$ and the prior matters. Using the extreme case of a
uniform prior for $R_V \geq 1.6$ (implying we are completely ignorant
of the relative likelihood of these different kinds of dust
distributions) yields the dotted blue histogram and points in Figure
\ref{ch4-fig-rv}. Even this is not completely satisfactory, because
three objects favor $R_V$ at the (arbitrary) lower limit of the prior,
which becomes increasingly unlikely (and perhaps unphysical; Draine
2003 and references therein): lower $R_V$ corresponds to dust that
reddens more and more for a given attenuation, favoring smaller and
smaller grains.

Nevertheless, even with this significant dependence on the prior, some
results regarding $R_V$ are robust.  The effect of the prior is less
important for objects which have very high extinction ($A_V^0 \gtrsim
2$ mag). In particular, the two most heavily extinguished objects in
the sample, SN 1996ai and SN 1999cl, both firmly require $R_V < 3.1$.
In addition, the posterior pdf has important differences from the
prior pdf (which is displayed normalized to the same total area).
There is a distinct deficit of SN Ia which yield $R_V$ significantly
greater than 3.1, whereas many such sight lines exist in the Milky Way
(particularly in dense clouds, with $R_V \simeq$ 4 to 5). The SN Ia
$R_V$ distribution may be more similar to lines of sight in the
Magellanic Clouds (Gordon \& Clayton 1998; Misselt, Clayton, \& Gordon
1999) which typically have lower $R_V$, and it might be worthwhile to
investigate priors on $R_V$ derived from more sophisticated models and
larger samples of stars, including LMC and SMC lines of sight
(Reichart 2001). An important possibility to consider is that the dust
around heavily extinguished SN Ia may be local to the SN environment;
SN 1995E, for example, with $A_V^0 = 2.24 \pm 0.18$ mag, occurred in a
relatively face-on host galaxy, suggesting our line of sight to the SN
does not have a large path length in the galaxy disk. If ``local''
dust plays a significant role, the $R_V$ distribution could be quite
different for moderately and heavily extinguished SN Ia, and such an
effect may be difficult to discern using the method which correlates
residuals with color because the most reddened objects have the
largest lever arm.

Constraints on $R_V$ could also be strengthened if we included
photometry in other passbands. Observations further in the UV would be
increasingly sensitive to variations in $R_V$, but they are difficult
to obtain (requiring space-based data for nearby extinguished SN Ia
or very deep ground-based data for distant extinguished SN Ia for
which the desired wavelengths redshift past the atmospheric cutoff),
and the intrinsic SN Ia magnitudes and dispersions are unexplored at
these wavelengths. On the other hand, near-infrared data will be
extremely valuable in providing a long wavelength baseline to
constrain both the extinction and the extinction law (e.g., Krisciunas
et al.~2006 and references therein). We are currently developing a
method to incorporate infrared photometry into MLCS2k2 by using
observations in $JHK$ to provide a joint prior constraint on $R_V$ and
$A_V^0$, that is then used to fit the optical data (Jha, Prieto, \&
Krisciunas, in preparation).

What are the effects of possible variations in $R_V$ on our derived
MLCS2k2 distances? The results from the extinguished objects with a
uniform prior have a mean $\langle R_V \rangle = 2.7$ (with the
default MLCS2k2 prior, the mean is 2.9). The majority of the Hubble
flow sample has low extinction, where $R_V$ variation has nearly
negligible effects, and we had fixed $R_V \equiv 3.1$ for objects with
$A_V^0 < 0.5$ mag. We have refit the full Hubble flow sample using a
new prior with the same shape but $\langle R_V \rangle = 2.7$, as well
as a fixed $R_V \equiv 2.7$ for objects with $A_V^0 < 0.5$ mag. On
average, the distance moduli increase by 0.019 mag, with an average
increase for the fixed $R_V$ subsample of 0.014 mag (both
corresponding to less than a 1\% increase in distance). The effect of
a smaller $R_V$ on differential measurements (such as determining
$H_0$ from the Hubble flow sample and a Cepheid calibrated sample)
would be less, because all distances would be slightly increased. At
present, we find no compelling evidence to favor a constant $R_V =
2.7$ over $R_V = 3.1$ \emph{for low-extinction SN Ia samples} such as
are typically used in cosmological applications.  Allowing for $R_V$
variations over this range could yield a plausible estimate of the
systematic uncertainty, but more problematic for precision cosmology
would be a systematic change in the $R_V$ distribution with redshift,

\acknowledgements

We thank Alex Filippenko, Weidong Li, Kevin Krisciunas, and Nick
Suntzeff for providing access to unpublished spectroscopy and
photometry. We also thank Peter Challis, Peter Garnavich, Peter
Nugent, Steve Furlanetto, and Tom Matheson for helpful discussions, as
well as the anonymous referee for useful suggestions. SJ gratefully
acknowledges support at UC Berkeley via a Miller Research Fellowship
and NSF grant AST-0307894 to A. V. Filippenko, as well as the Panofsky
Fellowship at KIPAC/SLAC, supported in part by the Department of
Energy contract DE-AC02-76SF00515. Research on supernovae at Harvard
University is supported by NSF Grant AST-0205808.

This research has made use of the NASA/IPAC Extragalactic Database
(NED) which is operated by the Jet Propulsion Laboratory, California
Institute of Technology, under contract with the National Aeronautics
and Space Administration.

\appendix

\section{Extinction Priors}

Measurements of the luminosity distances to type Ia supernovae need to
account for the line-of-sight absorption by dust to yield a precise
and accurate estimate.  The most common approach is to measure the
color excess resulting from selective absorption, and with the use of
a reddening law, estimate the extinction.  Unfortunately, uncertainty
in the fiducial color frequently dominates the individual distance
uncertainty, arising from the uncertainty (added in quadrature) of two
true flux measurements and the intrinsic dispersion of SN Ia colors.
Frequent and high precision monitoring of nearby SN Ia can remove the
contribution of measurement uncertainty, revealing the intrinsic color
dispersion to be approximately 0.05 mag for optical colors as shown in
Figure \ref{ch4-fig-zerocol}.  Multiplying this intrinsic uncertainty
by a standard reddening law, $R_V=3.1$, yields the observed SN Ia
distance dispersion of 0.15-0.20 mag.  However, failure to measure
colors to better precision than the intrinsic dispersion may result in
greatly degraded distance precision.

A significant improvement in the extinction estimate can be realized
by the use of a Bayesian prior on the extinction parameter because a
good deal is known about the likely values of extinction, \emph{a
priori}.  The most basic and unassailable truths about extinction are
that it cannot be negative\footnote{For the case of rare light echoes
in the presence of excessive quantities of circumstellar dust,
scattering can add some additional light to the observer's view,
though the net effect is still positive extinction, albeit with a
different ratio of selective to total absorption as discussed by Wang
(2005).} and that increasing values are decreasingly likely due to the
unlikely viewing angles required for large extinction, and for some
surveys, magnitude limits.  Models by Hatano, Branch, \& Deaton
(1998), Commins (2004), and Riello \& Patat (2005) show that for late
type galaxies, the likelihood distribution for extinguished lines of
sight follows an exponential function with a maximum at zero
extinction. In Figure \ref{ch4-fig-zerocol} we use this function
constrained by the a posteriori color distribution to determine a
decay constant for the exponential of $\sim$0.5 mag in $A_V$.  The use
of such an extinction prior represents great potential improvement to
this precision-limiting measure of SN Ia distances.  This approach was
first used by Riess, Press, and Kirshner (1996a) and later by Riess et
al.~(1998a) and Phillips et al.~(1999) as well as in the present work.
However, the well known danger of using a priori information is that
it may be in error and thus propagate systematic errors into the final
distance estimate.  In addition, one must be careful to use an
unbiased estimator of extinction from the posterior likelihood
distribution to avoid additional systematic error.  To weigh the
advantages and disadvantages of using extinction priors for
\emph{realistic} measurements of SN Ia, we have undertaken a set of
Monte Carlo simulations.

Our simulations model 10$^4$ SN Ia, each with a fixed intrinsic \bv\
color drawn from a Gaussian distribution (with zero mean, and standard
deviation $\sigma_{\rm int}$) and an extinction $A_V$ drawn from an
exponential distribution (with scale length $\tau$). We then simulate
$n_{\rm obs}$ measurements of the SN \bv\ color, each with a Gaussian
error $\sigma_{\rm meas}$, and subsequently ``fit'' for the
extinction, determining its posterior probability distribution
function with the application of a prior (either the true one
describing the input extinction distribution, a more conservative
prior, none at all, or an incorrect one). For each posterior pdf we
extract both the mean and the mode, calculate the error between the
posterior estimate and the input extinction, and analyze this error
distribution over the whole simulated sample.  These quantities may
also be calculated analytically using Bayes's Theorem, but we prefer
the numerical approach for ease of the incorporation of a variety of
functional forms for the extinction prior.\footnote{The simulation
routine is available at
\url{http://astro.berkeley.edu/~saurabh/mlcs2k2/}}

\begin{figure}
\begin{center}
\includegraphics[height=6.5in]{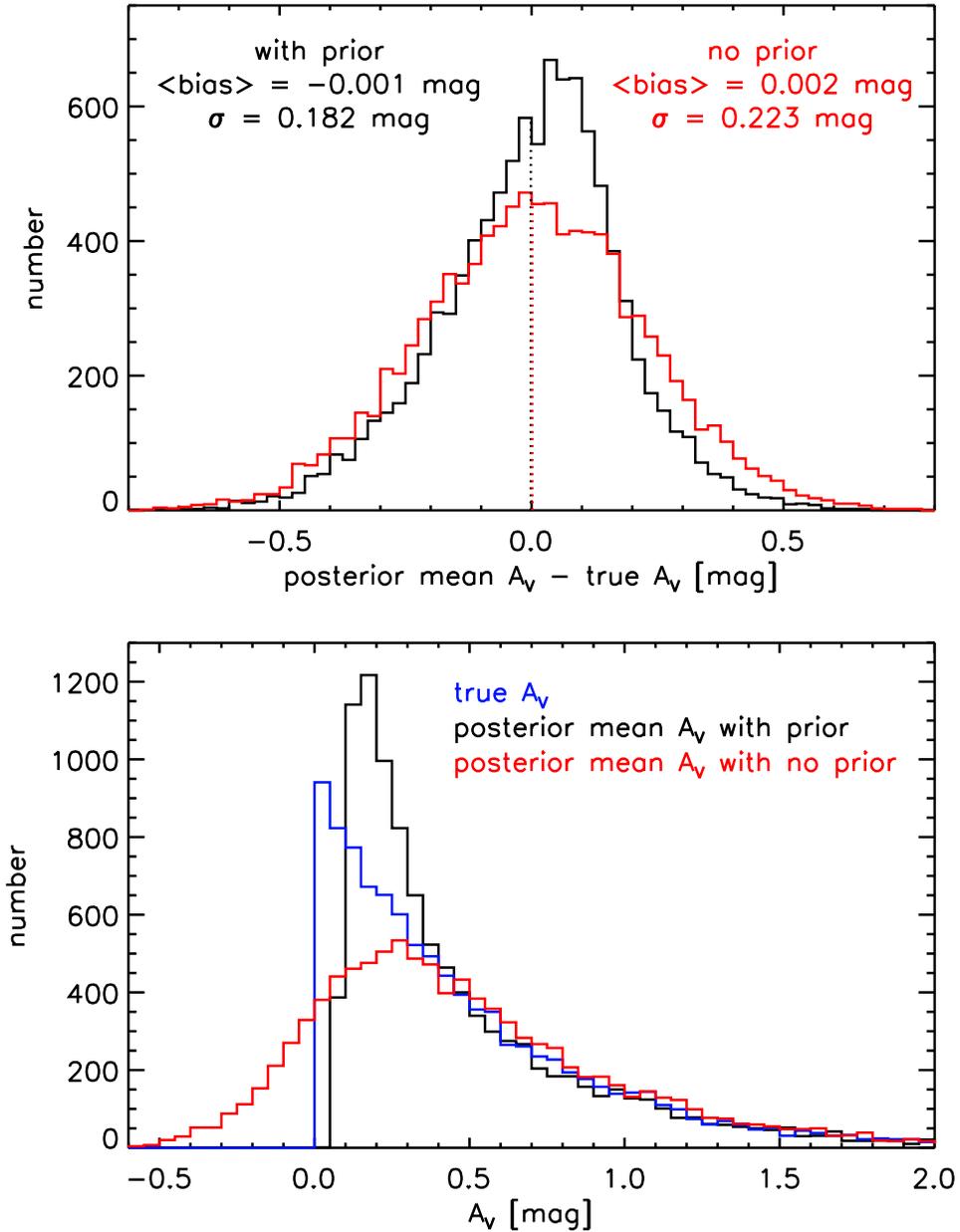}
\end{center}
\caption[Extinction Prior Simulation]{\singlespace Distributions of
the extinction bias (top panel) and input and posterior mean
extinction (bottom panel) from a simulation of 10$^4$ SN Ia, with a
Gaussian intrinsic \bv\ color scatter of $\sigma_{\rm int} = 0.05$
mag, ``observed'' on 5 epochs with a precision of $\sigma_{\rm meas} =
0.1$ mag in each color measurement. The SN are drawn from an
exponential $A_V$ distribution with a scale length $\tau = 0.5$ mag
(blue histogram), and then fit with a prior matching the extinction
distribution (black), as well as without any prior (red). For each
fit, the mean of the posterior extinction distribution (with and
without a prior) is used as the extinction estimate, and these
distributions are shown in the lower panel. The upper panel then shows
the difference between the posterior mean extinction and the true
extinction. Use of the correct extinction prior shows no mean bias
(dotted lines), and improves the precision of the extinction
measurement. \label{fig-app1}}
\end{figure}

As an example, we show in Figure \ref{fig-app1} the results for
$n_{\rm obs}$ = 5, $\sigma_{\rm meas}$ = 0.1 mag, $\sigma_{\rm int}$
=0.05 mag, and $\tau$ = 0.5 mag, parameters chosen to correspond to
adequate but minimal color measurements, such as those used in Riess
et al. (1998a) for the detection of cosmic acceleration, and we have
fit the simulated sample with both the correct prior ($\hat{p}(A_V)
\propto \exp (-A_V/\tau)$, with $\tau$ = 0.5 mag) and no prior at all
(allowing the fit $A_V$ to be negative as well).

As shown, use of the mean of the individual extinction likelihoods
results in an unbiased estimate of the extinction and an increased
precision relative to the absence of an extinction prior.  Because the
weight of distance estimates for cosmological parameter estimation is
proportional to the square of the uncertainty, the precision of 0.18
mag per SN Ia using the prior represents a 50\% improvement over the
use of no prior for this realistic case.  For more poorly measured
colors ($\sigma_{\rm meas}=0.2$ mag) or fewer independent
measurements, the improvement approaches a factor of two, akin to
doubling the cosmological sample.

\begin{figure}
\includegraphics[width=6in]{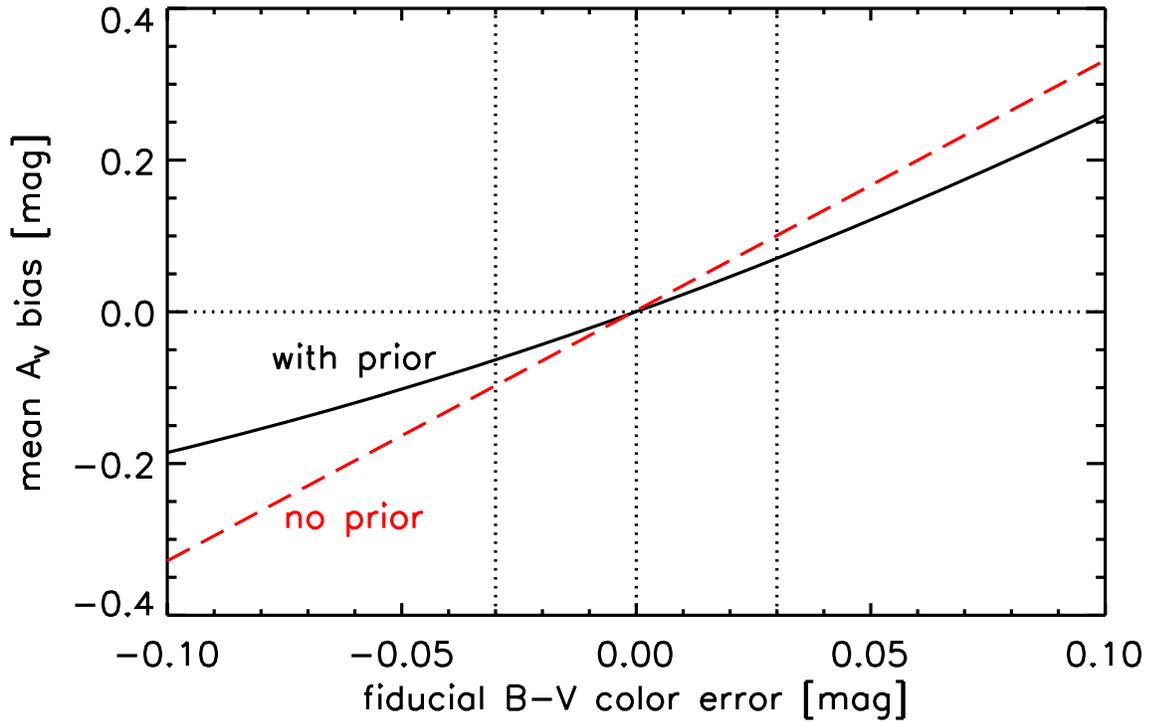}
\caption[Fiducial Color Evolution]{\singlespace Mitigating effect of
the extinction prior given an error or evolution in the mean intrinsic
color. The use of the prior (solid) reduces the extinction bias by a
factor of one-third relative to no prior (dashed, red in the
electronic edition). The dotted lines show a plausible range of
$\pm$0.03 mag in the mean intrinsic color. \label{fig-app2}}
\end{figure}

Another way extinction priors yield benefits besides improved distance
precision is by reining in errors in the knowledge of the intrinsic
properties of SN Ia.  The most relevant example is our estimate of
the fiducial, unreddened color of an SN Ia, used to estimate the color
excess and the net extinction.  As an example we consider in Figure
\ref{fig-app2} the extinction bias resulting from a misestimate of the
unreddened color (with the previous minimal but adequate color
measurements).  Without a prior the systematic extinction error is
exactly the color error multiplied by the reddening law. With the
prior, the extinction error becomes smaller because the valuable
knowledge about the likely extinction weighs in, reducing the effect
of the measurement inaccuracy.  Taking a realistic example of an 0.03
mag mis-estimate in the unreddened color, the bias is reduced from
0.10 mag without a prior to 0.07 mag with the prior.  However, the net
bias for most cosmological measurements will be negligible if the
color error (and extinction distribution) is independent of redshift.
For cosmological measurements, this color error may be usefully
thought of as a \emph{color evolution}.  Again, for such a color
evolution, the net bias is reduced by one-third with the use of an
extinction prior.

\begin{figure}
\includegraphics[width=6in]{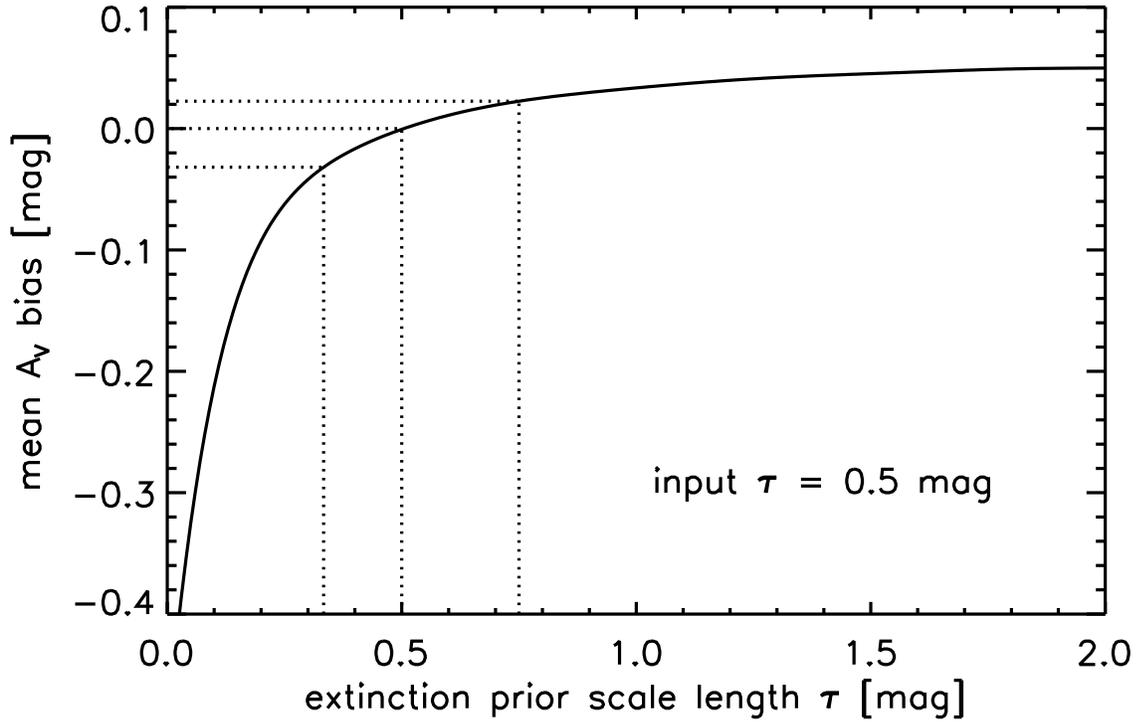}
\caption[Incorrect Prior Bias]{\singlespace Effect of using an
incorrect prior. The solid line shows the result of fitting the input
distribution (created with $\tau = 0.5$ mag) with different priors,
varying the prior scale length. The dotted lines highlight the results
of using the correct prior ($\tau = 0.5$ mag, no bias) and priors with
$\tau$ larger or smaller by a factor of 1.5. This plausible range of
variation (particularly in comparing different SN samples) could lead
to a bias of $\sim$0.03 mag. \label{fig-app3}}
\end{figure}

Nonetheless, a basic disadvantage of using priors is that biases may
result from the use of an incorrect prior.\footnote{A bias can also
occur when a moment other than the mean of the individual extinction
likelihood function is used as discussed by Perlmutter et al.~(1999).
For example, in Riess et al.~(1998a) the maximum likelihood (i.e., the
mode of posterior distribution) was initially used, which results in
an 0.06 mag bias as determined from our simulation.  However, because
the bias occurs for both low and high redshift SN Ia similarly, the
net bias (i.e., in the difference) was much smaller.  Modeling the
low-redshift SN Ia as having twice as many observations each with
twice the individual precision ($n_{\rm obs}$=10, $\sigma_{\rm
meas}=0.05$ mag) shows that the net bias is 0.02 mag in the distance
difference at low and high redshifts, much smaller than the 0.05 mag
error in the mean of the high-redshift sample.} As an example we show
in Figure \ref{fig-app3} the extinction bias as a function of the
exponential decay scale length assumed for the prior.  Values
differing from the input of $\tau=0.5$ mag result in an under- or
over-estimate of the mean extinction.  Using the results from Riello
\& Patat (2005) we can guess the approximate size of an error in
$\tau$ by imagining it was determined from galaxy-based simulations
and in so doing we have mis-estimated the relative importance of the
bulge to the total luminosity in late-type galaxies.  The dotted lines
show that a 30\% error in this ratio (and thus reducing or augmenting
the scale length by a factor of 1.5) results in a $\sim$0.03 mag bias.
In practice (i.e., for cosmological applications) a net bias would
only result from unaccounted for differences in $\tau$ between high
and low redshift such as may result from evolution or sample
selection.  Interestingly, we also find that the effect of taking a
so-called ``weak'' prior, that $\hat{p}(A_V)$ is constant for $A_V
\geq 0$, and zero for $A_V < 0$ is a worse choice, resulting in a bias
of 0.07 mag.  The reason is that the mean of such a prior is
unnaturally large, biasing the extinction estimate in that direction.
Likewise, assuming negligible extinction will result in an
underestimate, equal to the mean of the true distribution (which, for
an exponential, is the value of the scale length itself).  The lesson
is intuitively clear, when using a prior it is important to take the
best estimate (and estimate a systematic error from the range of good
guesses) as a seemingly "conservative" or "weak" version can be worse.

For a specific SN sample it is probably best to simulate the sources
and sizes of uncertainty depending on the data quality before
determining whether the use of an extinction prior offers more to gain
than to lose. Our simulations reveal nothing that intuition does not:
a good measurement is better than a bad prior, and a good prior will
improve imperfect measurements.

\end{document}